\documentclass[%
 reprint,
 amsmath,amssymb,
 aps,
 onecolumn,
 showkeys,
]{revtex4-2}

\usepackage{graphicx}
\usepackage{dcolumn}
\usepackage{bm}
\usepackage{amsmath, cancel}
\usepackage{fixmath}

\usepackage{tikz}
\usepackage{tikz-3dplot}
\usepackage{pgfplots}
\usetikzlibrary{decorations.markings,arrows.meta,calc,shapes,intersections,backgrounds,patterns.meta,bending,angles,decorations.pathreplacing,decorations.pathmorphing,pgfplots.colorbrewer,math}
\usepgfplotslibrary{fillbetween}
\usepackage{xifthen}
\usepackage[normalem]{ulem}
\usepackage{xcolor}
\usepackage[export]{adjustbox}
\newcommand\notsotiny{\@setfontsize\notsotiny\@vipt\@viipt}
\newcommand\notsosmall{\@setfontsize\notsosmall\@xpt\@xipt}

\definecolor{myc1}{HTML}{003049}
\definecolor{myc2}{HTML}{d62828}
\definecolor{myc3}{HTML}{f77f00}
\definecolor{myc4}{HTML}{6ca13b}
\definecolor{pnas}{HTML}{3273C3}

\definecolor{Q2}{HTML}{6ca13b}
\definecolor{Q1}{HTML}{6ca13b}
\definecolor{Q3}{HTML}{6ca13b}

\usepackage{pifont}

\let\tw\textwidth

\def\beq{\begin{equation}}
\def\eeq{\end{equation}}

\def\beq{\begin{equation}}
\def\eeq{\end{equation}}

\definecolor{olivegreen}{rgb}{0,0.6,0}

\def\drawline#1#2{\raise 2.5pt\vbox{\hrule width #1pt height #2pt}}

\def\trian{\raise 1.25pt\hbox{$\scriptstyle\triangle$}\nobreak}

\def\dtrian{\raise 1.25pt\hbox%
{$\scriptscriptstyle\bigtriangledown$}\nobreak}

\def\squar{\raise 1.25pt\hbox{$\scriptstyle\Box$}\nobreak}

\def\diamon{\raise 1.25pt\hbox{$\scriptstyle\diamond$}\nobreak}

\def\beq{\begin{equation}}
\def\eeq{\end{equation}}

\definecolor{C0}{HTML}{1F77B4} 
\definecolor{C1}{HTML}{FF7F0E}
\definecolor{C2}{HTML}{2CA02C}
\definecolor{C3}{HTML}{D62728}
\definecolor{C4}{HTML}{9467BD}
\definecolor{C5}{HTML}{8C564B}
\definecolor{C6}{HTML}{E377C2}
\definecolor{C7}{HTML}{7F7F7F}
\definecolor{C8}{HTML}{BCBD22}
\definecolor{C9}{HTML}{17BECF}

\let\bs\mathbold
\let\provc\providecommand
\newcommand{\mybq}{\bs{q}}        
\newcommand{\bq}{\bs{q}}        
\newcommand{\bQ}{\bs{Q}}        
\newcommand{\bi}{\bs{i}}        
\newcommand{\Is}{\tilde{\imath}}        
\newcommand{\Iss}{\Delta \mathcal{C}}                
\newcommand{\Inn}{\Delta \mathcal{N}}                

\newcommand{\myindexvar}[3]{%
  \ifthenelse{\isempty{#2}\and\isempty{#3}}%
  {#1}{#1_{#2}^{#3}}} 

\provc{\sq}[2]{\myindexvar{q}{#1}{#2}}       
\provc{\bsq}[2]{\myindexvar{\mybq}{#1}{#2}}  

\newcommand{\mybS}{\bs{S}}        
\newcommand{\mybA}{\bs{A}}        
\newcommand{\mybth}{\bs{\theta}}  
\newcommand{\mybJ}{\bs{J}}        

\provc{\bS}[2]{\myindexvar{\mybS}{#1}{#2}}     
\provc{\bA}[2]{\myindexvar{\mybA}{#1}{#2}}     
\provc{\bth}[2]{\myindexvar{\mybth}{#1}{#2}}   

\provc{\bWs}[2]{\myindexvar{\bs{W}}{#1}{#2}}               
\provc{\bWa}[2]{\myindexvar{\bs{V}}{#1}{#2}}   

\provc{\bJ}[2]{\myindexvar{\mybJ}{#1}{#2}}                  
\provc{\bJtilde}[2]{\myindexvar{\widetilde{\mybJ}}{#1}{#2}} 
\provc{\bJhat}[2]{\myindexvar{\widehat{\mybJ}}{#1}{#2}}     

\provc{\J}[2]{\myindexvar{J}{#1}{#2}}                  
\provc{\Jtilde}[2]{\myindexvar{\widetilde{J}}{#1}{#2}} 
\provc{\Jhat}[2]{\myindexvar{\widehat{J}}{#1}{#2}}     

\provc{\mun}{\mu}
\provc{\sgn}{\Xi}
\provc{\bmun}{\bs{\mu}}
\provc{\bsgn}{\bs{\Xi}}
\provc{\muopt}{\mu^*}
\provc{\sgopt}{\Xi^*}
\provc{\bmuopt}{\bs{\mu}^*}
\provc{\bsgopt}{\bs{\Xi}^*}
\provc{\mutar}{\hat{\mu}}
\provc{\sgtar}{\widehat{\Xi}}
\provc{\bmutar}{\hat{\bs{\mu}}}
\provc{\bsgtar}{\widehat{\bs{\Xi}}}

\provc{\relf}{\alpha}
\provc{\relfmu}{\relf_\mu}
\provc{\relfsg}{\relf_\xi}

\provc{\bths}{\mybth_s} 
\provc{\bthpa}{\mybth_p}
\provc{\bthaa}{\mybth_a}

\provc{\ths}{\theta_s} 
\provc{\thpa}{\theta_p}
\provc{\thaa}{\theta_a}

\provc{\shiftM}{\bs{a}}
\provc{\shiftmi}{a}

\definecolor{myc1}{HTML}{003049}
\definecolor{myc2}{HTML}{d62828}
\definecolor{myc3}{HTML}{f77f00}
\definecolor{myc4}{HTML}{6ca13b}

\usepackage[caption=false]{subfig}
\usepackage{hyperref}

\begin{document}

\preprint{APS/123-QED}

\title{Observational causality by states and interaction type for scientific discovery}

\author{\'Alvaro Mart\'inez-S\'anchez\textsuperscript{1,\dag}}
\author{Adri\'an Lozano-Dur\'an\textsuperscript{1,2}}

\affiliation{\textsuperscript{1}Department of Aeronautics and Astronautics, Massachusetts Institute of Technology, Cambridge, MA 02139, USA}
\affiliation{\textsuperscript{2}Graduate Aerospace Laboratories, California Institute of Technology, Pasadena, CA 91125, USA}
\affiliation{\vspace{0.2em}\rm{\textsuperscript{\dag}Corresponding author: \href{mailto:alvaroms@mit.edu}{alvaroms@mit.edu}}}





\begin{abstract}
Causality plays a central role in understanding interactions between
variables in complex systems. These systems often exhibit
state-dependent causal relationships, where both the strength and
direction of causality vary with the value of the interacting
variables. In this work, we introduce a state-aware causal inference
method that quantifies causality in terms of information gain about
future states. The effectiveness of the proposed approach stems from
two key features: its ability to characterize causal influence as a
function of system state, and its capacity to distinguish between
redundant and synergistic interactions. The method is validated across
a range of benchmark cases in which the direction and strength of
causality evolve in a prescribed manner with the state of the system.
We further demonstrate the applicability of our approach in two real
scenarios: the interaction between motions across scales in a
turbulent boundary layer, and the Walker circulation phenomenon in
tropical Pacific climate dynamics. Our results show that, without
accounting for state-dependent causality as well as redundant and
synergistic effects, traditional approaches to causal inference may
lead to incomplete or misleading conclusions.
\end{abstract}

\maketitle


\section*{Introduction}


Causality is a fundamental concept in understanding the interactions
between variables in complex systems. It provides insights into how
one variable influences another and guides the implementation of
meaningful changes within these systems~\cite{bunge1979}. In many
real-world scenarios, the nature of these interactions is not uniform
across all states of the system; they often vary depending on the
value of the variables involved. For example, in climate science, the
direction and magnitude of causality between atmospheric variables can
shift dramatically under different conditions, such as the wind flow
direction in the Pacific Region during El Niño or La Niña
events~\cite{elnino2019}. Similarly, in neuroscience, the causality
among brain signals often depends on the activation intensity of
surrounding regions~\cite{cordes2000}.  These examples highlight the
importance of developing methods that not only detect causal
relationships on average, but also capture how these interactions vary
with the state of the system---and consequently, over time. By
addressing this need, state-aware causal analysis can drive progress
in a wide range of scientific and engineering fields, including
climate science~\cite{Runge2023},
neuroscience~\cite{neuroscience2016}, economics~\cite{economic2008},
epidemiology~\cite{epidemiology2005}, social
sciences~\cite{social2010}, and fluid dynamics~\cite{lozano2020,
  martinez2023}.

To date, most causal inference methods offer an estimate of the
average causal strength across all system states, without providing
insight into how causality varies at the level of individual
states. One such method is Convergent Cross Mapping
(CCM)~\cite{sugihara2012} and its variants~\cite{extendedccm2015,
  mccm2015, lccm2021, causalccm}, which infer causality by analyzing
connections in the reconstructed attractor of the underlying dynamical
system using different combinations of observed variables. A similar
limitation arises in the Peter--Clark Momentary Conditional
Independence (PCMCI) framework~\cite{runge2019pcmci} and its
extensions~\cite{runge2020pcmci+, Gerhardus2020LPCMCI,
  saggioro2020RPCMCI}, which focus on identifying a minimal
conditioning set that includes the causal parents of the target
variable. Traditional approaches, such as Granger
causality~\cite{granger1969}, also suffer from this limitation, as
they assess causality by testing whether the past values of a source
variable improve the prediction of a target variable in the future,
but only in an average sense over all system states. Feature-ranking
methods, such as the distributed information
bottleneck~\cite{Murphy2024}, have gained popularity for isolating the
most informative variables in systems to enhance predictive
accuracy. However, these approaches do not explicitly address
causality, let alone how causal influence may vary with the state of
the system.

Among the methods that offer a decomposition of causality by states,
one of the most intuitive formulations relies on the concept of
interventions~\cite{pearl2000, Eichler2013}. Interventions actively
modify one variable to observe its effect on another. A family of
approaches within this framework is based on the concept of effective
information~\cite{Tononi2003, Hoel2013, hoel2016}, which quantifies
the strength of causal interactions by measuring the change in the
state of the system when a variable is replaced with maximally
uncertain noise. These methods provide a state-dependent decomposition
of causality by selectively intervening under specific conditions
(e.g., intervening on a variable $A$ only when $A=1$ to measure the
causal effect from that state).  However, when data are gathered from
physical experiments, establishing causality through interventions may
be highly challenging, impractical~\cite{surd}, or even unethical in
fields such as neuroscience or climate
science~\cite{runge2019pcmci}. Furthermore, the notion of causality
based on interventions raises important questions about the type of
intervention that should be introduced, and whether such interventions
might affect the outcome by forcing the system out of its natural
attractor.

Information theory~\cite{shannon1948} has also become a foundational
framework for quantifying causality. The motivation for employing
information-theoretic approaches stems from the recognition of
information as a fundamental property of physical
systems~\cite{landauer1996, lozano2022}. Early applications of
information-theoretic causality involved the use of conditional
entropies to define directed information~\cite{massey1990,
  kramer1998}. One of the most recognized developments in this area is
the introduction of transfer entropy (TE)~\cite{schreiber2000}, which
quantifies how knowledge of the past states of one variable reduces
uncertainty about the future state of another.  Subsequent efforts to
refine TE led to the development of conditional transfer entropy
(CTE)~\cite{verdes2005, lizier2008, barnett2009, lizier2010, cte2016},
aimed at addressing multivariate analyses. Other information-theoretic
approaches, inspired by dynamical systems theory~\cite{liang2006,
  liang2016if, liang2008if, liang2013if}, quantify causality as the
amount of information flowing from one process to another, as dictated
by the governing equations of the system. Similar to previous methods,
these information-based approaches provide only an average measure of
causal strength and do not offer insight into which specific states
contribute most to the observed causality.

Several extensions of TE have been developed to account for
state-dependent information transfer~\cite{Wang2022}, although many
are formulated primarily to quantify information dynamics rather than
to perform causal inference. One of these extensions is local
TE~\cite{lizier2008, lizier2010, jidt}, which evaluates the
information transfer between specific states of the (past) source and
(future) target variables by analyzing the expectation term in the
original TE formulation. This method preserves several desirable
properties, including the ability to recover the standard TE when
averaged over all states.  Local TE can take negative values,
reflecting a combination of contributions from both informative states
(positive values) and misinformative states (negative
values). Specific TE was later introduced~\cite{Darmon2017specific},
refining the interpretation of state-dependent information transfer by
projecting local TE onto the states of the target variable. This
projection yields a non-negative decomposition, enhancing
interpretability; however, it considers only the past states of the
source variables.  Other approaches, such as ensemble
TE~\cite{Martini2011, Wibral2013, Wollstadt2014, GomezHerrero2015,
  Zhu2022} or time-varying Liang--Kleeman information transfer
\cite{tvlk, zhou2024}, explore information transfer across varying
time windows to detect shifts in causal structure over time. More
recently, a local version of Granger
causality~\cite{Stramaglia2021local}---viewed as the linear
counterpart to local TE---has been proposed. This formulation offers a
computationally efficient alternative for analyzing state-dependent
causal interactions in linear systems with limited data.

While the methods above have advanced our understanding of
state-dependent interactions, they are unable to distinguish between
synergistic, unique, and redundant causal influences. It has been
shown that the inability to disentangle these components can obscure
the true causal structure of a system, often leading to spurious or
misleading conclusions~\cite{surd}.  For example, consider the effect
of two genes, $A$ and $B$, on the expression of eye color, $C$. These
interactions can manifest in several ways: \emph{synergistic}, where
both genes must be active together to produce a specific eye
color---such as green---that neither gene can induce on its own;
\emph{unique}, where gene $A$ contributes to blue eye color via a
pathway independent of gene $B$; and \emph{redundant}, where either
gene alone is sufficient to produce brown eyes due to overlapping
biological pathways.  Without a method capable of isolating these
different types of causal contributions, such complexities remain
hidden, and critical insights into causal relationships are missed.
Recently, the synergistic-unique-redundant decomposition of causality
(SURD)~\cite{surd} was introduced, an information-theoretic method for
causal inference that explicitly accounts for the distinct
contributions of synergistic, unique, and redundant causal
influences. Other information-theoretic approaches have also been
recently developed to disentangle synergistic and redundant effects in
transfer entropy \cite{stramaglia2024,korenek2025}. However, neither
SURD nor these earlier methods were designed to capture
state-dependent variations in causality within the systems.  As a
result, these methods cannot distinguish between scenarios in which a
specific value of gene $A =$ `$a$' exerts a unique causal influence on
a particular state `$c$' of the condition $C$, and those in which its
effect is redundant or synergistic with a specific value `$b$' of gene
$B$.

In this work, we present a causal inference method that simultaneously
accounts for synergistic, unique, and redundant causal influences,
while also decomposing these contributions at the level of individual
system states.  We show that this approach outperforms existing causal
inference methods, including the original SURD formulation, by
providing a more detailed, state-aware characterization of causal
interactions.  We validate the method across a range of benchmark
problems in which causal pathways depend on the system state in a
controlled and known manner.  We also demonstrate the applicability of
our approach in two real-world scenarios---fluid dynamics and climate
science---highlighting its ability to uncover state-specific causal
relationships.

\section*{Results}

\subsection*{State-dependent and interaction-specific causality}

Consider the collection of $N$ time-evolving source variables given by
the vector $\bQ = [Q_1(t), Q_2(t), \ldots, Q_N(t)]$. For example,
$Q_i$ may represent the activity level of a specific brain region over
time, a binary variable indicating the presence or absence of an event
(e.g., whether it rained on a particular day), or the daily average of
a stock market index (e.g., S\&P 500). The components of $\bQ$ are
observables and are treated as random variables. A particular
realization of $\bQ$ is denoted by $\bq$, representing a specific
state (i.e., value).  Following the examples above, $q_i$ may
correspond to a particular level of neural activity, one of the two
states (0 or 1) of the binary variable indicating whether it rained or
not, or a specific value of the stock market index (e.g., the closing
value on a given day). These states are defined by the user according
to the specific requirements of the problem. Our objective is to
quantify the causal influence of the states of the source, $\bq$, on
the future state of a target variable, denoted as $q_j^+ = q_j(t +
\Delta T)$, where $\Delta T > 0$ represents the future time interval
over which causal influence is assessed. The variable $Q_j$ is often
included in $\bQ$, but this is not required in general.

Our approach is formulated in three steps. First, following the
principle of \emph{forward-in-time information propagation}---i.e.,
information flows only toward the future~\cite{lozano2022}---we
quantify causality among variables in terms of increments of
information. Second, we separate these increments into causal and
non-causal components. Third, we further decompose the causal
increments into distinct types of interactions: synergistic, unique,
and redundant contributions.

For step one, we adopt the definition of causality proposed in
Ref.~\cite{surd}. In this framework, causality is quantified as the
increase in information about each future state $q_j^+$ gained by
observing individual or groups of past states $\bq$. The information
content in $Q_j^+$ is measured by Shannon information (or
entropy)~\cite{shannon1948}, denoted as $H(Q_j^+)$, which represents
the average number of bits needed to determine the value of $Q_j^+$
unambiguously. Shannon information can also be interpreted as a
measure of uncertainty: highly uncertain processes (high entropy)
yield greater information gain when their outcomes are revealed,
whereas completely deterministic processes (zero entropy) provide no
new information upon observation.

For the second step, we decompose the information in $H(Q_j^+)$ into a
sum of information increments contributed by each past state of the
observable vector $\bQ$:
\begin{equation}
  \label{eq:conservation_info}
  H(Q_j^+) = \sum_{q_j^+\in Q_j^+}\sum_{\bq \in \bQ} \left[ \Iss (q_j^+;\bq)
  + \Inn(q_j^+;\bq) \right]  + \Delta I_{\text{leak}\rightarrow j},
\end{equation}
where the terms $\Iss(q_j^+; \bq)>0$ and $\Inn(q_j^+; \bq)\leq0$
represent the causal and non-causal contributions, respectively, for a
given source and target state.  The term $\Delta I_{\text{leak}
  \rightarrow j}$ is referred to as the causality
leak~\cite{lozano2022}. The mathematical expression for each of the
terms in Eq.~\eqref{eq:conservation_info} is given in Methods and its
derivation is detailed in Supplementary Materials. Here, we provide an
interpretation of each term:
\begin{itemize}
\item The causal contribution $\Iss(q_j^+; \bq)>0$ is the
  \emph{positive} increment of information that a source state $\bq$
  provides about a future target state $q_j^+$, resulting in a
  decrease in uncertainty about $q_j^+$.
\item The non-causal contribution $\Inn(q_j^+; \bq)\leq0$ is the
  \emph{negative} increment of information obtained by observing
  $\bq$, which leads to increased uncertainty about the future state
  $q_j^+$.
\item The causality leak $\Delta I_{\text{leak} \rightarrow j}\geq 0$
  is the effect of \emph{unobserved} variables that influence $Q_j^+$
  but are not contained in $\bQ$, i.e., the information of $Q_j^+$
  that remains unexplained after having observed $\bQ$.
\end{itemize}

To illustrate the distinction between $\Iss(q_j^+; \bq)$ and
$\Inn(q_j^+; \bq)$, consider the task of forecasting whether it will
rain tomorrow, denoted by $Q_\text{rain}^+$, where $q_\text{rain}^+ =
1$ indicates rain and $q_\text{rain}^+ = 0$ indicates no rain. Suppose
we use an atmospheric pressure sensor as the observable, $Q_p$, which
can take on two values: low ($q_p = 0$) or high ($q_p = 1$).
Within our causality framework, if observing low pressure ($q_p =
0$) provides a positive increment of information (i.e., reduces
uncertainty) about the state $q_\text{rain}^+ = 1$, then $q_p = 0$ is
considered causal and contributes to $\Iss(q_\text{rain}^+; q_p)$.
Conversely, if observing $q_p = 0$ yields a negative increment of
information (i.e., increases uncertainty) about $q_\text{rain}^+ = 0$,
then $q_p = 0$ is regarded as non-causal and contributes to
$\Inn(q_\text{rain}^+; q_p)$.

The final step is to decompose $\Iss(q_j^+; \bq)$ and $\Inn(q_j^+;
\bq)$ into distinct types of interactions---namely, redundant, unique,
and synergistic components~\cite{surd2024code}:
\begin{equation} \label{eq:surd}
    \Iss(q_j^+;\bq) =
    \sum_{\bi\in\mathcal{P}}\Iss_{\boldsymbol{i}\rightarrow j}^R +
    \sum_{i=1}^N\Iss_{i\rightarrow j}^U +
    \sum_{\bi\in\mathcal{P}} \Iss_{\boldsymbol{i}\rightarrow j}^S,
\end{equation}
where $\Iss_{\bi \rightarrow j}^R$, $\Iss_{i\rightarrow j}^U$, and
$\Iss_{\bi \rightarrow j}^S$ are the redundant, unique, and
synergistic causalities, respectively, from the source states $\bq$ to
target state $q_j^+$.  Unique causalities are associated with
individual components of $\bq$, whereas redundant and synergistic
causalities arise from groups of components from $\bq$. Consequently,
the set $\mathcal{P}$ contains all combinations involving more than
one variable. For instance, for $N=2$, Eq.~\eqref{eq:surd} reduces to
$\Iss(q_j^+;\bq) = \Iss_{12\rightarrow j}^R + \Iss_{1\rightarrow j}^U
+ \Iss_{2\rightarrow j}^U + \Iss_{12 \rightarrow j}^S$.  Similar to
the causal and non-causal contributions, we provide an interpretation
of redundant, unique and synergistic causalities, with the
mathematical details available in the Methods section and the
Supplementary Materials.
\begin{itemize}
  \item Redundant causality from $\bq_{\bi} = [q_{i_1}, q_{i_2},
    \ldots]$ to $q_j^+$ (denoted by $\Iss^R_{\bi \rightarrow j}$)
    refers to the causal influence shared among all individual states
    in $\bq_{\bi}$, where $\bq_{\bi}$ is a subset of $\bq$. Redundant
    causality occurs when observing any of the states $q_{i_1}$,
    $q_{i_2}$, etc., provides identical amount of information about
    the outcome $q_j^+$. Consider again the example of rain
    forecasting: we might use two pressure sensors placed in close
    proximity, both offering redundant causal information---i.e., each
    is equally useful for predicting whether it will rain or not.
  \item Unique causality from $q_i$ to $q_j^+$ (denoted by $\Iss^U_{i
    \rightarrow j}$) is the causal influence from $q_i$ that cannot be
    obtained from any other individual variable different from
    $Q_i$. This occurs when observing $q_i$ yields more information
    about $q_j^+$ than observing all the states of any other individual
    variable. In the previous example, this will occur when one
    pressure sensor provides more information about the likelihood of
    rain than the other.
  \item Synergistic causality from $\bq_{\bi} = [q_{i_1}, q_{i_2},
    \ldots]$ to $q_j^+$ (denoted by $\Iss^S_{\bi \rightarrow j}$)
    refers to the causal influence that arises from the joint effect
    of the states $\bq_{\bi}$. Synergistic causality occurs when more
    information about $q_j^+$ is gained by observing the collection of
    states $[q_{i_1}$, $q_{i_2},...]$ simultaneously than by observing
    all the states of subsets of $\bq_\bi$ individually. In the rain
    forecasting scenario, a synergy would take place when the
    information gained by measuring with two pressure sensors
    simultaneously is greater than the information provided by each
    sensor individually.
\end{itemize}
An analogous decomposition (redundant, unique, and synergistic)
applies to the non-causal term $\Inn(q_j^+; \bq)$. However, in this
case, the information gain is negative—acquiring this knowledge
actually increases uncertainty about future outcomes. Here, we focus
on the causal contribution $\Iss(q_j^+; \bq)$. The non-causal
components also offer valuable insight, with further discussion and
examples provided in the Supplementary Materials.

\begin{figure}
    \centering
    \includegraphics[width=\linewidth]{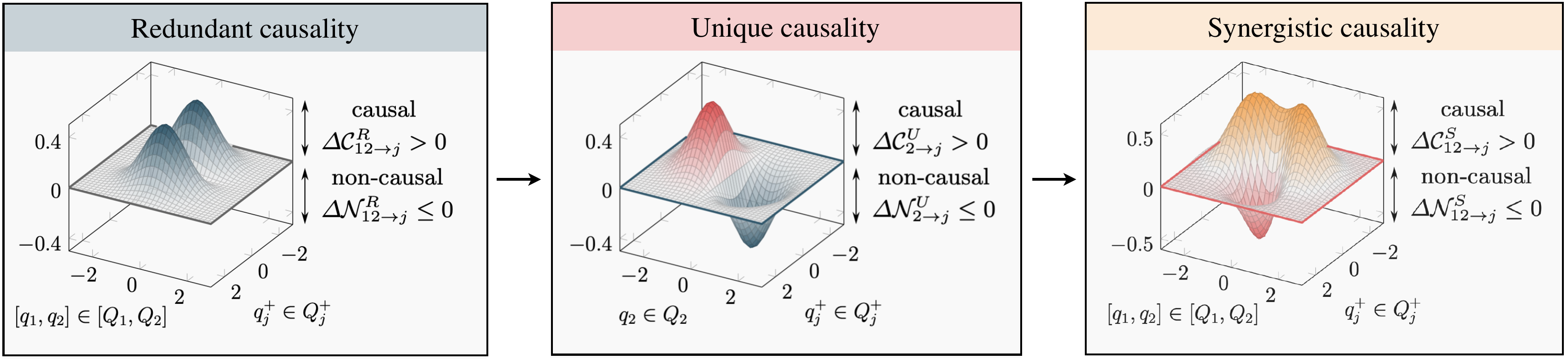}
    \caption{Diagram of the decomposition of causal dependencies
      between a vector of observed states $\bq = [q_1, q_2]$ and a
      future target state $q_j^+$ into synergistic (S), unique (U),
      and redundant (R) components---shown in orange, red, and blue,
      respectively.  Note that redundant and synergistic interactions
      depend on all three states, $[q_1, q_2, q_j^+]$, but the source
      $[q_1, q_2]$ is shown along a single axis for schematic
      simplicity.  The causal contributions are assigned following the
      order: redundant $\rightarrow$ unique $\rightarrow$ synergistic.
      As a result, causal states are color-coded according to the type
      of causality they contribute to, whereas non-causal states are
      shown using the color of the preceding causality level. For
      example, in the unique causality panel, causal states are shown
      in red, while non-causal states are shown in blue. This approach
      prevents the double counting of causal influences.
        } 
    \label{fig:surd-states}
\end{figure}

By definition, causal contributions are assigned following the order:
redundant $\rightarrow$ unique $\rightarrow$ synergistic. Redundant
causalities are identified first. The next increments of information
that are not redundant are attributed to unique causality. Finally,
the remaining causal influence is classified as synergistic. This
implies that when contributions are labeled as causal (information
gain) or non-causal (information loss), the gain/loss is referred to
the previously identified redundant, unique, or synergistic
causality. For example, in a system with one redundant, one unique,
and one synergistic contribution: redundant causality reflects a gain
relative to the information in the target $q_j^+$; unique causality
reflects a gain relative to the redundant component; and synergistic
causality reflects a gain relative to the unique one. This process is
illustrated in the diagrams in Fig.~\ref{fig:surd-states}, where the
causal and non-causal contributions between the source states $q_1$
and $q_2$ and the target state $q_j^+$ are decomposed into its
redundant, unique, and synergistic components. This approach ensures
that causal contributions are not double-counted if they have already
been accounted for in the previous type of interaction.

Our approach also enables the representation of causality as a
function of time. The instantaneous value of causality at each time
step is obtained by identifying the current state of the system and
extracting the corresponding state-dependent value of $\Iss^R_{\bi
  \rightarrow j}$, $\Iss^U_{i \rightarrow j}$, and $\Iss^S_{\bi
  \rightarrow j}$.  As a result, it becomes possible to visualize the
dynamic evolution of causal interactions, capturing temporal reversals
and highlighting periods dominated by different types of causal
contributions.

Finally, we define the cumulative causality over all possible states,
after accounting for the uncertainty introduced by non-causal
contributions, as $\Delta I^{\alpha}_{\boldsymbol{i}\rightarrow j} =
\sum_{q_j^+\in Q_j^+}\sum_{\bq \in \bQ} \left[
  \Iss^{\alpha}_{\boldsymbol{i}\rightarrow j} +
  \Inn^{\alpha}_{\boldsymbol{i}\rightarrow j} \right]$, for $\alpha\in
    [R,U,S]$.
These cumulative increments of information directly yield the
causalities from SURD proposed in Ref.~\cite{surd}. This is a key
feature of our method, as SURD causality has been shown to outperform
other definitions of causality across a wide range of benchmark
cases. 
%

\subsection*{Validation}
\label{subsec:validation}

We demonstrate our causal inference approach using validation cases
representative of two common types of causal relationships:
source-dependent and target-dependent causality. Each system consists
of two source variables, $Q_1$ and $Q_2$, and a target variable,
$Q_2^+$, designed such that the direction of causality varies with the
intensity of either the source or the target. These benchmark cases
are constructed so that the direction of causality for each state is
known \emph{a priori}, allowing us to assess the accuracy and
reliability of the method.  Importantly, the cases are also designed
to exhibit qualitatively similar SURD causalities---i.e., comparable
cumulative increments $\Delta I^{\alpha}_{\boldsymbol{i}\rightarrow
  2}$---while the underlying causal flow reveals an entirely opposite
structure as a function of the system state, a distinction that can
only be revealed by a causality-by-state decomposition.  
Additional
validation cases are presented in the Supplementary Materials.


\subsubsection*{Source-dependent causality}

\begin{figure}
    \centering
    \begin{minipage}{\textwidth}
        $
          q_2(n+1) =  
          \begin{cases}
            \sin\left[q_1(n)\right] + 0.1 w_2(n) & \text{if } q_1(n) > 0 \\
            q_2(n) + 0.1 w_2(n) & \text{otherwise}
          \end{cases}
        $
    \end{minipage}
    \vspace{0.0075\textwidth}

    \scalebox{1.25}{ 
    \hspace{-2.1cm}
    \begin{minipage}{\textwidth}
    \hspace{-0.3cm}
    \begin{minipage}{0.24\textwidth}
    \begin{tikzpicture}
        \node[anchor=north west] (img) at (0, 0) {\includegraphics[width=\linewidth]{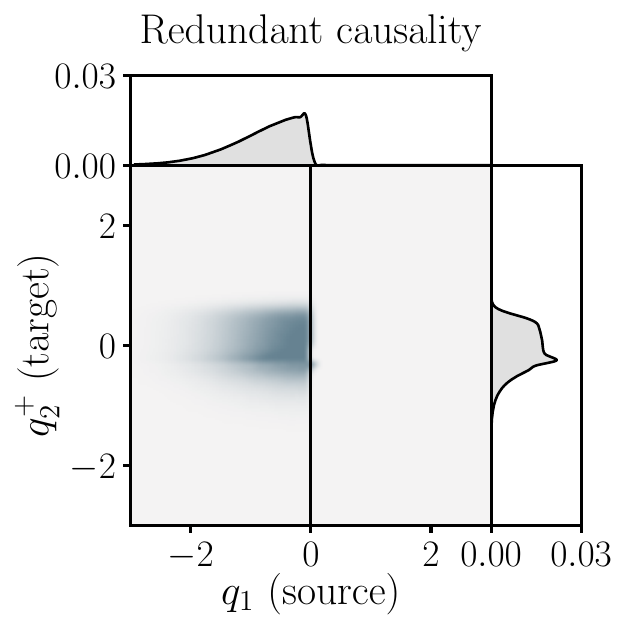}};
        \node[anchor=north east, align=center] at (2.1, -3.1) {\tiny{causal}};
        \node[anchor=north east, align=center] at (3.55, -3.1) {\tiny{non-causal}};
        \node[anchor=south, font=\small, fill=white, text=black, inner sep=2pt, rounded corners, minimum width=0.2\textwidth,minimum height=6mm] at ([yshift=-5.6mm,xshift=-1mm]img.north) {$\quad$Redundant R12$\quad$};
    \end{tikzpicture}
    \end{minipage}
    \hspace{-0.3cm}
    \begin{minipage}{0.23\textwidth}
    \begin{tikzpicture}
        \node[anchor=north west] (img) at (0, 0) {\includegraphics[width=\linewidth, trim=0 0 0 0, clip]{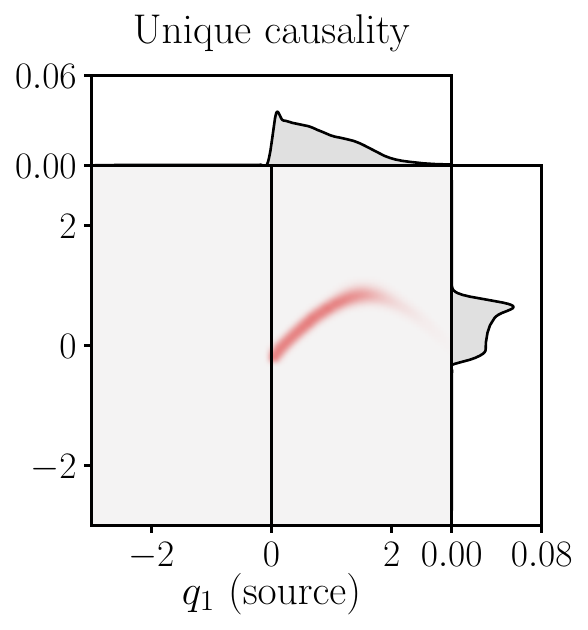}};
        \node[anchor=north east, align=center] at (2.075, -3.15) {\tiny{non-causal}};
        \node[anchor=north east, align=center] at (3.1, -3.15) {\tiny{causal}};
        \node[anchor=south, font=\small, fill=white, text=black, inner sep=2pt, rounded corners, minimum width=0.2\textwidth,minimum height=6mm] at ([yshift=-5.8mm,xshift=-3mm]img.north) {$\quad$Unique U1$\quad$};
    \end{tikzpicture}
    \end{minipage}
    \hspace{-0.3cm}
    \begin{minipage}{0.23\textwidth}
    \begin{tikzpicture}
        \node[anchor=north west] (img) at (0, 0) {\includegraphics[width=\linewidth, trim=0 0 0 0, clip]{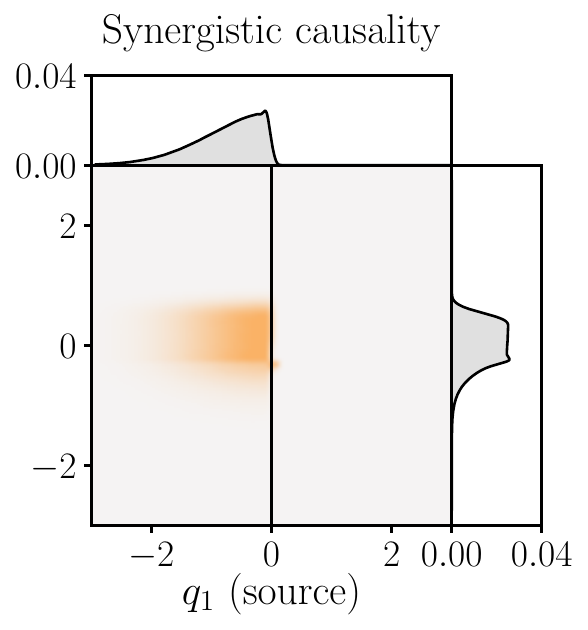}};
        \node[anchor=north east, align=center] at (1.8, -3.15) {\tiny{causal}};
        \node[anchor=north east, align=center] at (3.35, -3.15) {\tiny{non-causal}};
        \node[anchor=south, font=\small, fill=white, text=black, inner sep=2pt, rounded corners, minimum width=0.2\textwidth,minimum height=6mm] at ([yshift=-5.8mm,xshift=-2mm]img.north) {$\quad$Synergistic S12$\quad$};
    \end{tikzpicture}
    \end{minipage}
    \hspace{-0.3cm}
    \begin{minipage}{0.112\textwidth}
    \vspace{-0.16\textwidth}
    \begin{tikzpicture}
        \node[anchor=north west] (img) at (0, 0) {\includegraphics[width=\linewidth]{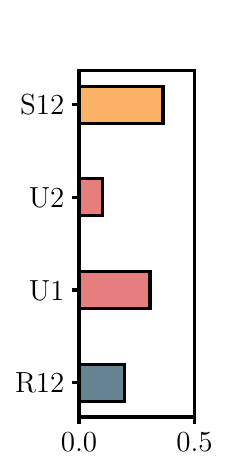}};
        \node[anchor=south, font=\small, fill=white, text=black, inner sep=2pt, rounded corners, minimum width=0.2\textwidth,minimum height=6mm] at ([yshift=-6.5mm,xshift=2mm]img.north) {SURD};
    \end{tikzpicture}
    \end{minipage}
    \vspace{0.01\textwidth}
    \begin{minipage}{\textwidth}
        \hspace{0.025\linewidth}
        \includegraphics[height=0.0945\linewidth,trim=0 45 0 0, clip]{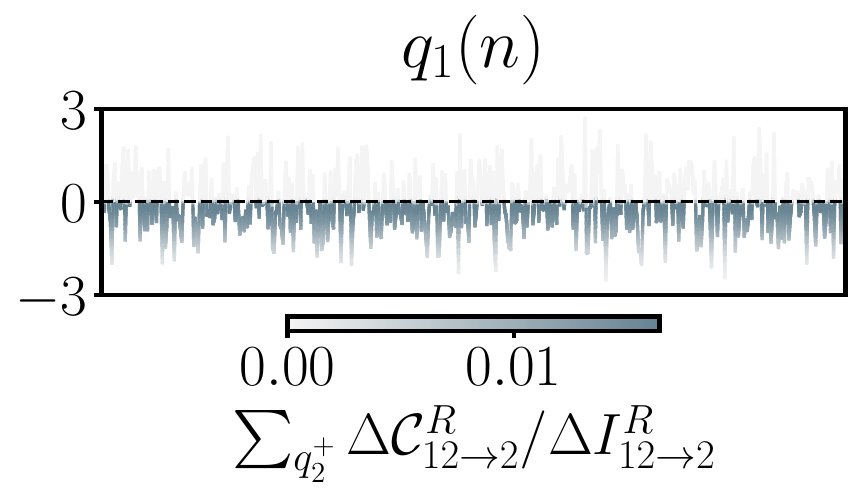}
        \hspace{0.01\linewidth}
        \includegraphics[height=0.0945\linewidth,trim=0 45 0 0, clip]{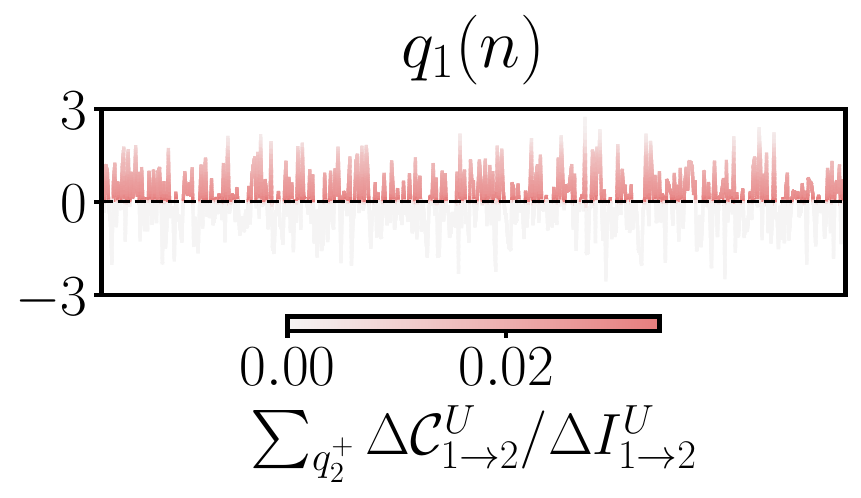}
        \hspace{0.01\linewidth}
        \includegraphics[height=0.0945\linewidth,trim=0 45 0 0, clip]{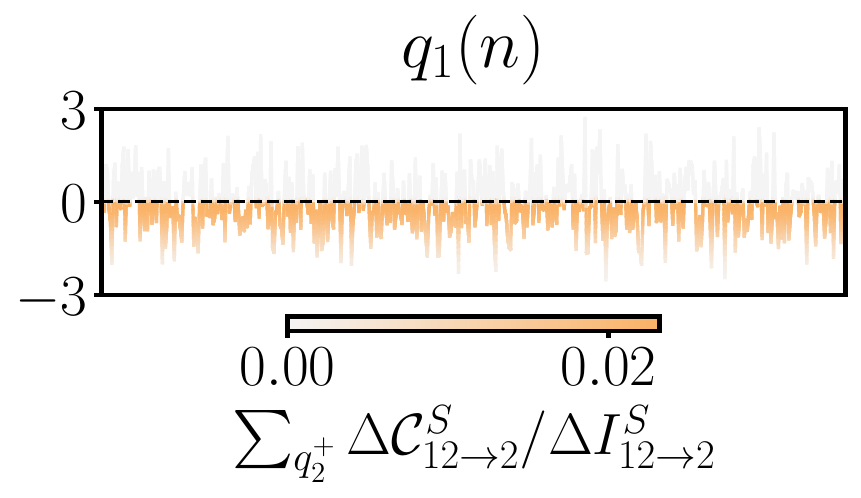}
        \hspace{0.12\linewidth}
    \end{minipage}
    \end{minipage}
    }

    \caption{State-dependent decomposition of causality for the
      source-dependent case. From left to right, the panels show:
      state-dependent redundant ($\Iss^R_{12\to 2}$), unique
      ($\Iss^U_{1\to 2}$), and synergistic ($\Iss^S_{12\to 2}$) causal
      contributions in blue, red, and orange, respectively; and the SURD
      causalities to the target variable $Q_2^+$.  The
      notation follows: U1 denotes the unique SURD causality from
      $Q_1$ to $Q_2^+$, i.e., $\Delta I^U_{1\to 2}$, with analogous
      definitions for other components (i.e., R and S).  All SURD and
      state-dependent causal contributions are normalized by the
      mutual information $I(Q_2^+; Q_1, Q_2)$.  The causal maps for
      redundant and synergistic components show averages over all
      states of $q_2$.  In each panel, the bottom row displays the
      temporal evolution of $q_1(n)$, color-coded according to the
      corresponding instantaneous state-dependent causal
      contribution.}
    \label{fig:example-contributions}
\end{figure}

\begin{figure}
    \begin{minipage}{\textwidth}
        $
          q_2(n+1) =  
          \begin{cases}
            q_1(n)\sin\left[q_1(n)\right]+0.1w_2(n) & \text{if } q_1(n)\sin\left[q_1(n)\right]+0.1w_2(n) > 0 \\
            q_2(n) + 0.1 w_2(n) & \text{otherwise}
          \end{cases}
        $
    \end{minipage}
    \vspace{0.0075\textwidth}

    \scalebox{1.25}{ 
    \hspace{-2.1cm}
    \begin{minipage}{\textwidth}
    \hspace{-0.3cm}
    \begin{minipage}{0.24\textwidth}
    \begin{tikzpicture}
        \node[anchor=north west] (img) at (0, 0) {\includegraphics[width=\linewidth]{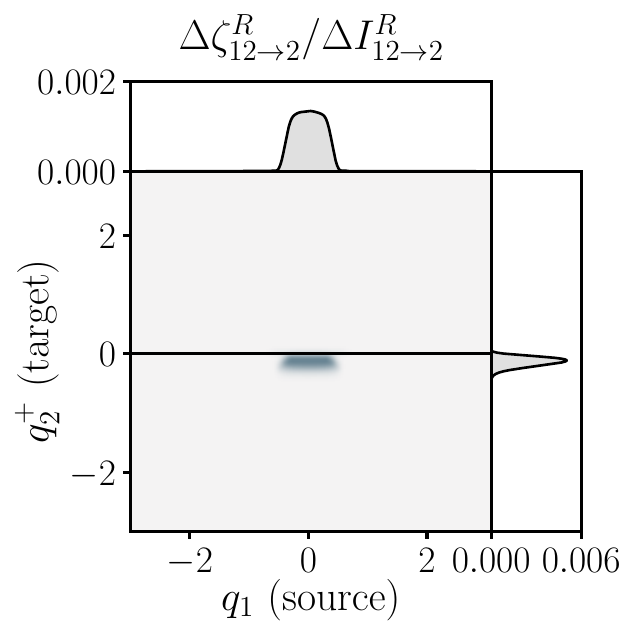}};
        \node[anchor=north east, align=center] at (2.7, -3.1) {\notsotiny{causal}};
        \node[anchor=north east, align=center] at (2.95, -1.5) {\notsotiny{non-causal}};
        \node[anchor=south, font=\small, fill=white, text=black, inner sep=2pt, rounded corners, minimum width=0.2\textwidth,minimum height=6mm] at ([yshift=-5.6mm,xshift=-1mm]img.north) {$\quad$Redundant R12$\quad$};
    \end{tikzpicture}
    \end{minipage}
    \hspace{-0.3cm}
    \begin{minipage}{0.23\textwidth}
    \begin{tikzpicture}
        \node[anchor=north west] (img) at (0, 0) {\includegraphics[width=\linewidth, trim=0 0 0 0, clip]{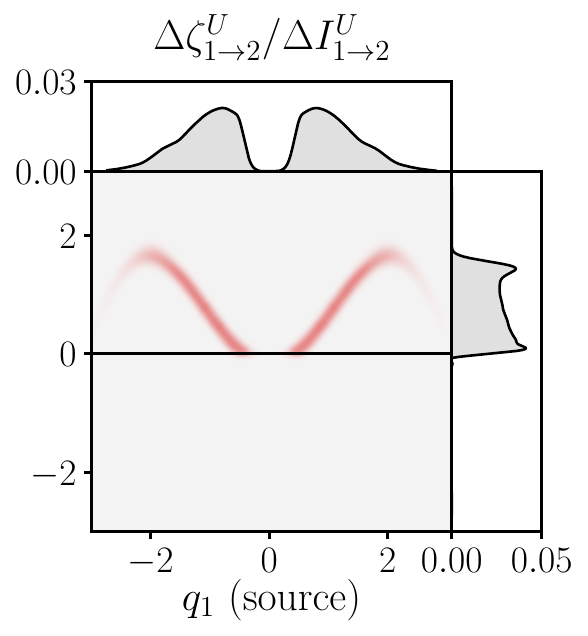}};
        \node[anchor=north east, align=center] at (2.55, -3.09) {\notsotiny{non-causal}};
        \node[anchor=north east, align=center] at (2.3, -1.49) {\notsotiny{causal}};
        \node[anchor=south, font=\small, fill=white, text=black, inner sep=2pt, rounded corners, minimum width=0.2\textwidth,minimum height=6mm] at ([yshift=-5.8mm,xshift=-3mm]img.north) {$\quad$Unique U1$\quad$};
    \end{tikzpicture}
    \end{minipage}
    \hspace{-0.3cm}
    \begin{minipage}{0.23\textwidth}
    \begin{tikzpicture}
        \node[anchor=north west] (img) at (0, 0) {\includegraphics[width=\linewidth, trim=0 0 0 0, clip]{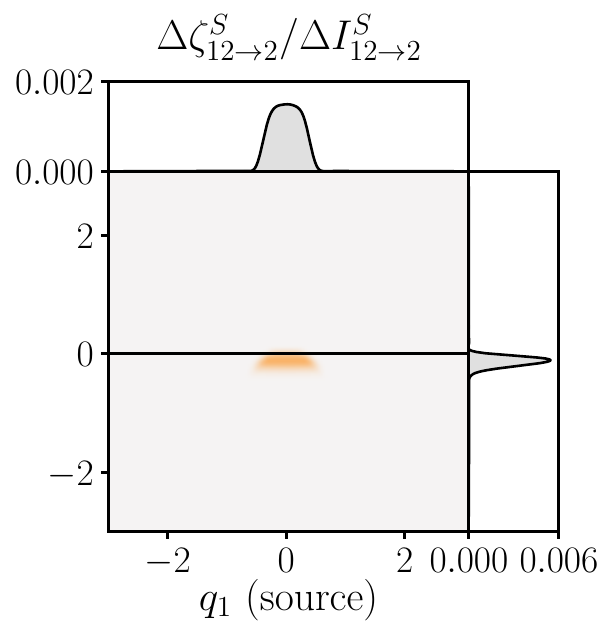}};
        \node[anchor=north east, align=center] at (2.3, -3.0) {\notsotiny{causal}};
        \node[anchor=north east, align=center] at (2.55, -1.49) {\notsotiny{non-causal}};
        \node[anchor=south, font=\small, fill=white, text=black, inner sep=2pt, rounded corners, minimum width=0.2\textwidth,minimum height=6mm] at ([yshift=-5.8mm,xshift=-2mm]img.north) {$\quad$Synergistic S12$\quad$};
        \node[anchor=south, font=\small, fill=white, text=black, inner sep=2pt, minimum width=0.2\textwidth,minimum height=2mm, rotate=90] at ([yshift=-23mm,xshift=-18mm]img.north) {$\quad\quad\quad\quad\quad$};
    \end{tikzpicture}
    \end{minipage}
    \hspace{-0.3cm}
    \begin{minipage}{0.095\textwidth}
    \vspace{-0.16\textwidth}
    \begin{tikzpicture}
        \node[anchor=north west] (img) at (0, 0) {\includegraphics[width=\linewidth]{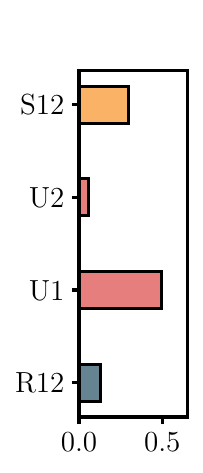}};
        \node[anchor=south, font=\small, fill=white, text=black, inner sep=2pt, rounded corners, minimum width=0.2\textwidth,minimum height=6mm] at ([yshift=-6.5mm,xshift=2mm]img.north) {SURD};
    \end{tikzpicture}
    \end{minipage}
    \vspace{0.01\textwidth}
    \begin{minipage}{\textwidth}
        \hspace{0.0375\linewidth}
        \includegraphics[height=0.094\linewidth,trim=0 45 0 0, clip]{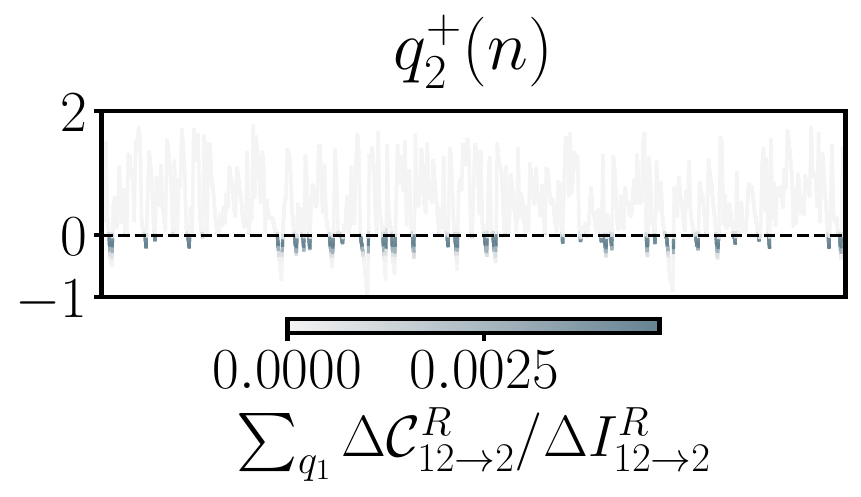}
        \hspace{0.01\linewidth}
        \includegraphics[height=0.094\linewidth,trim=0 45 0 0, clip]{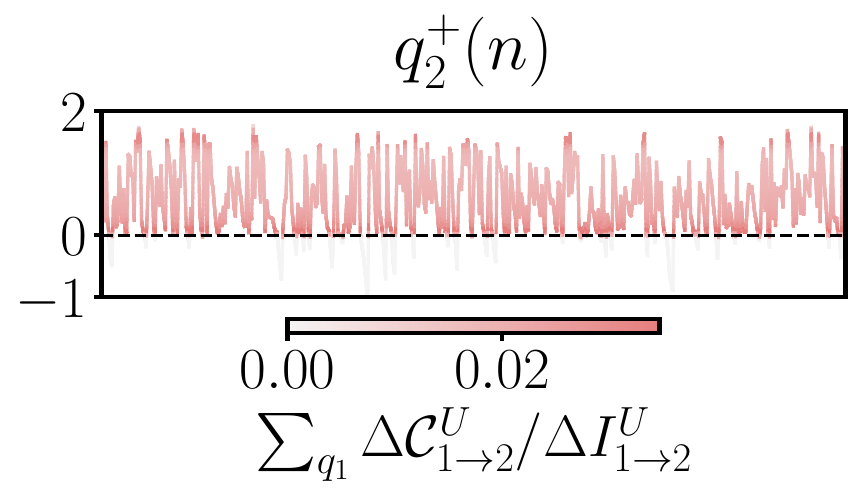}
        \hspace{0.01\linewidth}
        \includegraphics[height=0.094\linewidth,trim=0 45 0 0, clip]{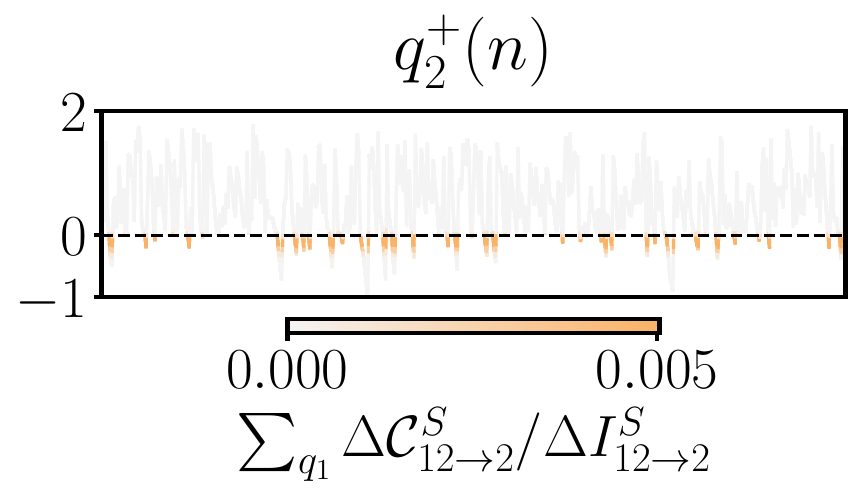}
        \hspace{0.12\linewidth}
    \end{minipage}
    \end{minipage}
    }

    \caption{State-dependent decomposition of causality for the
      target-dependent case. From left to right, the panels show:
      state-dependent redundant ($\Iss^R_{12\to 2}$), unique
      ($\Iss^U_{1\to 2}$), and synergistic ($\Iss^S_{12\to 2}$)
      causal contributions in blue, red, and orange, respectively; and the
      SURD causalities to the target variable $Q_2^+$.  The
      notation follows: U1 denotes the unique SURD causality from
      $Q_1$ to $Q_2^+$, i.e., $\Delta I^U_{1\to 2}$, with analogous
      definitions for other components (i.e., R and S).  All SURD and
      state-dependent causal contributions are normalized by the
      mutual information $I(Q_2^+; Q_1, Q_2)$.  The causal maps for
      redundant and synergistic causalities are shown for the $q_2$
      that maximizes the total sum of individual contributions.  In
      each panel, the bottom row displays the temporal evolution of
      $q_2^+(n)$, color-coded according to the corresponding
      instantaneous state-dependent causal contribution.}
    \label{fig:example-contributions-target}
\end{figure}

We consider the two-variable system, $[Q_1, Q_2]$, in which the
direction of causality toward the future state $Q_2^+$ depends on the
value of the source variable $Q_1$. Specifically, $Q_1 \rightarrow
Q_2^+$ when $q_1 > 0$, and $Q_2 \rightarrow Q_2^+$ when $q_1 \leq
0$. The equation governing the system is provided in
Fig.~\ref{fig:example-contributions}.  The variable $Q_1$ is sampled
from a normal distribution with zero mean and unit variance. The
variable $Q_2$ is influenced by an external noise term, $W_2$, drawn
from a normal distribution with mean $-2$ and unit standard
deviation. This noise term is not included in the vector of
observables and serves to mimic unobserved perturbations commonly
encountered in real-world systems.

The state-dependent redundant, unique, and synergistic causalities for
the target variable $Q_2^+$ are shown in
Fig.~\ref{fig:example-contributions}. The figure also includes the
time evolution of the causalities, which is obtained by mapping the
state-aware causality to the corresponding time instances.

For states with $q_1 > 0$, the results clearly show a nonzero
contribution from unique causality ($\Iss_{1 \to 2}^U$). In this
regime, $Q_1$ acts as the sole driver of $Q_2^+$, providing unique
information that is not contained in $Q_2$.
For states with $q_1 < 0$, both redundant ($\Iss^R_{12 \to 2}$) and
synergistic ($\Iss^S_{12 \to 2}$) causalities coexist. 
Redundancy arises because, in the region $q_1 < 0$, the information
contained in $q_1$ is already provided (i.e., rendered redundant) by
$q_2$, which fully determines the outcome. This leads to duplicated
information about $Q_2^+$ from both $Q_1$ and $Q_2$.
Synergistic causality emerges because two pieces of information are
simultaneously needed to predict $Q_2^+$, i.e., the value of $Q_1$, to
assess the condition $q_1 < 0$, and the value of $Q_2$, to determine
the outcome. Thus, $Q_1$ and $Q_2$ jointly provide information about
$Q_2^+$ that neither can supply independently.

Fig.~\ref{fig:example-contributions} shows the cumulative causal
contributions to $Q_2^+$ based on the SURD causalities ($\Delta
I^{\alpha}_{\boldsymbol{i}\rightarrow 2}$). These results, which are
consistent with the discussion above, will serve as a reference for
the next benchmark case, where we demonstrate that the underlying
state-dependent causal structures follow opposite trends despite
exhibiting similar SURD causalities.


\subsubsection*{Target-dependent causality}

The second benchmark case involves a system in which the direction of
causality depends on the state of the target variable. As shown in
Fig.~\ref{fig:example-contributions-target}, the system again consists
of two variables, $[Q_1, Q_2]$, such that $f(Q_1) \to Q_2^+$ when
$f(q_1) < 0$, and $Q_2 \to Q_2^+$ when $f(q_1) \geq 0$. Here,
$f(\cdot)$ denotes a nonlinear function that maps the past of the
source variable to the future of the target. The variable $Q_1$ is
sampled from a normal distribution with zero mean and unit variance,
while $Q_2$ is influenced by an external noise term, $W_2$, drawn from
a normal distribution with unit mean and unit standard deviation.

The state-dependent redundant, unique, and synergistic causalities for
the target variable $Q_2^+$ are shown in
Fig.~\ref{fig:example-contributions-target}, along with the
corresponding time evolution of the causalities.
The analysis clearly shows that the unique causality, $\Iss^U_{1
  \rightarrow 2}$, is primarily determined by the state of the target
variable rather than the source. In particular, positive target states
($q_2^+ > 0$) exhibit the strongest unique influence from $Q_1$,
consistent with the fact that $Q_1$ acts as the sole driver of $Q_2^+$
in this regime.
In contrast, negative values of $q_2^+$ give rise to both synergistic
($\Iss^S_{12 \rightarrow 2}$) and redundant ($\Iss^R_{12 \rightarrow
  2}$) causalities, as both $Q_1$ and $Q_2$ are required to predict
$Q_2^+$ when $f(q_1) \leq 0$. This behavior mirrors the reasoning
observed in the source-dependent benchmark case.

It is worth noting that the cumulative SURD causality shown in
Fig.~\ref{fig:example-contributions-target} resembles that of the
source-dependent case discussed earlier, despite the fundamentally
different interactions between the system variables.  This illustrates
a clear case where a state-aware decomposition of causality is
essential to uncover causal pathways that would be overlooked by a
state-averaged inference approach.


\subsection*{Application to flow interactions in wall turbulence}

\begin{figure}
    \centering
    \includegraphics[width=\linewidth]{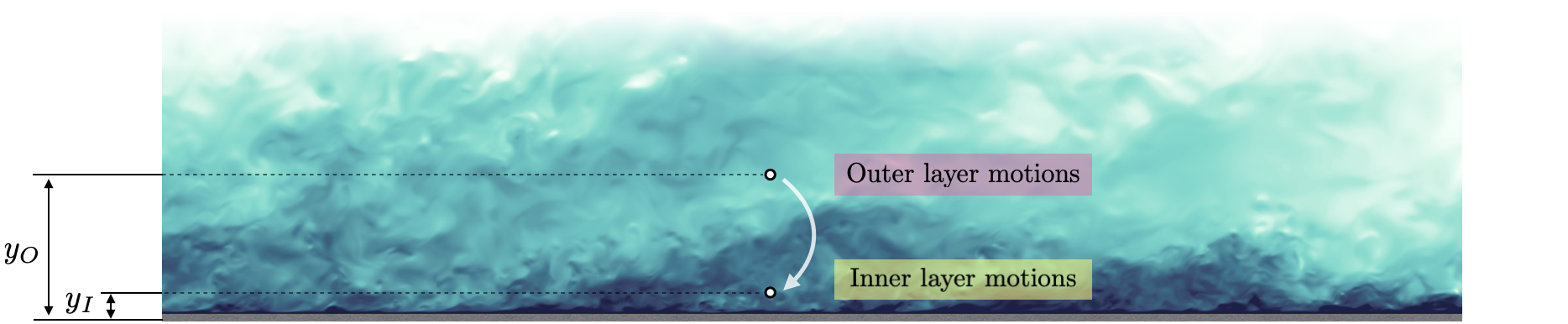}
    \vspace{0.001cm}

    \begin{minipage}{\textwidth}
    \includegraphics[width=0.29\textwidth]{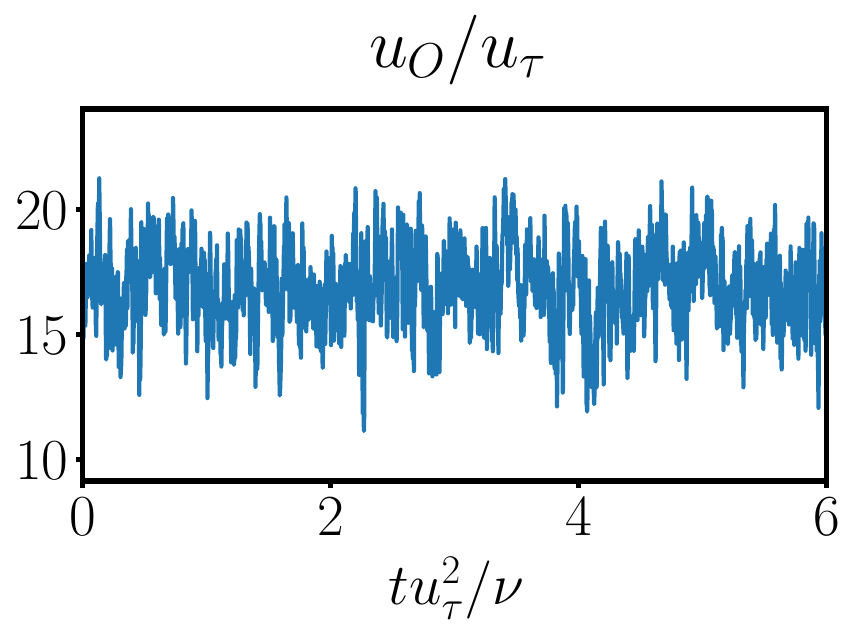}
    \hspace{0.04\textwidth}
    \includegraphics[width=0.29\textwidth]{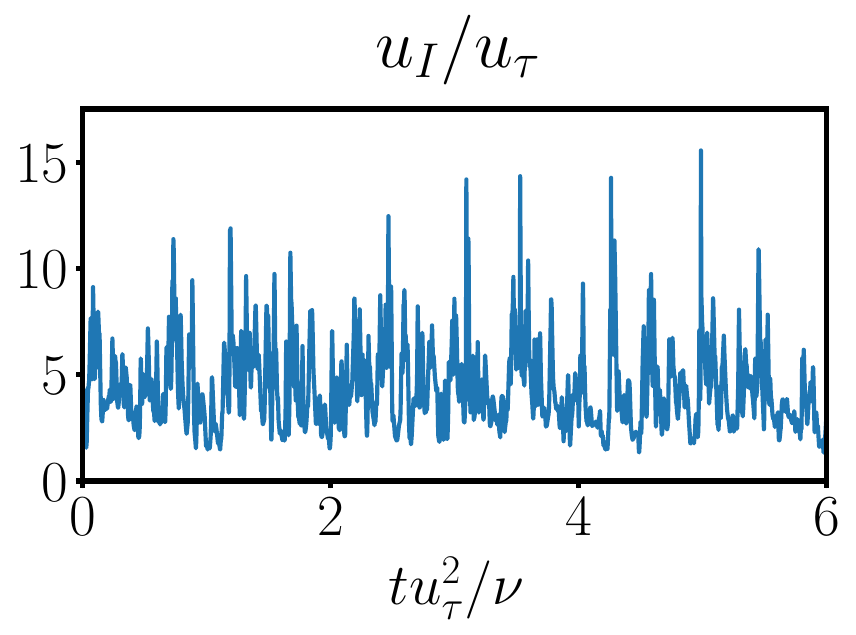}
    \hspace{0.012\textwidth}
    \begin{minipage}{0.27\textwidth}
    \vspace{-3.5cm}
    \includegraphics[width=\textwidth,trim=0 0 52 135, clip]{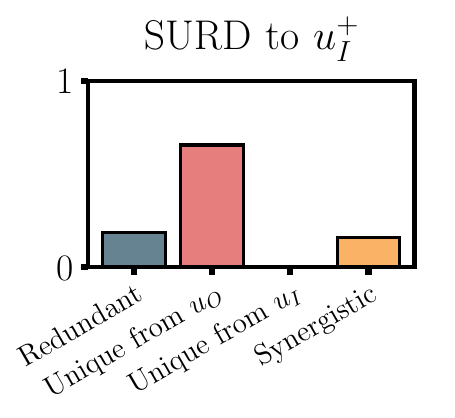}
    \end{minipage}
    \end{minipage}

    \begin{minipage}{\textwidth}
        \vspace{-0.0075\textwidth}
        \hspace{-0.03\textwidth}
        {\includegraphics[width=0.33\textwidth]{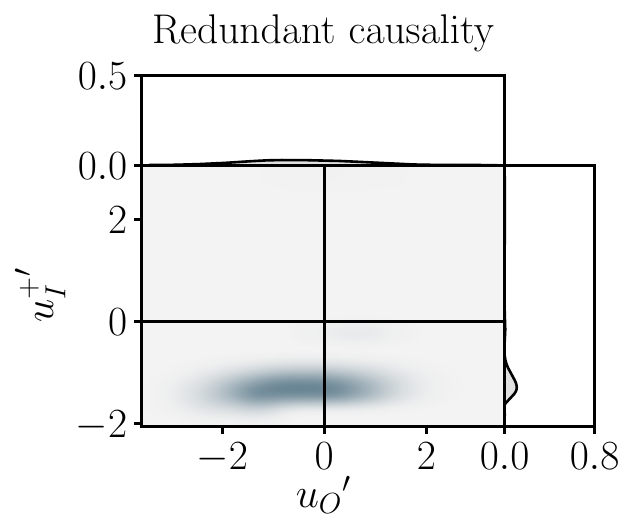}}
        \hspace{-0.02\textwidth}
        {\includegraphics[width=0.33\textwidth]{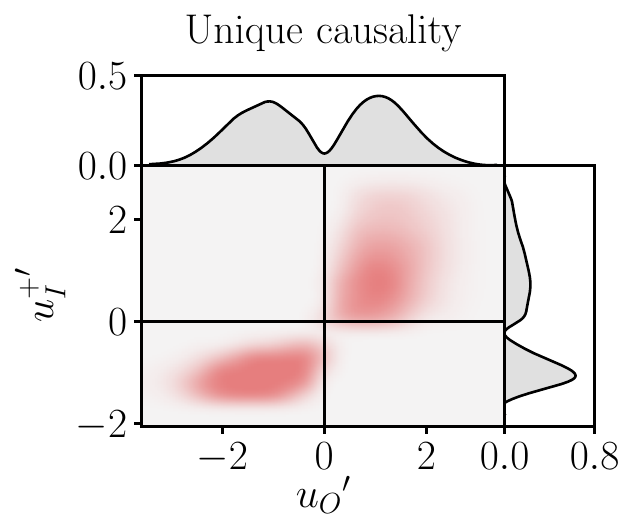}}
        \hspace{-0.02\textwidth}
        {\includegraphics[width=0.33\textwidth]{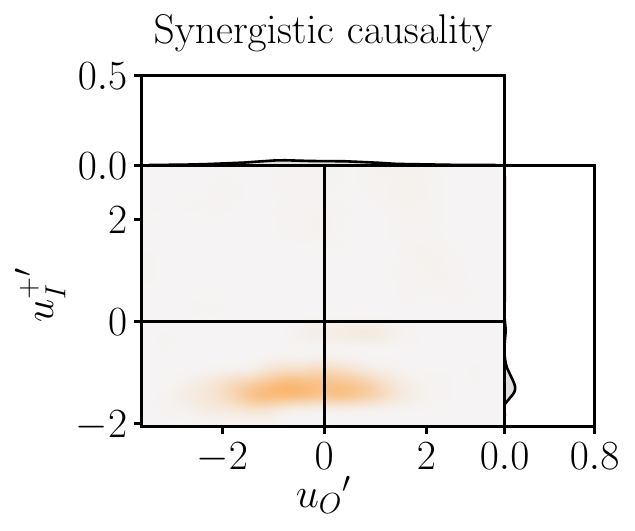}}
    \end{minipage}

    \vspace{-0.15cm}
    \begin{minipage}{\textwidth}
        \hspace{-0.03\textwidth}
        \includegraphics[width=0.31\textwidth, trim=0 40 0 0, clip]{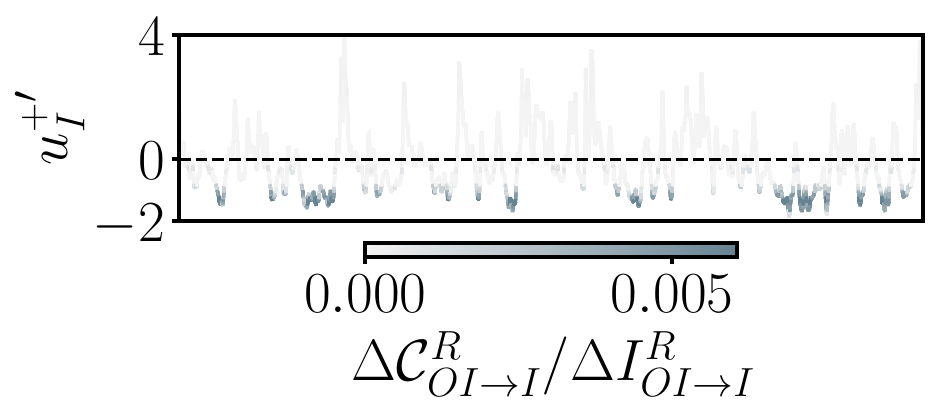}
        \hspace{0.005\textwidth}
        \includegraphics[width=0.31\textwidth, trim=0 40 0 0, clip]{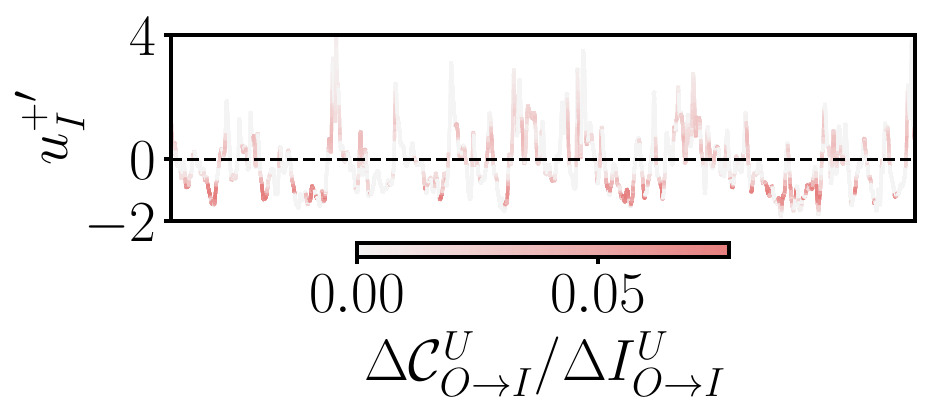}
        \hspace{0.005\textwidth}
        \includegraphics[width=0.31\textwidth, trim=0 40 0 0, clip]{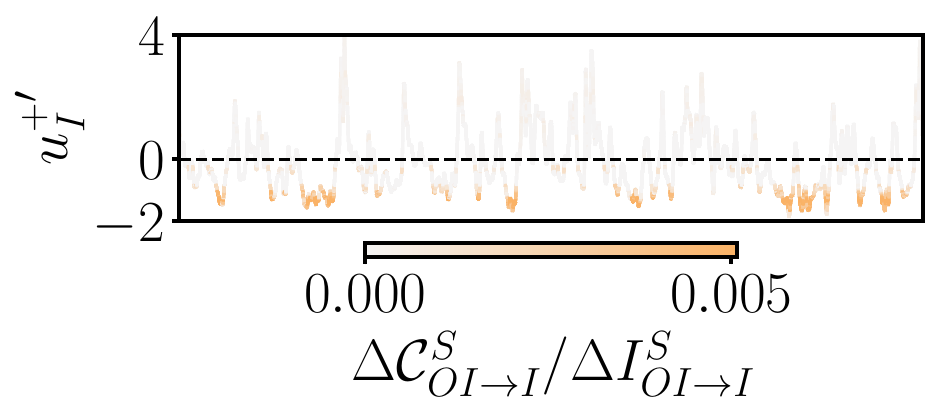}
        \vspace{0.005\textwidth}
    \end{minipage}
    \vspace{0.1cm}


    \vspace{-0.2cm}
    \begin{tikzpicture}
        \node[anchor=south west, inner sep=0] (image) at (0,0) {\includegraphics[height=0.174\textwidth]{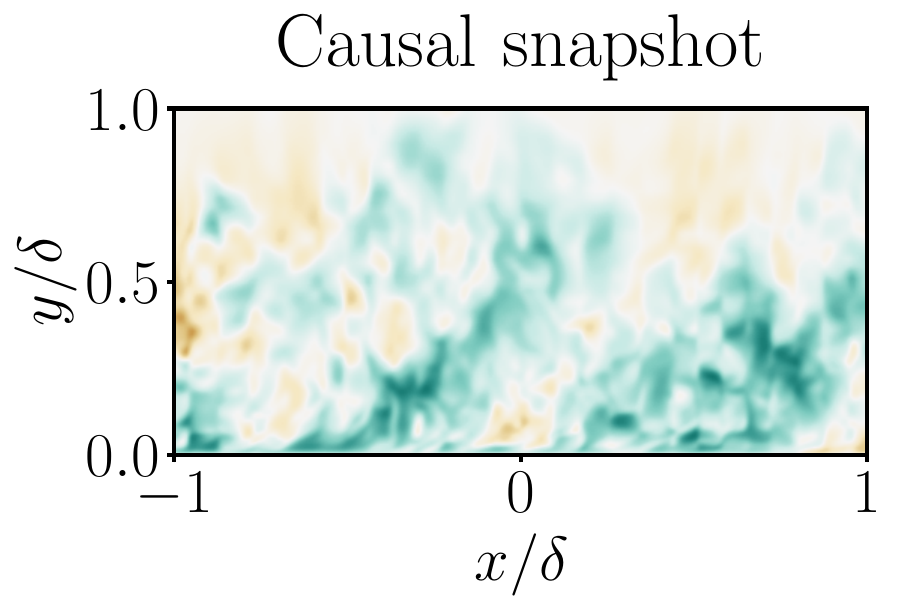}};
        \node[anchor=south, font=\normalsize, fill=white, text=black, inner sep=2pt, rounded corners, minimum width=0.15\textwidth,minimum height=6mm] at ([yshift=-5mm,xshift=4mm]image.north) {Causal state};
    \end{tikzpicture}
    \begin{tikzpicture}
        \node[anchor=south west, inner sep=0] (image) at (0,0) {\includegraphics[height=0.174\textwidth]{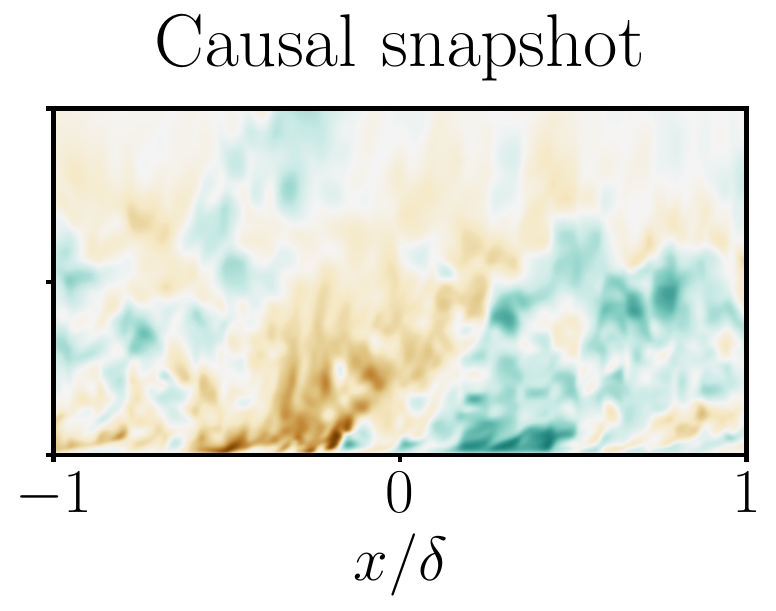}};
        \node[anchor=south, font=\normalsize, fill=white, text=black, inner sep=2pt, rounded corners, minimum width=0.15\textwidth,minimum height=6mm] at ([yshift=-5mm,xshift=1mm]image.north) {Causal state};
    \end{tikzpicture}
    \begin{tikzpicture}
        \node[anchor=south west, inner sep=0] (image) at (0,0) {\includegraphics[height=0.174\textwidth]{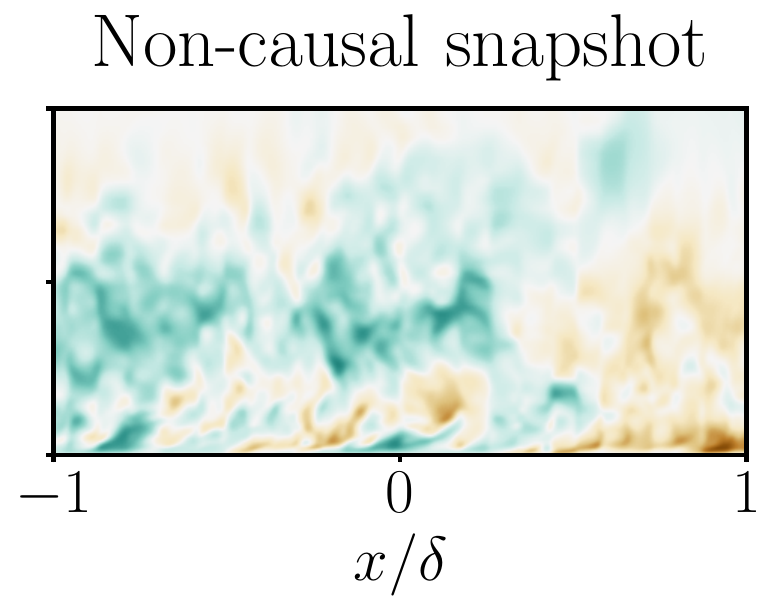}};
        \node[anchor=south, font=\normalsize, fill=white, text=black, inner sep=2pt, rounded corners, minimum width=0.18\textwidth,minimum height=6mm] at ([yshift=-5mm,xshift=0.5mm]image.north) {Non-causal state};
    \end{tikzpicture}
    \begin{tikzpicture}
        \node[anchor=south west, inner sep=0] (image) at (0,0) {\includegraphics[height=0.174\textwidth]{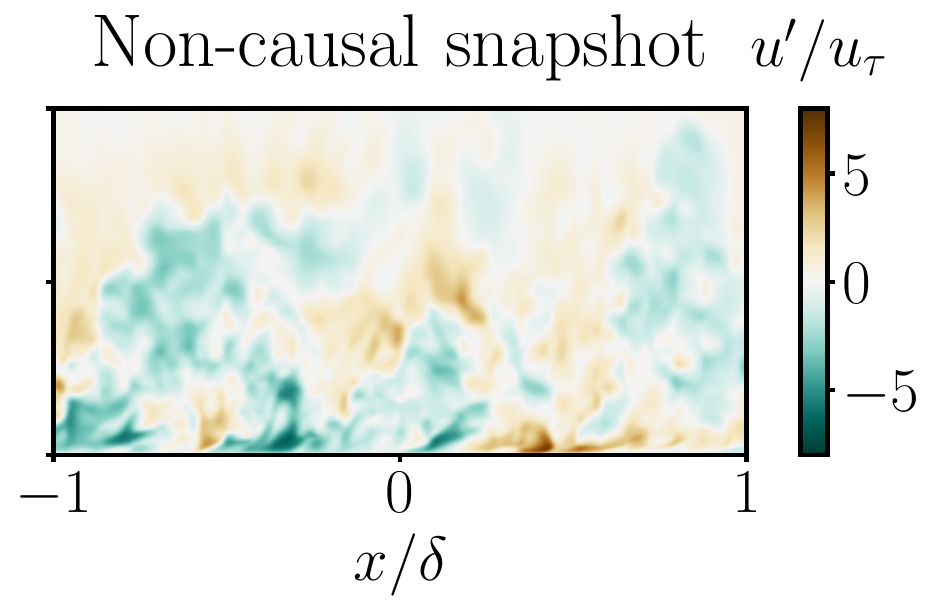}};
        \node[anchor=south, font=\normalsize, fill=white, text=black, inner sep=2pt, rounded corners, minimum width=0.18\textwidth,minimum height=6mm] at ([yshift=-5mm,xshift=-3.5mm]image.north) {Non-causal state};
    \end{tikzpicture}
    
    \vspace{-0.45cm}
    \begin{tikzpicture}
        \node[anchor=south west, inner sep=0] (image) at (0,0) {\includegraphics[height=0.174\textwidth]{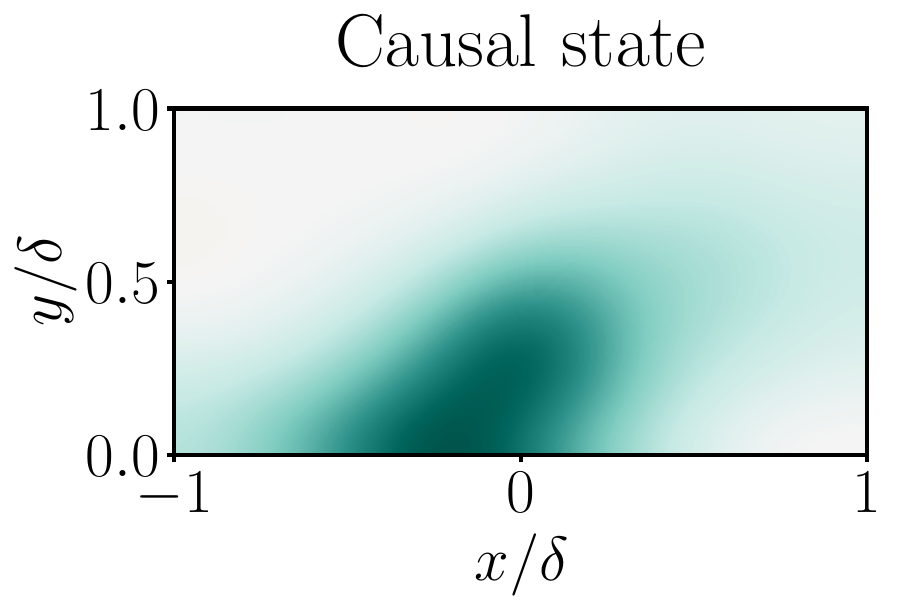}};
        \node[anchor=south, font=\normalsize, fill=white, text=black, inner sep=2pt, rounded corners, minimum width=0.15\textwidth,minimum height=6mm] at ([yshift=-5mm,xshift=4mm]image.north) {Causal average};
    \end{tikzpicture}
    \begin{tikzpicture}
        \node[anchor=south west, inner sep=0] (image) at (0,0) {\includegraphics[height=0.174\textwidth]{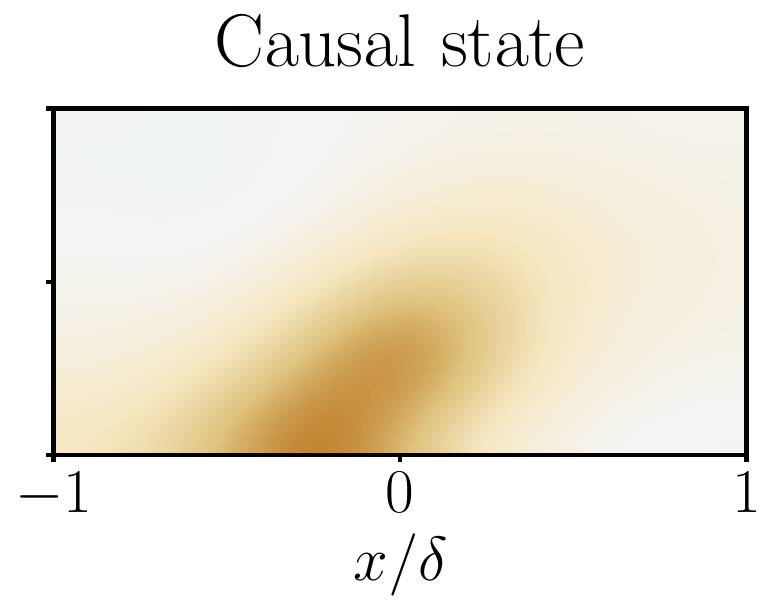}};
        \node[anchor=south, font=\normalsize, fill=white, text=black, inner sep=2pt, rounded corners, minimum width=0.15\textwidth,minimum height=6mm] at ([yshift=-5mm,xshift=1mm]image.north) {Causal average};
    \end{tikzpicture}
    \begin{tikzpicture}
        \node[anchor=south west, inner sep=0] (image) at (0,0) {\includegraphics[height=0.174\textwidth]{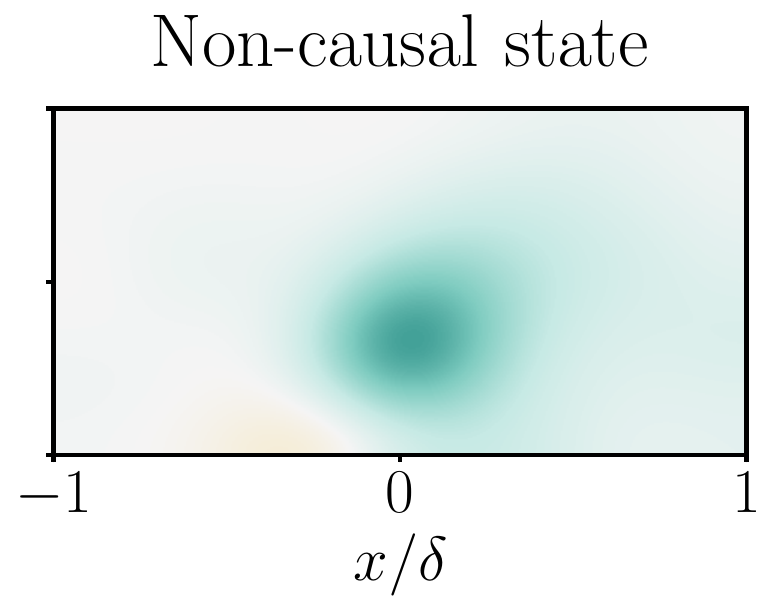}};
        \node[anchor=south, font=\normalsize, fill=white, text=black, inner sep=2pt, rounded corners, minimum width=0.18\textwidth,minimum height=6mm] at ([yshift=-5mm,xshift=0.5mm]image.north) {Non-causal average};
    \end{tikzpicture}
    \begin{tikzpicture}
        \node[anchor=south west, inner sep=0] (image) at (0,0) {\includegraphics[height=0.174\textwidth]{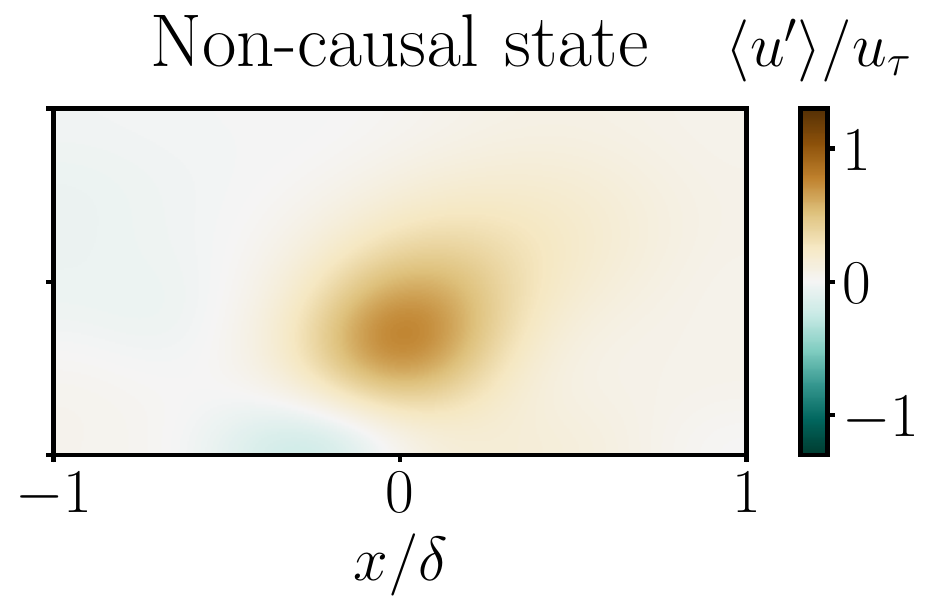}};
        \node[anchor=south, font=\normalsize, fill=white, text=black, inner sep=2pt, rounded corners, minimum width=0.18\textwidth,minimum height=6mm] at ([yshift=-5mm,xshift=-3.5mm]image.north) {Non-causal average};
    \end{tikzpicture}

    \vspace{-0.25cm}
    \caption{Causality between inner and outer flow motions in a
      turbulent boundary layer.  (a) Instantaneous visualization of
      the streamwise velocity field.  (b) Temporal evolution of the
      streamwise velocity at two fixed wall-normal locations: the
      outer layer ($y_O = 0.3\delta$) and the inner layer ($y_I =
      4\nu/u_\tau$).  (c) Redundant ($\Delta I^R_{OI \to I}$), unique
      ($\Delta I^U_{(\cdot) \to I}$), and synergistic ($\Delta I^S_{OI
        \to I}$) SURD causalities to the future inner-layer velocity
      $u_I^+ = u_I(t + \Delta T)$. The time lag $\Delta T$ is chosen
      to maximize cross-induced unique causality.  (d) State-dependent
      redundant ($\Iss^R_{OI \to I}$), unique ($\Iss^U_{O \to
        I}$), and synergistic ($\Iss^S_{OI \to I}$) causalities as
        functions of the instantaneous streamwise velocities $u_O$
      (outer layer) and $u_I^+$ (inner layer).  Note that
      for some of the panels, the color scale has been intentionally
      saturated to enhance visual contrast; however, the causalities
      are of small magnitude, as evidenced by the horizontal and
      vertical projections. (e) Time series of
      $u_I^+$ color-coded by the dominant state-dependent causal
      component at each time step.  (f) Instantaneous and (g)
      conditionally averaged visualizations of flow fields
      corresponding to the causal and non-causal states,
      respectively.}
    \label{fig:inn-out-intensities}
\end{figure}

We apply our state-dependent causal inference method to identify the
most causal flow states within a turbulent boundary layer, i.e., the
chaotic fluid motion within a thin region adjacent to solid
surfaces. Turbulent boundary layers are responsible for nearly 50\% of
the aerodynamic drag on modern airliners and play a critical role in
the lower atmosphere, particularly within the first hundred meters,
where they influence broader meteorological
phenomena~\cite{marusic2010}.  Understanding and modeling the
interactions among turbulent motions of different scales within these
layers remains a major challenge due to strong nonlinearities and high
dimensionality of the system. Prior studies have documented the impact
of large-scale motions in the outer layer on smaller-scale motions
near the wall~\cite{flack2005, flores2006, Hutchins2007, Mathis2009,
  marusic2010, busse2012, Mizuno2013, Chung2014, lozano2019x, surd},
supporting the notion of top-down causality (a.k.a. Townsend’s
outer-layer similarity hypothesis~\cite{townsend1976}).  In this work,
we focus on identifying the specific flow states that contribute to
the causal influence from the outer layer flow (far from the wall) to
the inner layer flow (near the wall).

We utilize data from a high-fidelity numerical simulation of turbulent
flow over a flat plate with zero mean-pressure gradient, which
provides a realistic representation of turbulent boundary layers
encountered in engineering
applications~\cite{Towne2023}. Fig.~\ref{fig:inn-out-intensities}
shows an instantaneous visualization of the streamwise velocity. The
flow conditions are characterized by the friction Reynolds number,
$Re_{\tau} = u_\tau \delta / \nu$, which ranges from around $300$
(inflow) to $700$ (outflow), where $\delta$ is the boundary-layer
thickness, $\nu$ is the kinematic viscosity, and $u_\tau$ is the
friction velocity at the wall. The time signals analyzed here
correspond to the streamwise velocity measured at two wall-normal
locations, representing the inner and outer layers of the boundary
layer. These velocity signals, denoted as $u_I(t)$ and $u_O(t)$, are
located at wall-normal distances of $y_I=4 \nu/u_\tau$ and
$y_O=0.3\delta$, respectively. Fig.~\ref{fig:inn-out-intensities}
presents a sample of these velocity signals at $Re_\tau = 500$.

The objective is to identify the states that contribute most
significantly to the causal influence of the outer-layer velocity
${u_O}(t)$ on the future inner-layer velocity ${u_I}^{+}(t) = u_I(t +
\Delta T)$. The time lag $\Delta T$ for the causal evaluation is
selected to maximize the unique causality from $u_O(t)$ to $u_I^+(t)$,
resulting in an optimal delay of approximately $\Delta T \approx 60
\nu/u_\tau^2$. The causal influence in the reverse direction, from
${u_I}(t)$ to ${u_O}(t+\Delta T)$, is discussed in the Methods
section. We adopt the notation $(\cdot)^\prime$ to indicate
standardization of the signals by their respective mean and standard
deviation. Thus, negative values of ${u_O}^\prime$ represent
below-average deviations of $u_O$, while positive values correspond to
above-average deviations, with the same convention applied to
${u_I}^\prime$.

We begin by analyzing the SURD causalities from $[u_I, u_O]$ to
$u_I^+$, as shown in Fig.~\ref{fig:inn-out-intensities}. The
redundant, unique, and synergistic components are denoted by $\Delta
I^R_{OI \to I}$, $\Delta I^U_{O \to I}$, and $\Delta I^S_{OI \to I}$,
respectively, where the subscripts indicate contributions from the
outer-layer ($O$) and inner-layer ($I$) velocities. The results
demonstrate that the unique causality from $u_O$ to $u_I^+$ dominates,
followed by the synergistic and redundant components. In contrast, the
unique causality from $u_I$ to $u_I^+$ is negligible at the current
time lag $\Delta T$, since this influence occurs on timescales shorter
than $\Delta T$~\cite{jimenez1999}.

We now examine the state-dependent decomposition of causality,
presented in Fig.~\ref{fig:inn-out-intensities}, which also
illustrates the temporal evolution of the individual causal
components. The dominant form of causality is the unique 
component, which exceeds the redundant and synergistic 
components by more than an order of magnitude.
%
The small redundant and synergistic causalities are  concentrated
in regions where ${u_I^+}^\prime < 0$. This indicates that, during
low-speed motions in the inner layer (i.e., velocities below the mean
flow), part of the causal influence arises from overlapping
information simultaneously conveyed by the outer- and inner-layer
histories, while another portion is jointly determined by concurrent
fluctuations in both layers.

The unique causal contribution is predominantly distributed within the
first (${u_O^\prime > 0}, {u_I^+}^\prime > 0$) and third (${u_O^\prime
  < 0}, {u_I^+}^\prime < 0$) quadrants of the
${u_O}^\prime$–${u_I^+}^\prime$ space. These quadrants represent
states where outer- and inner-layer motions concurrently exceed or
fall below the mean velocity. Moreover, the region of causal states 
$\Iss^U_{O \rightarrow I}$ in the first quadrant is wider than the
third, consistent with previous studies~\cite{Mathis2009} showing that
large positive fluctuations in the outer layer exert a stronger
influence on small-scale inner-layer structures compared to negative
fluctuations. This observation supports the well-known modulation
effect, whereby high-speed outer-layer events leave a stronger causal
imprint on the inner-layer dynamics~\cite{Mathis2009}. In contrast,
the second and fourth quadrants do not contain  causal
contributions.  Overall, these findings indicate that high-speed
streaks propagating towards the wall act as dominant drivers of unique
causality. Conversely, low-speed streaks are predominantly associated
with redundant or synergistic causal interactions.

We conclude this section by examining the flow structures associated
with causal and non-causal states. To this end, we analyze instantaneous
realizations of the streamwise velocity field, $u(x,z,t)$, within a
spatial region of size $2\delta$, conditioned on states of $u_O(t)$ and
$u_I^+(t)$ corresponding to the four quadrants of the unique causality
map $\Iss^U_{O \rightarrow I}$ discussed
previously. Fig.~\ref{fig:inn-out-intensities} shows example snapshots
of $u(x,z,t)$ representative of each quadrant. To highlight
characteristic flow patterns for each quadrant, we compute
ensemble-averaged velocity fields, $\langle u(x,z,t) \rangle$, over
all instances belonging to the respective quadrant. For instance, the
flow field in the first quadrant is averaged over all time instants
for which $u_O^\prime > 0$ and $ {u_I^+}^\prime > 0$, and similarly
for the other quadrants.  The results (bottom of
Fig.~\ref{fig:inn-out-intensities}) reveal that causal states
correspond to coherent velocity structures attached to the wall,
extending beyond $y = 0.3\delta$, the wall-normal location of
$u_O$. These findings indicate that causal interactions are associated
with coherent flow structures---such as high- or low-speed
streaks---that span across the inner and outer layers. Furthermore,
the shape of these structures is in agreement with other averaged flow
structures found in wall-bounded turbulence~\cite{jimenez2012}. In
contrast, the non-causal states (second and fourth quadrant) do not
exhibit such structural coherence, with velocity fields appearing
fragmented and detached from the wall. This fragmentation shows a
breakdown of spatial coherence and diminished influence from the outer
to the inner layer.

\subsection*{Application to Walker circulation}

\begin{figure}
    \centering
    \begin{minipage}{\tw}
    \centering
        \hspace{0.01\textwidth}
        \includegraphics[width=0.925\textwidth]{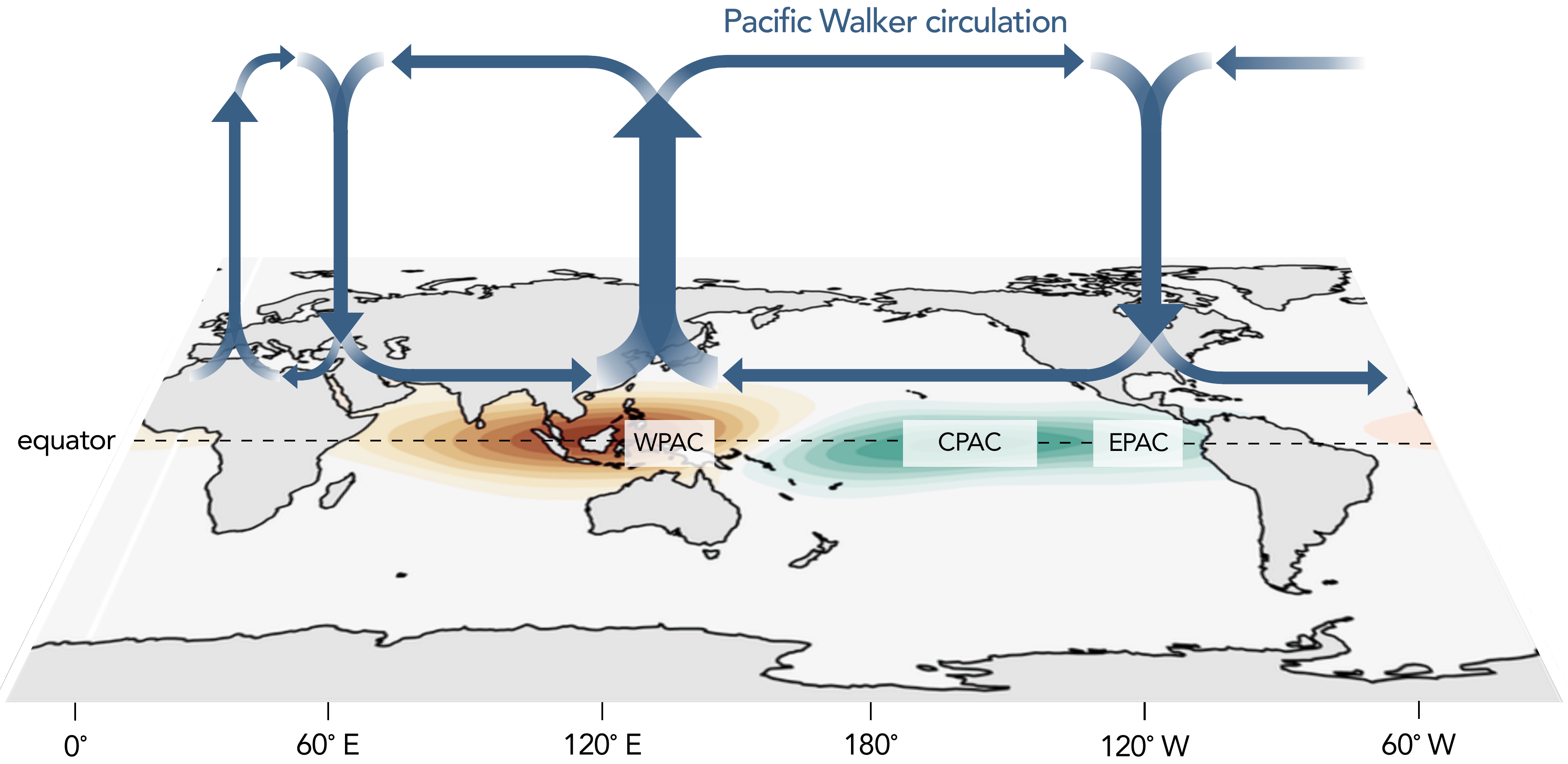}
    \end{minipage}

    \begin{minipage}{0.74\textwidth}
        \vspace{0.01\textwidth}
        \hspace{-0.02\textwidth}
        \includegraphics[height=0.22\textwidth]{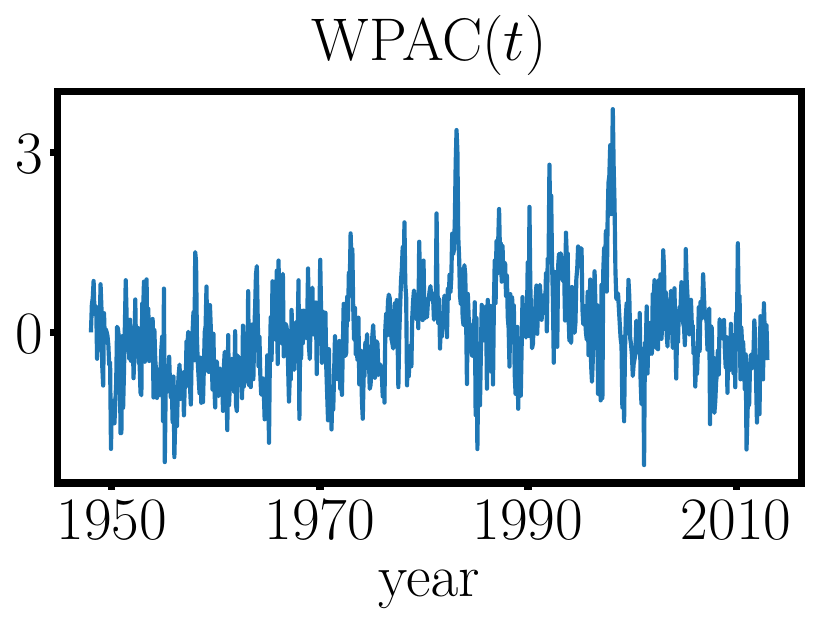}
        \hspace{0.01\textwidth}
        \includegraphics[height=0.22\textwidth]{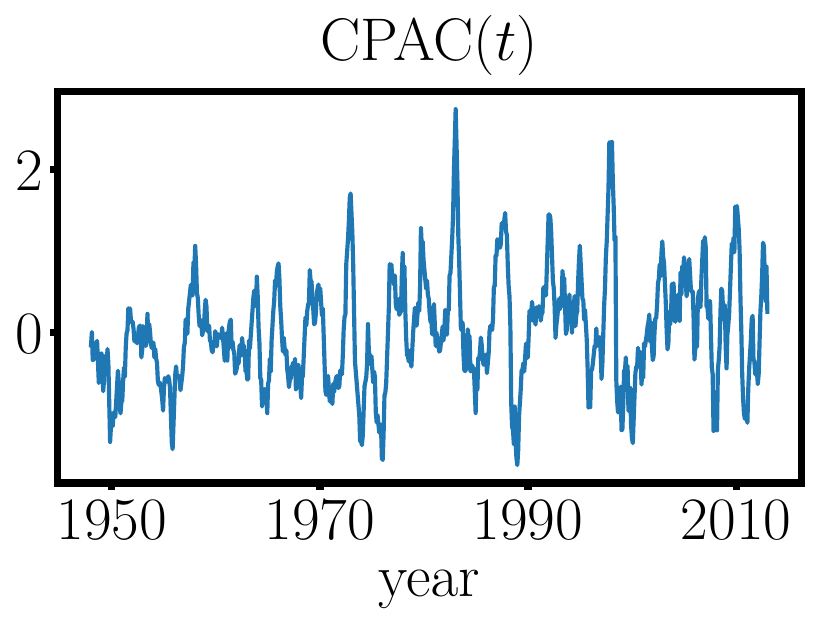}
        \hspace{0.01\textwidth}
        \includegraphics[height=0.22\textwidth]{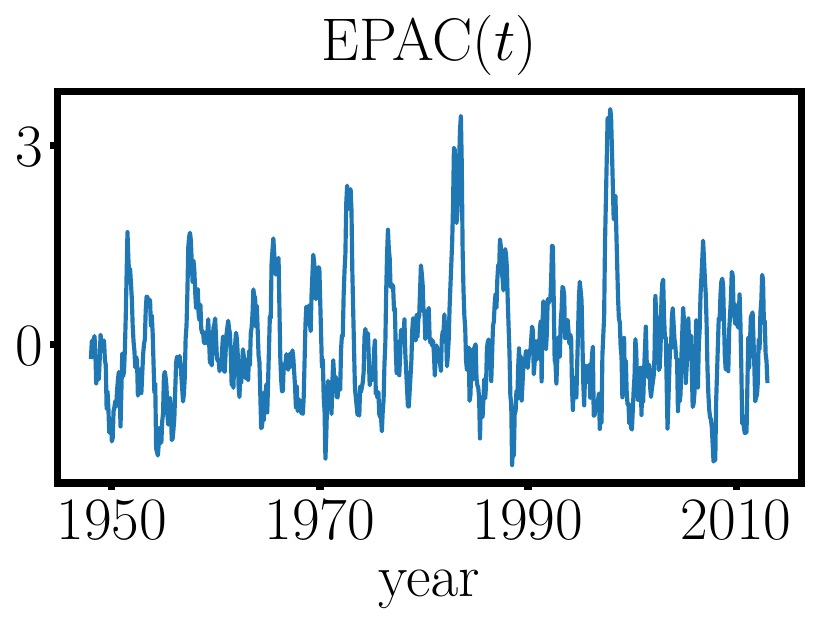}
    \end{minipage}
    \begin{minipage}{0.225\textwidth}
    \vspace{0.02\textwidth}
    \hspace{-0.06\textwidth}
    \includegraphics[width=\linewidth, trim=0 0 70 0, clip]{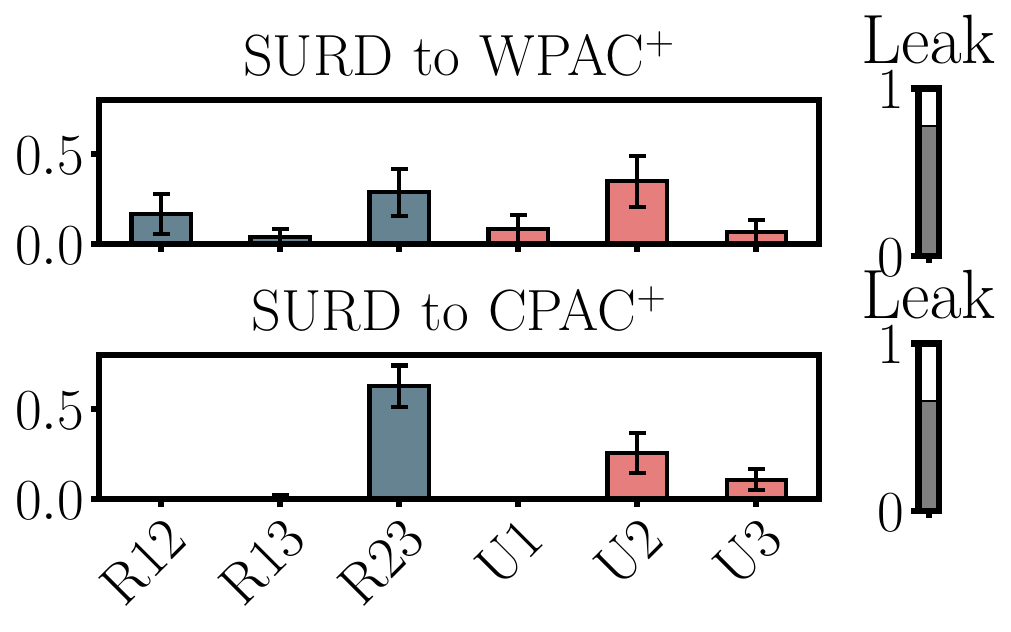}
    \end{minipage}

    \begin{minipage}{\tw}
    \centering
    \hspace{-0.01\textwidth}
    \begin{minipage}{0.24\textwidth}
    \begin{tikzpicture}
        \node[anchor=north west] (img) at (0, 0) {\includegraphics[width=\textwidth]{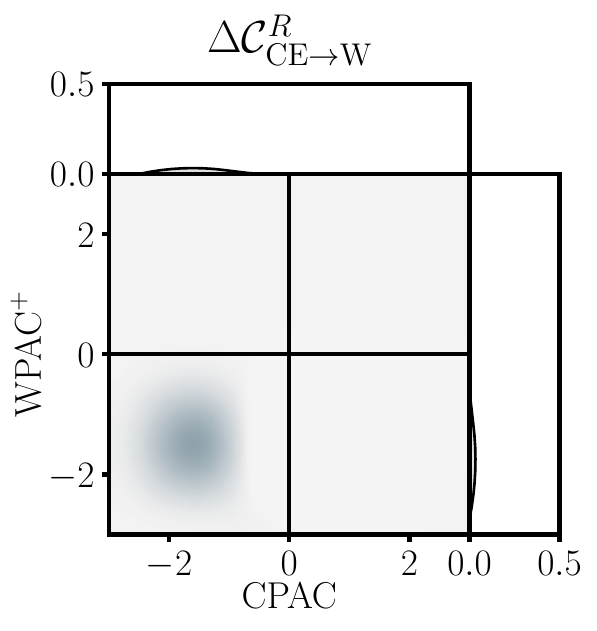}};
        \node[anchor=south, font=\small, fill=white, text=black, inner sep=2pt, rounded corners, minimum width=0.2\textwidth,minimum height=6mm] at ([yshift=-6.25mm,xshift=-1mm]img.north) {$\quad$[CPAC,EPAC]$\to$WPAC$^+\quad$};
        \node[anchor=south, font=\small, fill=white, text=black, inner sep=2pt, rounded corners, minimum width=0.2\textwidth,minimum height=6mm] at ([yshift=-2mm,xshift=-1mm]img.north) {Redundant causality};
    \end{tikzpicture}
    \end{minipage}
    \begin{minipage}{0.24\textwidth}
    \begin{tikzpicture}
        \node[anchor=north west] (img) at (0, 0) {\includegraphics[width=\textwidth]{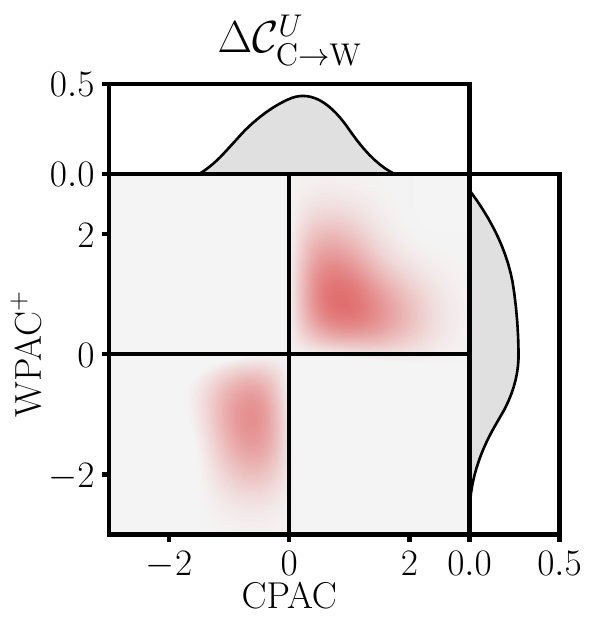}};
        \node[anchor=south, font=\small, fill=white, text=black, inner sep=2pt, rounded corners, minimum width=0.2\textwidth,minimum height=6mm] at ([yshift=-6.25mm,xshift=-1mm]img.north) {$\quad$CPAC$\to$WPAC$^+\quad$};
        \node[anchor=south, font=\small, fill=white, text=black, inner sep=2pt, rounded corners, minimum width=0.2\textwidth,minimum height=6mm] at ([yshift=-2mm,xshift=-1mm]img.north) {Unique causality};
    \end{tikzpicture}
    \end{minipage}
    \begin{minipage}{0.24\textwidth}
    \begin{tikzpicture}
        \node[anchor=north west] (img) at (0, 0) {\includegraphics[width=\textwidth]{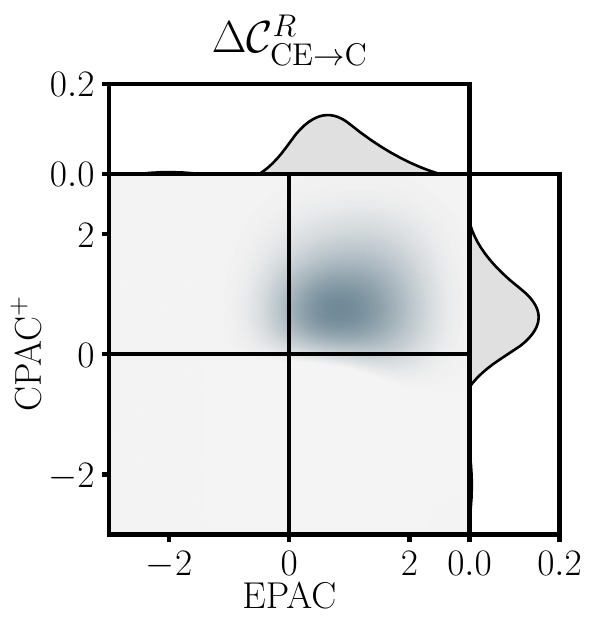}};
        \node[anchor=south, font=\small, fill=white, text=black, inner sep=2pt, rounded corners, minimum width=0.2\textwidth,minimum height=6mm] at ([yshift=-6.25mm,xshift=-1mm]img.north) {$\quad$[CPAC,EPAC]$\to$CPAC$^+\quad$};
        \node[anchor=south, font=\small, fill=white, text=black, inner sep=2pt, rounded corners, minimum width=0.2\textwidth,minimum height=6mm] at ([yshift=-2mm,xshift=-1mm]img.north) {Redundant causality};
    \end{tikzpicture}
    \end{minipage}
    \begin{minipage}{0.24\textwidth}
    \begin{tikzpicture}
        \node[anchor=north west] (img) at (0, 0) {\includegraphics[width=\textwidth]{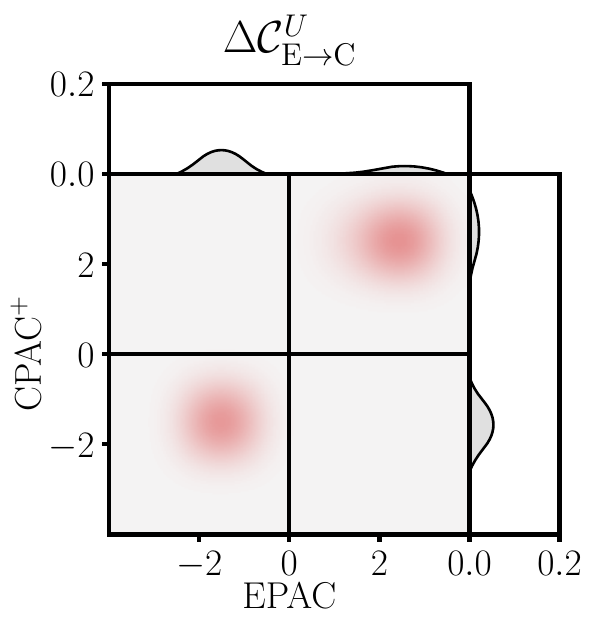}};
        \node[anchor=south, font=\small, fill=white, text=black, inner sep=2pt, rounded corners, minimum width=0.2\textwidth,minimum height=6mm] at ([yshift=-6.25mm,xshift=-1mm]img.north) {EPAC$\to$CPAC$^+$};
        \node[anchor=south, font=\small, fill=white, text=black, inner sep=2pt, rounded corners, minimum width=0.2\textwidth,minimum height=6mm] at ([yshift=-2mm,xshift=-1mm]img.north) {Unique causality};
    \end{tikzpicture}
    \end{minipage}
    \end{minipage}

    \begin{minipage}{\textwidth}
        {\includegraphics[width=0.23\textwidth, trim=0 40 0 0, clip]{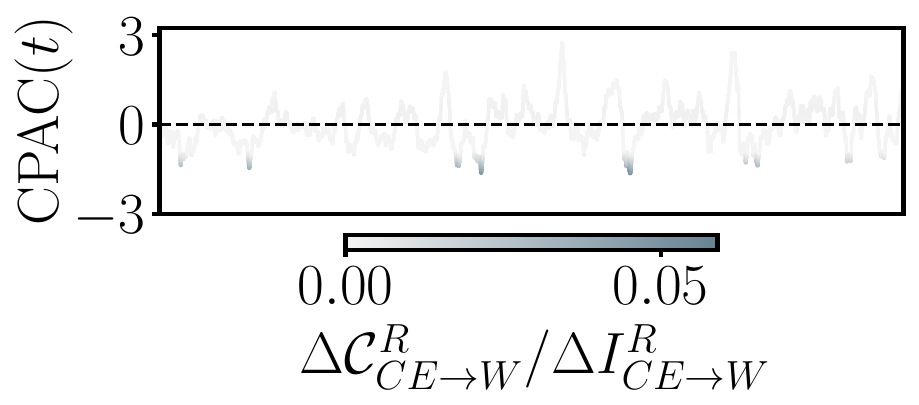}}
        \hspace{0.005\textwidth}
        {\includegraphics[width=0.23\textwidth, trim=0 40 0 0, clip]{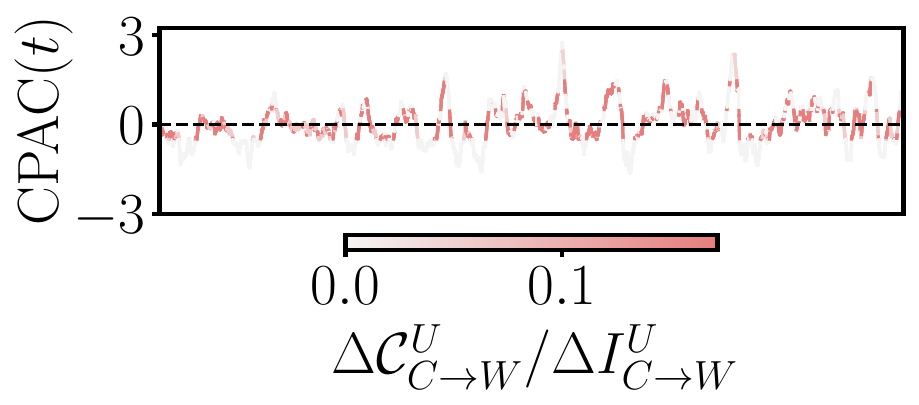}}
        \hspace{0.005\textwidth}
        {\includegraphics[width=0.23\textwidth, trim=0 40 0 0, clip]{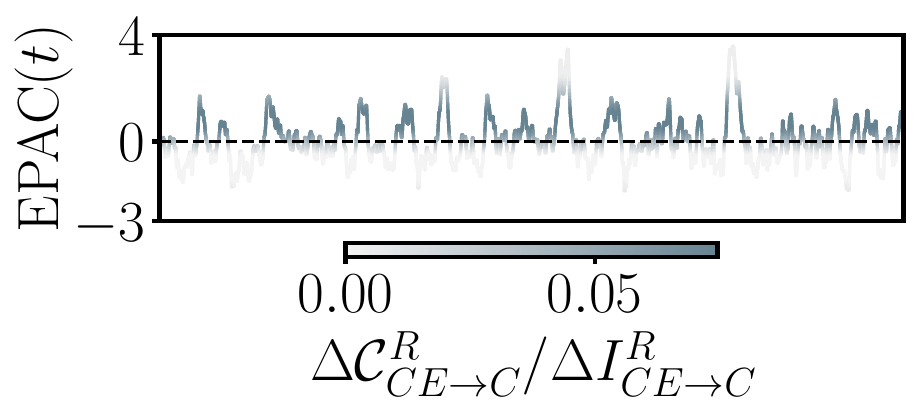}}
        \hspace{0.005\textwidth}
        {\includegraphics[width=0.23\textwidth, trim=0 40 0 0, clip]{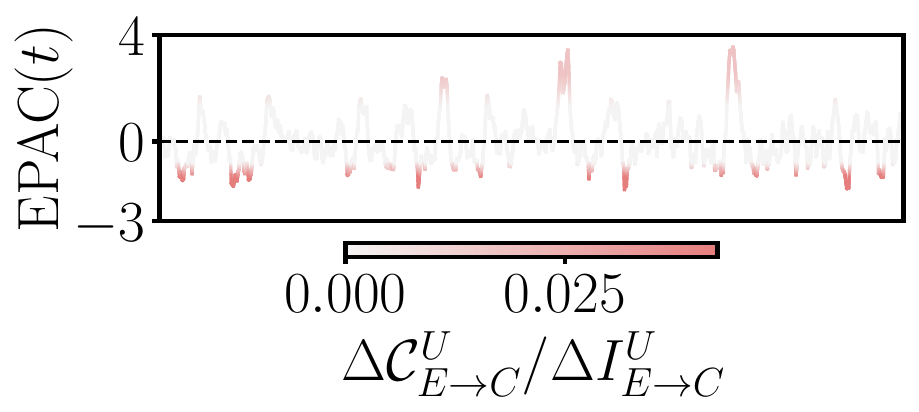}}
    \end{minipage}


    \caption{Causality in the Pacific Walker circulation. Top:
      Schematic representation of the Pacific Walker circulation
      mechanism, showing the regions where regional averages were
      computed for the period 1948–2012 ($T = 780$ months): surface
      pressure anomalies in the West Pacific (WPAC) and surface air
      temperature anomalies in the Central Pacific (CPAC) and East
      Pacific (EPAC). The shaded boxes indicate the specific regions
      of interest. The schematic diagram was adapted from
      Ref.~\cite{noaa2024} and the data was taken from Watanabe et
      al.~\cite{Watanabe2023}, which illustrates how local sea surface
      temperature (SST) warming affects the strength of the Pacific
      Walker circulation, with positive (orange) values indicating
      strengthening and negative (green) values indicating weakening.
      Middle top: Time evolution of WPAC, CPAC, and EPAC, along with
      redundant (R) and unique (U) SURD causalities to WPAC$^+$ =
      WPAC$(t+\Delta T)$ and CPAC$^+$ = CPAC$(t+\Delta T)$. The
      variables WPAC, CPAC, and EPAC are denoted as 1, 2, and 3,
      respectively. For example, the label R13 $\to$ CPAC$^+$
      indicates the redundant causality from WPAC and EPAC to
      CPAC$^+$. The causal analysis is performed with a time lag of
      $\Delta T = 4$ months, which corresponds to the maximum
      cross-induced causality. Error bars on the SURD causalities
      represent the standard deviation across 50 bootstrap
      resamplings. Statistical significance was assessed via 1000
      random permutations, yielding a $p$-value of zero for both
      redundant and unique components.
      Middle bottom: Decomposition of state-dependent causalities:
      redundant [CPAC, EPAC] $\to$ WPAC$^+$, unique CPAC $\to$
      WPAC$^+$, redundant [CPAC, EPAC] $\to$ CPAC$^+$, and unique EPAC
      $\to$ CPAC$^+$ as a function of the source and target variable
      states. Note that in the left-most panel, the color scale has
      been intentionally saturated to enhance visual contrast;
      however, the actual causalities are of small magnitude, as
      indicated by the horizontal and vertical projections.
      Bottom: Temporal evolution of CPAC$(t)$ and EPAC$(t)$
      color-coded according to the state-dependent causal contribution
      at that instant.}
    \label{fig:climate}
\end{figure}

We apply our state-dependent causal inference method to analyze the
Walker circulation in the tropical Pacific---a key atmospheric process
with a well-established physical mechanism~\cite{Bjerknes1969}, and a
subject of previous causal investigations~\cite{runge2019pcmci,
  Runge2023}. However, a state-aware analysis that accounts for
redundant and unique causalities has not yet been performed.  The
Walker circulation is a large-scale atmospheric circulation pattern in
the tropical Pacific, driven by the temperature gradient between the
warm waters of the Western Pacific and the cooler Eastern
Pacific. Under normal conditions, strong trade winds transport warm
surface waters westward, leading to moist air rising over the Western
Pacific due to enhanced convection. This air ascends, moves eastward
aloft, cools, and descends over the Eastern Pacific. The circulation
is completed as the cooler, drier air flows westward near the surface,
forming a closed loop across the tropical Pacific.
Fig.~\ref{fig:climate} illustrates this circulation pattern, which
governs critical atmospheric and oceanic interactions in the Pacific
basin and plays a central role in driving global weather patterns and
climate variability.

We use data consisting of regional averages of surface pressure
anomalies in the West Pacific (WPAC) and surface air temperature
anomalies in the Central Pacific (CPAC) and East Pacific (EPAC)
regions for the period 1948–2012, comprising 780
months~\cite{runge2019pcmci}. The regions used for these averages are
illustrated as shaded boxes in Fig.~\ref{fig:climate}. Here, we focus
on understanding what particular states of each variable are
responsible for the Walker circulation phenomenon from the East to
West Pacific close to the sea surface. To achieve this, we analyze the
most relevant total redundant and cross-induced unique causalities to
CPAC and WPAC involving at most two variables, which are shown in
Fig.~\ref{fig:climate}. The number of samples in the dataset was
insufficient to reliably estimate synergistic causalities, which are
therefore not included in the analysis.

The decomposition for these causalities is shown in
Fig.~\ref{fig:climate}, as a function of the values of the source and
target variables. First, we focus on the redundant and cross-induced
unique causalities to WPAC$^+$. In this case, unique causalities
dominate and redundant causalities are small.  The small redundant
causality shows causal states for negative values of CPAC and
WPAC$^+$, whereas the higher causal states for the unique causality
occur when both CPAC and WPAC$^+$ are either above or below their
respective means. Notably, states above the mean exhibit stronger
causal influence compared to those below \cite{cai2020}. For the rest
of the cases, i.e., where CPAC and WPAC$^+$ have opposite signs, the
state-dependent unique causality is zero, which indicates that
causality is already explained by the redundant components. These
results are consistent with the observation that positive CPAC
temperature anomalies (associated with El Niño conditions) have a more
pronounced impact on Western Pacific surface pressure than negative
anomalies (associated with La Niña conditions)~\cite{Wunsch1990,
  cai2020}.

For CPAC$^+$, the most significant contributions are the redundant
[CPAC, EPAC]$\rightarrow$CPAC$^+$ and unique EPAC$\rightarrow$CPAC$^+$
causalities. We find that the values of EPAC and CPAC$^+$ contributing
most significantly to EPAC$\rightarrow$CPAC$^+$ come from the extreme
values of EPAC and CPAC$^+$, far from the mean. In contrast, EPAC
values near the mean are identified as non-causal, as their causality
to CPAC$^+$ is already explained by the redundant component [CPAC,
  EPAC]$\rightarrow$CPAC$^+$, with causal states concentrated around
positive values of EPAC and CPAC$^+$. This result supports previous
observations~\cite{Pan2024}, where it was reported that positive
temperature anomalies in the Central Pacific can originate from
positive temperature anomalies in the Eastern
Pacific~\cite{timmermann2018}.

Overall, our findings indicate that causal states for sea surface
temperatures in the Central Pacific (CPAC) and Eastern Pacific (EPAC),
as well as surface pressure in the Western Pacific (WPAC),
predominantly occur when all three variables are either simultaneously
above (EPAC $ > 0 $, CPAC $ > 0 $, WPAC $ > 0 $) or below (EPAC $ < 0
$, CPAC $ < 0 $, WPAC $ < 0 $) their respective means. In contrast,
non-causal states emerge when there is a mismatch in trends between
CPAC and WPAC, while EPAC remains near its mean. Specifically,
non-causal interactions are observed when CPAC is below the mean and
WPAC is above (CPAC $ < 0 $, WPAC $ > 0 $), or vice versa (CPAC $ > 0
$, WPAC $ < 0 $).

\section*{Discussion}





We have introduced an information-theoretic causal inference method
that decomposes causality according to both the system state and the
interaction type—namely, synergistic, unique, and redundant.  This
dual decomposition allows for the identification of causal system
configurations, distinguishing them from non-causal states. The
approach further enables visualization of causality over time,
facilitating the detection of causal pathways.  A key feature of the
method is that summing the state-dependent causal and non-causal
contributions recovers the SURD causality~\cite{surd}, which has
previously been shown to outperform existing methods for causal
inference across diverse scenarios.

Two benchmark cases were designed to represent common causal pathways:
source-dependent and target-dependent causality. In both systems, the
direction and strength of causal influence vary with the state of the
interacting variables. Although the resulting SURD causalities for the
two systems appear similar, their underlying causal flows are
fundamentally different---a distinction clearly captured by our
state-dependent causal inference method
(Fig.~\ref{fig:example-contributions}). We have also compared our
method to a state-dependent variant of conditional transfer entropy
(CTE), detailed in the Methods section. Although CTE enables a state-wise evaluation of causality, it
fails to detect source-dependent and target-dependent causal
structures. This limitation arises because CTE aggregates all causal
contributions into a single measure, without distinguishing between
synergistic, unique, and redundant components.  In general, we
anticipate that existing causal inference methods---such as Granger
causality, PCMCI, or convergent cross mapping---will likewise be
unable to differentiate between source- and target-dependent causal
pathways, since they rely on average causal strength and do not
disentangle redundant and synergistic interactions.

We have applied our method to two real-world systems that had
not previously been examined through the lens of state-level
causality: scale interactions in a turbulent boundary layer and the
Walker circulation in climate dynamics. In the turbulent boundary
layer, our analysis revealed causal flow structures from the outer to
the inner layer, demonstrating that high-speed streaks modulate
near-wall motions in a state-specific manner
(Fig.~\ref{fig:inn-out-intensities}). In the Walker circulation case,
we found that causal interactions between sea surface temperatures and
surface pressure anomalies in the Eastern, Central, and Western
Pacific emerge only when these variables deviate coherently from their
mean values—highlighting the asymmetry between El Niño and La Niña
regimes (Fig.~\ref{fig:climate}). These detailed causal relationships
are not captured by traditional causal inference methods that rely on
state-aggregated measures.

Finally, it is informative to compare our method with other approaches
proposed in the literature for analyzing the impact of source states
on a target variable. Among these methods are local
TE~\cite{lizier2008}, specific TE~\cite{Darmon2017specific}, local
GC~\cite{Stramaglia2021local}, and the time-varying Liang--Kleeman
(TvLK) information flow~\cite{tvlk}. It is important to note that
neither local TE nor specific TE are explicitly designed for causal
inference---as discussed by their respective authors---but rather
aim to quantify local information dynamics. Technical details of each
approach are provided in the Supplementary Materials, and results for
the two benchmark systems shown in
Fig.~\ref{fig:example-contributions} are discussed in Methods. Here,
we offer a brief summary.
%
Local TE decomposes information transfer across both source and target
states, yielding informative (positive) and misinformative (negative)
contributions, while local GC can be viewed as a variation of local TE
tailored for systems with linear relationships. Specific TE provides a
decomposition based solely on source states, which has the benefit of
yielding only positive contributions. Finally, TvLK relies on an
estimation of the original Liang--Kleeman information flow
formulation, employing a square-root Kalman filter to estimate
covariance matrices over time~\cite{tvlk}.  Although these approaches
have proven useful for analyzing dynamical systems, they are not well
suited to the problems considered here---namely, source-dependent and
target-dependent systems---as their outcomes diverge from the
intuitive causal pathways expected in such cases (see
\textit{Methods}).  These limitations may be due to assumptions of
linearity, difficulties in selecting an appropriate temporal scale,
and the inability to distinguish redundant, unique, and synergistic
contributions.

Overall, by providing a state-dependent, interaction-aware
decomposition of causality, our method contributes a new perspective
to the study of causal inference in complex systems. It offers a means
to disentangle pathways for causal relationships and to identify
specific conditions under which variables interact synergistically or
redundantly. We hope that this approach can complement existing
methods and help deepen our understanding of multiscale, nonlinear
systems in fields such as climate science, fluid mechanics,
neuroscience, and beyond.

\section*{Materials and methods}

\subsection*{Assumptions for causal discovery}

The causal inference method proposed in this study follows an
\textit{observational} (i.e., non-interventional) approach within a
probabilistic framework, inferring causal relationships from the
transition probabilities between system states under the assumption
that explicit interventions are not necessary. Specifically, we employ
the notion of \textit{causality} proposed in Ref.~\cite{surd}, in which causal influence is
quantified through the informational gain associated with observing
individual or groups of variables. The state-dependent causalities
identified by our framework are consistent with the total causal
contributions obtained from the SURD formulation~\cite{surd}.

We assume the \textit{causal Markov condition}, which states that all
relevant probabilistic information about a variable is contained in
its direct causes (i.e., parents), rather than in indirect
dependencies. This condition ensures that the inferred causal
structure accurately reflects the true generative mechanism of the
system. In addition, the method assumes \textit{faithfulness}, i.e.,
observed statistical dependencies (or independencies) arise from the
underlying causal structure itself, rather than from pathological
parameter configurations or coincidences.

We further impose the principle of \textit{forward-in-time information
  propagation}~\cite{lozano2022}, which prohibits backward-in-time
causation. To account for latent influences, our formulation
incorporates the concept of a \textit{causality
  leak}~\cite{lozano2022,surd}, which quantifies the portion of causal
influence that remains unexplained by the observed variables. This
allows us to relax the standard assumption of \textit{causal
  sufficiency}, which traditionally requires all common causes to be
explicitly measured.

Finally, the method is \emph{model-free}, requiring no prior knowledge
of the governing equations or dynamics of the system. This generality
makes the approach applicable to a wide variety of multivariate
systems, whether deterministic or stochastic, and irrespective of
whether the dependencies are linear or nonlinear. However, the
analysis assumes that statistical properties of the input time series
remain invariant over time. This assumption enables the estimation of
slowly varying non-stationary dynamics, provided there is sufficient
statistical support within those periods.

\subsection*{Decomposition of redundant, unique, and synergistic causalities by states}

To perform the decomposition proposed in Eq.~\eqref{eq:surd}, we start
with the SURD approach, which is based on a decomposition of mutual
information~\cite{shannon1948, kullback1951, Kreer1957}.  The mutual
information between a target variable $Q_j^+$ and a vector of source
variables $\bQ$ is given by:
\begin{equation}
    I(Q_j^+;\bQ) = \sum_{q_j^+\in Q_j^+} \sum_{\bq\in \bQ} p(q_j^+,\bq) \log_2 \left(\frac{p(q_j^+,\bq)}{p(q_j^+)p(\bq)}\right)  \geq 0,
\end{equation}
where $q_j^+$ and $\bq$ denote the states of the target $Q_j^+$ and
source $\bQ$ variables, respectively. Mutual information measures how
different the joint probability distribution $p(q_j^+,\bq)$ is from
the hypothetical distribution $p(q_j^+)p(\bq)$, where $q_j^+$ and
$\bq$ are assumed to be independent. For instance, if $Q_j^+$ and
$\bQ$ are not independent, then $p(q_j^+,\bq)$ will differ
significantly from $p(q_j^+)p(\bq)$. The portion of information in 
$Q^+_j$ that remains unexplained by the source variables $\bQ$ is 
referred to as the causality leak. This can be quantified in closed 
form as the conditional Shannon information $H(Q_j^+|\bQ)$ \cite{shannon1948}:
\begin{equation}
    H(Q_j^+\vert\bQ) = \sum_{q_j^+\in Q_j^+} \sum_{\bq\in \bQ} -p(q_j^+,\bq) \log_2 \left[p(q_j^+\vert\bq)\right]  \geq 0.
\end{equation}

We decompose the mutual information into its redundant, unique, and
synergistic components for each of the states of the target variable
$q_j^+$ using the \emph{specific} mutual information
$\Is(Q_j^+=q_j^+;\bQ)$, i.e., the mutual information between a fixed
target state $q_j^+$ and all the source states in $\bQ$
\cite{DeWeese1999, surd}:
\begin{equation}
  \label{eq:specific_mutual_info}
   \Is(q_j^+;\bQ) = \sum_{\bq\in\bQ} \frac{p(q_j^+,\bq)}{p(q_j^+)}
   \log_2\left( \frac{p(q_j^+,\bq)}{p(q_j^+)p(\bq)} \right) \geq 0.
\end{equation}
For a given state $q_j^+$ of the target variable $Q_j^+$, the specific
causalities $\Is$ are computed for all the possible combinations of
source variables. These components are organized in ascending
order. The increments of information between each $\Is$, denoted by
$\Delta \Is$, define the redundant $\Delta\Is_{\bi \to j}^R$, unique
$\Delta\Is_{i\to j}^R$, and synergistic $\Delta\Is_{\bi \to j}^S$
causalities.  Fig.~\ref{fig:diagram-states}(left) shows an example
for two source variables $\bQ=[Q_1,Q_2]$ and a given state of the
target variable $q_j^+$. The example illustrates the specific mutual
information $\Is_1$, $\Is_2$, $\Is_{12}$ and the corresponding
redundant $\Delta \Is_{12}^R$ and unique $\Delta \Is_{2}^U $
increments.

The redundant, unique, and synergistic information increments are
further decomposed into causal and non-causal contributions for each
state pair $q_j^+$--$\bq$. These causal redundant, unique, and
synergistic contributions are denoted as $\Iss^R_{\bi \rightarrow j}$,
$\Iss^U_{i \rightarrow j}$, and $\Iss^S_{\bi \rightarrow j}$,
respectively. The general mathematical expressions for $\Iss^\alpha_{i
  \rightarrow j}$ with $\alpha\in\{R,U,S\}$ are detailed in the
Supplementary Materials. Here, we provide the equations particularized
for a system with two source variables $\bQ=[Q_1,Q_2]$ and a target
variable $Q_j^+$, where $\Is_1 < \Is_2 < \Is_{12}$ is satisfied for
all possible states $q_j^+\in Q_j^+$:
\begin{equation} \label{eq:Iss-decomp}
\begin{aligned}
\Iss_{12 \rightarrow j}^{R} &= p(q_j^+, q_1, q_2) \max\left(0,\log_2 \frac{p(q_j^+|q_1)}{p(q_j^+)}\right), \\
\Iss_{2 \rightarrow j}^{U} &= \sum_{q_1 \in Q_1} p(q_j^+, q_1, q_2) \max\left(0,\log_2 \frac{p(q_j^+|q_2)}{p(q_j^+|q_1)}\right), \\
\Iss_{12 \rightarrow j}^{S} &= p(q_j^+, q_1, q_2) \max\left(0,\log_2 \frac{p(q_j^+|q_1, q_2)}{p(q_j^+|q_2)}\right).
\end{aligned}
\end{equation}
%
Similar expressions can be derived for the non-causal components $\Inn
_{\bi\to j}^\alpha$, which are detailed in the Supplementary Materials.
%

%
Fig.~\ref{fig:diagram-states} illustrates the interpretation of the
unique causal and non-causal states, denoted by $\Iss^U_{2 \rightarrow
  j}$ and $\Inn^U_{2 \rightarrow j}$, respectively. In this context,
causal states (in red) indicate that  $q_2$ provides more
information about the $q_j^+$ than $q_1$, i.e., $ \log_2 p(q_j^+|q_2)
> \log_2 p(q_j^+|q_1)$.  Conversely, non-causal states (in blue)
indicate that $q_2$ offers less or the same information about $q_j^+$
than $q_1$, i.e., $ \log_2 p(q_j^+|q_2) < \log_2 p(q_j^+|q_1)$.
\begin{figure}
    \centering
    \begin{minipage}{\textwidth}

    \hspace{0.015\textwidth}
    \begin{minipage}{0.22\textwidth}





\begin{tikzpicture}[
 			bar/.style={thick,black!70,fill=black!5,
            rounded corners=.1mm},
 			barline/.style={thick,black!70,rounded corners=.1mm},
 			myfill/.style={thick,rounded corners=.1mm},
            >={Latex[length=.2cm]},
            scale=1.35
    ]
    
    \begin{normalsize}


    \def\ths{{.8,1.3,1.8}}
    \def\lbs{{1,2,12}}


    \pgfmathsetmacro{\ypaneltwo}{0}


    \draw[thick,<->] (-.6,2.5) node[anchor=north east] {$\Is$} -- (-.6,0) --+ (3.25,0);

    \draw[barline,fill=myc1!50] (-.3,\ypaneltwo) --++ (.6,0) 
        --++ (0,\ths[0])  --++ (-.6,0) -- cycle; 
    \pgfmathsetmacro{\MyPgfMathResult}{{\lbs[0]}}
    \node[anchor=north] at (0,\ypaneltwo) {$\Is_{\MyPgfMathResult}$}; 

    \foreach \y in {1} {
        \draw[bar] (\y-.5,\ypaneltwo) --++ (.6,0) 
        --++ (0,\ths[\y])  --++ (-.6,0) 
        -- cycle; 
        \pgfmathsetmacro{\MyPgfMathResult}{{\lbs[\y]}}
        \node[anchor=north] at (\y-0.2,\ypaneltwo) {$\Is_{\MyPgfMathResult}$}; 
        }

    \foreach \y in {2} {
        \draw[bar] (\y-.6,\ypaneltwo) --++ (.6,0) 
        --++ (0,\ths[\y])  --++ (-.6,0) 
        -- cycle; 
        \pgfmathsetmacro{\MyPgfMathResult}{{\lbs[\y]}}
        \node[anchor=north] at (\y-0.2,\ypaneltwo) {$\Is_{\MyPgfMathResult}$}; 
        }



    \draw[myfill,myc2,fill=myc2!50] (1-.5,\ths[0]+\ypaneltwo) --++ (.6,0) --++ 
    (0,\ths[1]-\ths[0])  --++ (-.6,0) -- cycle;


    \node[anchor=south] at (0,\ths[0]+\ypaneltwo) {$\Delta \Is^R_{12}$};
    \node[anchor=south] at (0.8,\ths[1]+\ypaneltwo) {$\Delta \Is^U_{2}$};

	
    \draw[thin] ( 1+.25,0+\ypaneltwo) --++ (0,2.5);
    
    \end{normalsize}

\end{tikzpicture}

    \end{minipage}
    \hspace{0.005\textwidth}
    \begin{minipage}{0.74\textwidth}
        \vspace{-1.3cm}
\pgfmathdeclarefunction{gauss}{3}{%
  \pgfmathparse{1/(#3*sqrt(2*pi))*exp(-((#1-#2)^2)/(2*#3^2))}%
}
\tikzmath{%
  function h1(\x, \lx) { return (9*\lx + 3*((\lx)^2) + ((\lx)^3)/3 + 9); };
  function h2(\x, \lx) { return (3*\lx - ((\lx)^3)/3 + 4); };
  function h3(\x, \lx) { return (9*\lx - 3*((\lx)^2) + ((\lx)^3)/3 + 7); };
  function skewnorm(\x, \l) {
    \x = (\l < 0) ? -\x : \x;
    \l = abs(\l);
    \e = exp(-(\x^2)/2);
    return (\l == 0) ? 1 / sqrt(2 * pi) * \e: (
      (\x < -3/\l) ? 0 : (
      (\x < -1/\l) ? \e / (8 * sqrt(2 * pi)) * h1(\x, \x*\l) : (
      (\x <  1/\l) ? \e / (4 * sqrt(2 * pi)) * h2(\x, \x*\l) : (
      (\x <  3/\l) ? \e / (8 * sqrt(2 * pi)) * h3(\x, \x*\l) : (
      sqrt(2/pi) * \e)))));
  };
}

\begin{tikzpicture}[scale=1.5]

    \def\B{13.3};   
    \def\S{16};   
    \def\T{7}; 

    \def\Bs{3.10};  
    \def\Ss{3.20};  
    \def\Ts{4.40};  

    \def\alphaT{-0.3}; 
    \def\alphaS{1}; 
    \def\xIntersect{13.25};

    \def\xmax{25.5};
    \def\ymin{{-0.15*gauss(\B,\B,\Bs)}};
 
    \begin{axis}[every axis plot post/.append style={
                 mark=none,domain={-0.5*(\xmax)}:{1.2*\xmax},samples=80,smooth},
                 xmin={-0.4*(\xmax)}, xmax=\xmax,
                 ymin=\ymin, ymax={1.3*gauss(\B,\B,\Bs)},
                 axis lines=middle,
                 axis line style=thick,
                 enlargelimits=upper, 
                 ticks=none,
                 xlabel=$q_j^+$,
                 y axis line style={opacity=0},
                 x label style={at={(axis description cs:0.69,0.065)},anchor=east},
                 width=9cm, height=5cm,
                ]

      \addplot[name path=B, thick,black!10!black] {0.95*gauss(x,\B,\Bs)};

      \addplot[name path=T, thick, black!10!myc2] {skewnorm((x-\T)/\Ts, \alphaT) / \Ts};
      \addplot[name path=S, thick, black!10!myc1] {0.5*skewnorm((x-\S)/\Ss, \alphaS) / \Ss};
    
      \addplot [very thick, myc2!20, forget plot] fill between [
        of=T and S, soft clip={domain=-25:\xIntersect}];
      \addplot [very thick, myc1!20, forget plot] fill between [
        of=T and S, soft clip={domain=\xIntersect:50}];  
 
      \begin{huge}
      \node[above,black!20!black ] at (260,142) {$p(q_j^+)$};
      \node[above,black!20!myc2 ] at (185,112) {$p(q_j^+|q_2)$};
      \node[above,black!20!myc1 ] at (315,95) {$p(q_j^+|q_1)$};
      \draw[-, align=center] (150,60) --++ (-40,25) node[anchor=east, align=center, yshift=10] {causal \\[2pt] $\Iss^U_{2\to j}>0$};

      
      \draw[-, align=center] (317,60) --++ (32,25) node[anchor=west, align=center, yshift=10,xshift=-2] {non-causal \\[2pt] $\Inn^U_{2\to j}<0$};
      \end{huge}

    \end{axis}


\end{tikzpicture}
    \end{minipage}

    \end{minipage}
    \caption{ Left: Ranking of specific mutual information between a
      target state $q_j^+$ and all combinations of two source
      variables $\bQ = [Q_1, Q_2]$, namely $\Is_1$, $\Is_2$,
      and $\Is_{12}$. The redundant increment $\Delta \Is_{12}^R$ and
      the unique increment $\Delta \Is_{2}^U$ are computed based on
      this ranking.  Right: State-dependent unique causality
      $\Iss^U_{2 \to j}$ and non-causality $\Inn^U_{2 \to j}$ for a
      system with target variable $Q_j^+$ and source variables $Q_1$
      and $Q_2$, as defined by Equation~\ref{eq:Iss-decomp}.  In this
      example, it is assumed that the specific mutual information
      satisfies $\Is_2 > \Is_1$ for all $q_j^+ \in Q_j^+$.  Positive
      values of $\Iss^U_{2 \to j}$ correspond to states where
      knowledge of $q_2$ increases the likelihood of the target state
      $q_j^+$ relative to $q_1$, i.e., $p(q_j^+ \mid q_2) > p(q_j^+
      \mid q_1)$, indicating unique causality.  Conversely, negative
      or zero values (denoted by $\Inn^U_{2 \to j}$) identify
      non-causal states where $q_2$ provides less information about
      $q_j^+$ than $q_1$. i.e., $p(q_j^+ \mid q_2) < p(q_j^+ \mid
      q_1)$.  }
    \label{fig:diagram-states}
\end{figure}

The SURD causalities, $\Delta I_{\bi\to j}^\alpha$ where $\alpha\in
[R,U,S]$, are recovered by summing the causal and non-causal
contributions across all possible states:
\begin{equation}\label{eq:contribution-surd}
    \Delta I_{\bi\to j}^\alpha = \sum_{q_j^+\in Q_j^+} \sum_{\bq\in
      \bQ} \left[ \Iss_{\bi\to j}^\alpha (q_j^+;\bq) + \Inn_{\bi\to
        j}^\alpha(q_j^+;\bq) \right].
\end{equation}
%
Note that in SURD, the average causal effect of one variable on
another is obtained by accounting for both causal and non-causal
contributions. There are two main reasons for this
construction. First, including both causal and non-causal
contributions ensures consistency with the mutual information between
variables, as illustrated in Fig.~\ref{fig:surd-states}. This also
guarantees consistency with the forward-in-time propagation of
information condition from Eq.~\eqref{eq:conservation_info}.  The
second reason is the need to define an average causal measure that
discounts instances where knowledge of the source variable actually
increases uncertainty about the target. In practice, if only positive
(causal) contributions were considered, one might mistakenly deem a
variable $Q_1$ causal to another variable $Q^+$ based on a small
subset of states $q_1$ where the influence is strong. However, it
could occur that, in the vast majority of instances, knowledge of
$q_1$ does not increase, and may even decrease, certainty about
$Q^+$. Including non-causal (negative) contributions ensures that
these misleading instances are properly accounted for, leading to an
average measure that more faithfully reflects the overall causal
relationship between variables across all conditions.  This is
analogous to the distinction between informative and misinformative
components in pointwise mutual information~\cite{cover2006}. As a
result, SURD avoids overestimating causality based on rare or atypical
states and provides a more robust and interpretable measure of causal
influence.

Finally, it is worth noting that other approaches for decomposing
mutual information into redundant, unique, and synergistic components
have been proposed in the literature, such as Partial Information
Decomposition (PID)~\cite{williams2010} and its
variants~\cite{griffith2014, griffith2015, ince2017, gutknecht2021,
  lozano2022, kolchinsky2022}. While these approaches provide valuable
insights into the structure of mutual information, it was discussed in
Ref.~\cite{surd} that they either fail to satisfy all the properties
required by our method or result in an unmanageable number of
terms. Hence, our formulation does not follow PID, and the meanings of
redundancy, uniqueness, and synergy adopted in this work differ from
those in PID and related approaches.  Instead, our framework focuses
on quantifying the \textit{increments of information} gained about the
states of the target variable from different combinations of source
variable states. Although this definition departs from earlier notions
of redundancy, synergy, and uniqueness~\cite{williams2010}, it enables
us, for example, to group synergistic components by specific order or
to account for them through the concept of \textit{causality leak},
thanks to the additivity of causal components. In our view, this
simplifies the interpretation of scenarios involving many variables,
where the number of decomposed terms can become prohibitively large,
or situations where the number of observations is very low. For
instance, PID results in a number of terms that grows according to the
Dedekind numbers; in the case of nine variables, this decomposition
yields over $10^{23}$ terms, whereas our method produces 512 causal
terms (along with 512 non-causal counterparts)---which, although
still large, remains computationally feasible.  Despite our different
notion of redundancy, uniqueness, and synergy, by focusing on
incremental information, our method ensures that causality is not
double-counted and that the minimal set of variables containing
information about the target states can be systematically identified.

\subsection*{State-dependent conditional transfer entropy}

We use a state-dependent version of conditional transfer entropy (CTE)
to quantify directed information transfer between a target variable
$Q_j^+$ and a vector of source variables $\bQ$. The CTE from a source
variable $Q_i$ to the target $Q_j^+$ is defined as:
\begin{equation}
    \overline{\text{CTE}}_{i \to j} = H(Q_j^+ | \bQ_{\cancel{i}}) - H(Q_j^+ | \bQ),
\end{equation}
where $\bQ_{\cancel{i}}$ denotes the vector $\bQ$ excluding the
component $Q_i$, and $H(\cdot | \cdot)$ is the conditional Shannon
entropy \cite{shannon1948}.  This expression measures the additional
information that $Q_i$ provides about $Q_j^+$ beyond what is already
available from the remaining variables in $\bQ$. By expanding the
conditional Shannon entropies in the definition above, we obtain a
state-level representation of CTE:
\begin{equation} \label{eq:cte-states}
    \text{CTE}_{i \to j} = \sum_{\bq_{\cancel{i}} \in \bQ_{\cancel{i}}} p(q_j^+, q_i, \bq_{\cancel{i}}) \log_2 \left( \frac{p(q_j^+ | q_i, \bq_{\cancel{i}} )}{p(q_j^+ | \bq_{\cancel{i}})} \right).
\end{equation}

For a system consisting of two source variables \( Q_1 \) and \( Q_2
\), and a target variable \( Q_j^+ \) with joint probability
distribution \( p(q_j^+, q_1, q_2) \), the state-dependent conditional
transfer entropy from the source variable \( Q_2 \) to the target \(
Q_j^+ \) can be expressed as:
\begin{equation}
    {\text{CTE}}_{2 \rightarrow j} = \sum_{q_1 \in Q_1}  p(q_j^+, q_1, q_2) \log_2 \left( \frac{p(q_j^+ | q_1, q_2)}{p(q_j^+ | q_1)} \right),
    \label{eq:specific_cte}
\end{equation}
which compares the log of the conditional probabilities \( p(q_j^+ |
q_1, q_2) \) and \( p(q_j^+ | q_1) \), weighted by the joint
probability \( p(q_j^+, q_1, q_2) \). In contrast to the specific
unique causality defined in Eq.~\eqref{eq:Iss-decomp}, we
observe that here the comparison is made between \( p(q_j^+ | q_1,
q_2) \) and \( p(q_j^+ | q_1) \), while in the unique causality the
comparison is between \( p(q_j^+ | q_2) \) and \( p(q_j^+ | q_1) \).

\subsection*{Comparison with other methods}

\begin{figure}
    \centering
    \includegraphics[width=0.195\linewidth]{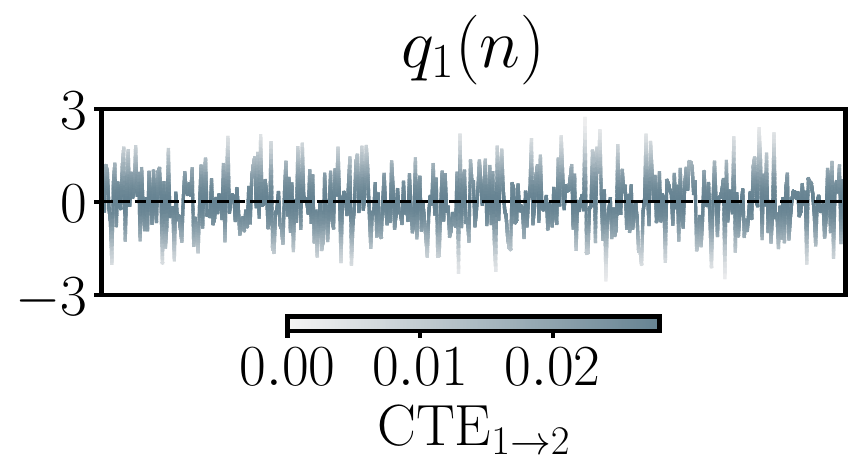}
    \includegraphics[width=0.195\linewidth]{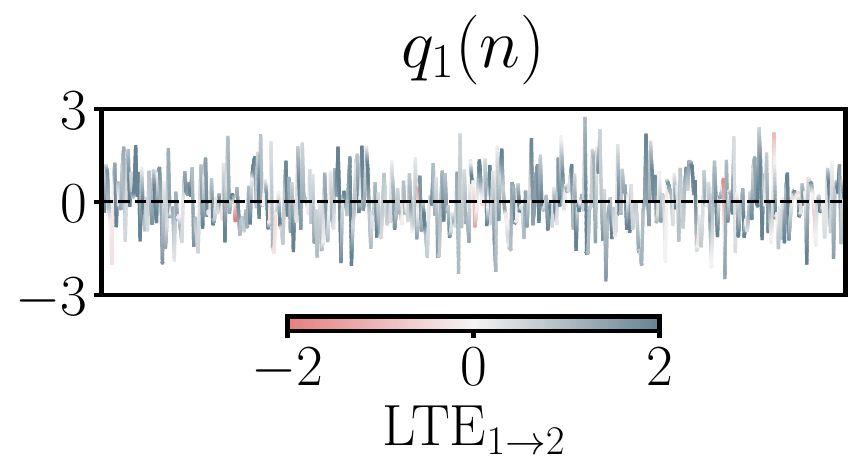}
    \includegraphics[width=0.195\linewidth]{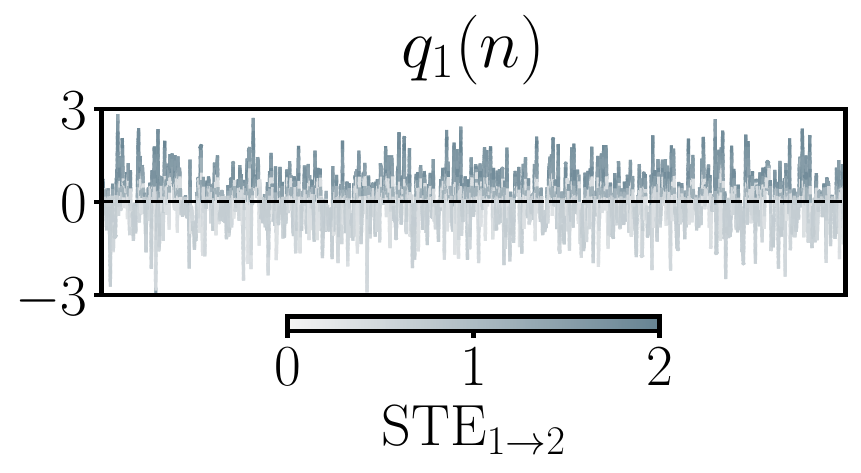}
    \includegraphics[width=0.195\linewidth]{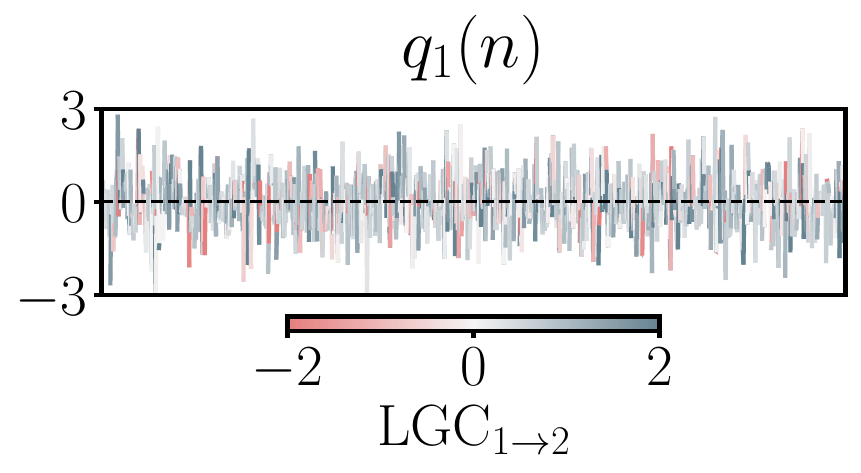}
    \includegraphics[width=0.195\linewidth]{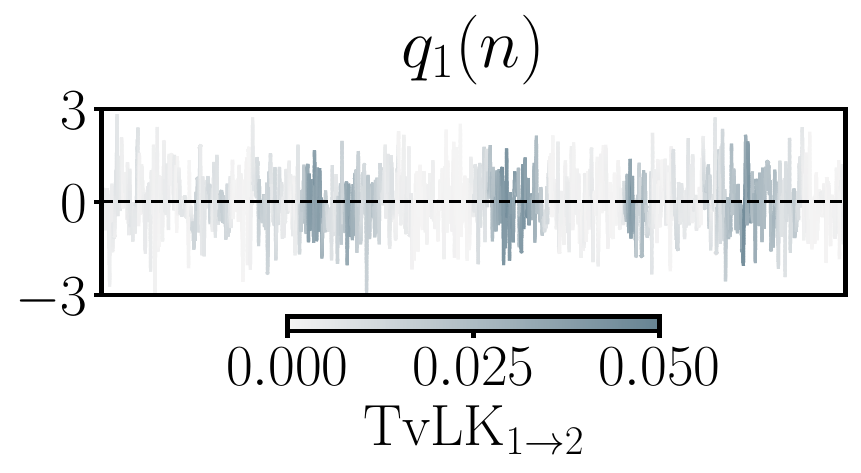}

    \includegraphics[width=0.195\linewidth]{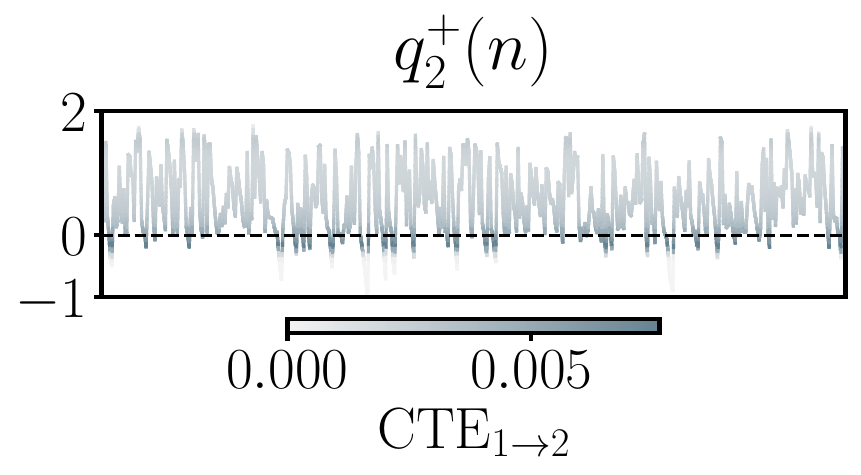}
    \includegraphics[width=0.195\linewidth]{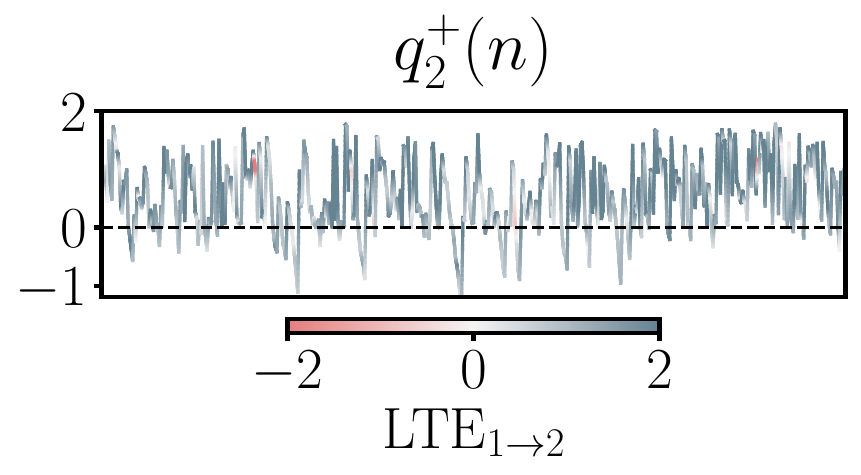}
    \includegraphics[width=0.195\linewidth]{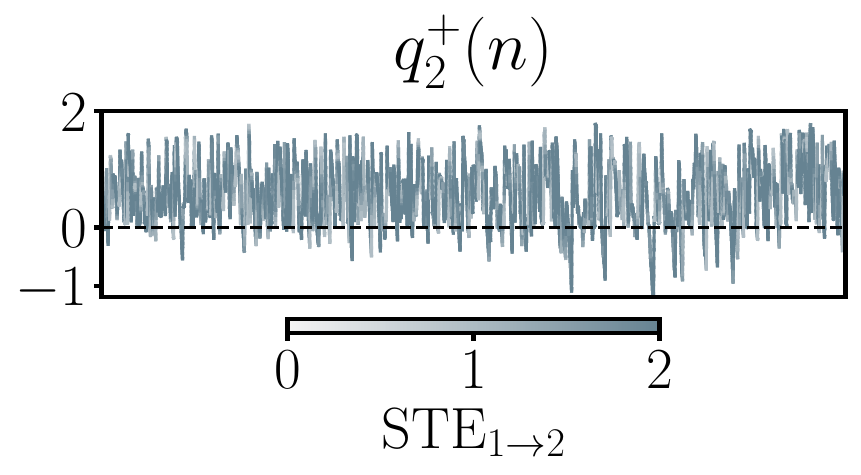}
    \includegraphics[width=0.195\linewidth]{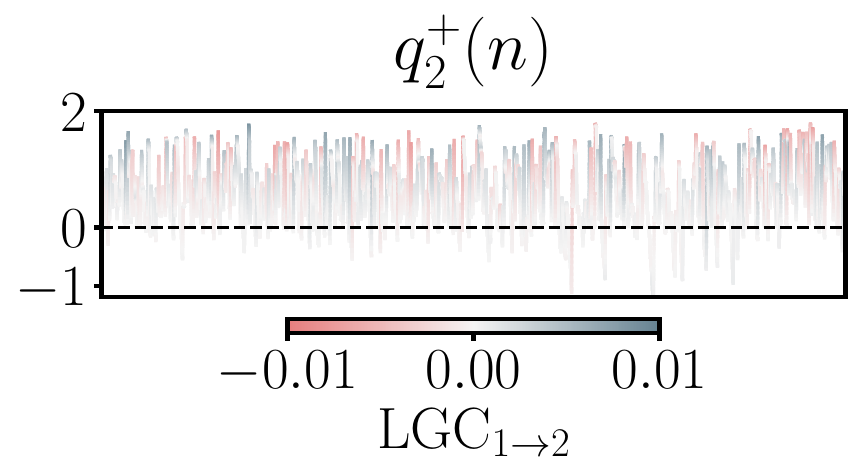}
    \includegraphics[width=0.195\linewidth]{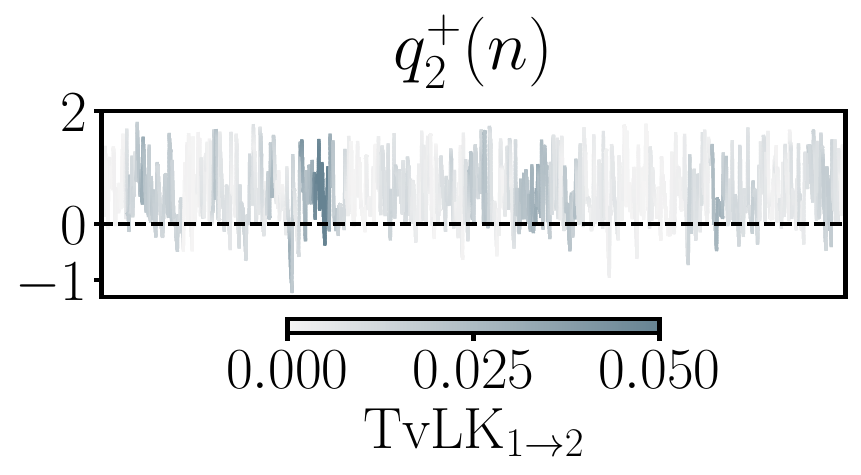}
    
    \caption{Comparison of time-varying causal inference methods,
      including state-dependent conditional transfer entropy (CTE),
      local transfer entropy (LTE) \cite{Lizier2014}, specific
      transfer entropy (STE) \cite{Darmon2017specific}, local Granger
      causality (LGC) \cite{Stramaglia2021local}, and time-varying
      Liang--Kleeman (TvLK) information flow \cite{tvlk}. Results are
      shown for the two benchmark cases from
      Fig.~\ref{fig:example-contributions}: (top) System with
      source-dependent causality and (bottom) system with
      target-dependent causality.}
    \label{fig:toy-methods}
\end{figure}

We compare the results for the two benchmark cases shown in
Fig.~\ref{fig:example-contributions} using five different methods that
estimate time-varying measures of influence between variables: a
state-dependent variant of conditional transfer entropy (CTE), local
transfer entropy (LTE)~\cite{Lizier2014}, specific transfer entropy
(STE)~\cite{Darmon2017specific}, local Granger causality
(LGC)~\cite{Stramaglia2021local}, and time-varying Liang--Kleeman
(TvLK) information flow~\cite{tvlk}. LTE and STE are designed to
quantify interactions among variables in dynamical systems without
necessarily asserting an underlying causal structure, whereas LGC and
TvLK are explicitly formulated as causal inference methods.

For LTE and STE, we employ the implementation by Darmon and
Rapp~\cite{Darmon2017specific}, where LTE estimates are obtained using
a $k$-nearest neighbors ($k$-NN) estimator with $k=5$ neighbors
\cite{ksg2004}. Following the authors' recommendations, the STE
estimates are derived from LTE using $k_{\text{reg}} = \lfloor
\sqrt{T} \rfloor$, where $T$ denotes the total number of samples. This
choice controls the degree of smoothing and balances the bias-variance
trade-off in the estimation of STE.  For LGC, we implement the
formulation proposed in Ref.~\cite{Stramaglia2021local} and validate
our implementation by reproducing the toy examples presented in that
study. For TvLK, we adopt the implementation described in
Ref.~\cite{tvlk}, applying a uniformly weighted moving average (UWMA)
filter with a window length of 10 points, and a sliding estimation
window of 100 points. We verified that varying the estimation window
length between 30 and 150 points did not affect the conclusions.
Further details on each method and their implementation are provided
in the Supplementary Materials.

We applied the five methods described above to the source- and
target-dependent causality cases considered in the validation
section. The results are visualized in Fig.~\ref{fig:toy-methods},
which shows the time evolution of $q_1$ (for source-dependent
causality) and $q_2^+$ (for target-dependent causality). The signals
are colored according to the value of CTE, LTE, STE, LGC, or TvLK.  The
expectation is that, for source-dependent causality, only instances
where $q_1>0$ should exhibit strong causal influence from $q_1$ to
$q_2^{+}$, while for target-dependent causality, strong causal
influence from $q_1$ to $q_2^{+}$ should be detected when $q_2^{+}>0$
However, the results reveal that none of the methods consistently
exhibit the expected trends.

Among the methods evaluated, STE performs best in the source-dependent
case, where larger values are observed for $q_1 > 0$, consistent with
the dependence $q_1 \to q_2^+$ when $q_1 > 0$. However, STE remains
nonzero even for $q_1 < 0$, where $q_2$ is also required to predict
$q_2^+$, indicating that STE does not fully resolve the dependence
structure of the system. In the case of target-dependent causality,
STE projects the results from LTE onto the future states of the target
variable. As a consequence, the results for STE in the
target-dependent case are not consistent with the functional
dependencies inherent to the system.

LTE, LGC, and TvLK do not offer clear evidence for either
source-dependent or target-dependent interaction. This outcome is not
surprising for LGC and TvLK, as these methods are constrained by the
assumptions of linearity and Gaussianity. In the present cases, the
dependencies between variables are nonlinear, complicating the
detection of the \textit{a priori} causal dependencies. TvLK estimates
time-varying causal relationships by applying a square-root Kalman
filter to track the evolution of covariance matrices based on the
original Liang--Kleeman information flow formulation. While this
approach can be effective when the underlying probability
distributions evolve smoothly over time, it may lack the sensitivity
required to detect abrupt, short-time-scale changes in causality, such
as those presented in the benchmark cases shown in
Fig.~\ref{fig:example-contributions}.

Finally, in CTE, all redundant, unique, and synergistic causal
contributions are aggregated into a single measure, obscuring the
interpretation of the interactions between variables. For example, CTE
fails to reveal in the source-dependent case that when $q_1 > 0$, the
variable $Q_1$ uniquely determines the future of $Q_2$, whereas when
$q_1 < 0$, both $Q_1$ and $Q_2$ are jointly required to predict
$Q_2^+$. As a result, the detailed structure of variable interactions
is lost when all forms of causality are conflated into a single
quantity.

In closing this comparison, it is important to emphasize that the
results produced by the methods discussed above are not incorrect, nor
do we suggest otherwise. Each of these approaches offers valuable
insights into interactions and information dynamics in complex
systems, within the scope of their respective assumptions and intended
applications, as demonstrated in prior studies. The key conclusion
here is that these methods are not specifically designed to quantify
the types of interaction and causality targeted in the present
work. In contrast, our framework is explicitly tailored to address
these challenges, particularly in systems characterized by source- or
target-dependent causal structures.

We conclude this section by comparing the current method with our
previous approach, SURD. The key similarities and differences are
summarized in Table~\ref{tab:surd-vs-proposed}. Both SURD and the
proposed method are grounded in the information-theoretic principle of
forward-in-time information propagation. Similarly, both frameworks
decompose causality according to the type of interaction---namely,
redundant, unique, and synergistic contributions.  However, SURD
computes these causal contributions as averages over all system
states, whereas the proposed method performs the decomposition at the
level of individual states, both for the source and target
variables. As a consequence, SURD may yield similar averaged causal
attributions for systems with fundamentally different underlying
state-dependent dynamics. This limitation is illustrated in
Figs.~\ref{fig:example-contributions} and
\ref{fig:example-contributions-target}, where source-dependent and
target-dependent causal effects are discernible only through the
proposed method.  In addition, the current approach distinguishes
between causal and non-causal components, with the latter
corresponding to states whose knowledge increases uncertainty about
the target. SURD does not offer this distinction. Another limitation
of SURD is its inability to capture the temporal evolution of
causality. In contrast, the proposed method allows for tracking such
evolution, provided sufficient statistical support exists across the
observed states.  Overall, the current method enables a more detailed
and state-resolved characterization of causal structure in complex
systems, going beyond what is possible with existing causal inference
frameworks.
\renewcommand{\arraystretch}{1.4}
\begin{table}[t]
\centering
\caption{Comparison of SURD and the proposed method. A checkmark
  (\ding{51}) indicates the feature is supported; a cross mark
  (\ding{55}) indicates it is not.}
\label{tab:surd-vs-proposed}
\begin{tabular}{p{0.75\textwidth}cc}
\hline
\textbf{Capability} & \textbf{SURD} & \textbf{Proposed method} \\ \hline
Based on forward-in-time propagation of information & \ding{51} & \ding{51} \\ \hline
Interaction-specific decomposition (redundant, unique, synergistic) & \ding{51} & \ding{51} \\\hline
State-dependent causality & \ding{55} & \ding{51} \\\hline
Causal vs. non-causal state distinction & \ding{55} & \ding{51} \\\hline
Temporal evolution of causality & \ding{55} & \ding{51} \\\hline
\end{tabular}
\end{table}
\renewcommand{\arraystretch}{1}

\subsection*{Limitations}

We discuss here some of the limitations of our method. First, the
proposed causal framework is observational in nature---i.e., it infers
causal relationships from statistical dependencies in time-resolved
data without requiring interventions. While this allows for broad
applicability to real-world systems where interventions are infeasible
or unethical, it also introduces limitations: observational causality
may not coincide with interventional or counterfactual definitions of
causality, and can be confounded by hidden variables or latent
dynamics.

Another key limitation is that the method is data-intensive, as it
requires the estimation of probability distributions, which becomes
increasingly challenging as the dimensionality of the dataset
grows. This challenge primarily affects the quantification of
synergistic causalities, which rely on higher-dimensional joint
distributions compared to redundant and unique contributions. To
mitigate this, our method permits restricting the order of synergistic
interactions (e.g., limiting to pairwise synergistic causalities in
systems with many source variables and limited data). The unestimated
higher-order contributions can then be interpreted as a causality
leak, thanks to the additive structure of our causal
decomposition. Furthermore, recent advances in distribution estimation
techniques, such as transport maps~\cite{baptista2023} and flow
matching~\cite{mathieu2024flow}, can enhance the applicability of our
approach to high-dimensional systems.

Another important consideration is that our method requires defining a
partition of the state space of the system. In this work, we employed
a uniform partition based on the values of the signals. However, in
other contexts, phase-space partitioning may be tailored to the
specific research question---for example, by distinguishing between
regimes with and without extreme events. We believe that recent
advances in quantized deep autoencoders, such as vector-quantized
variational autoencoders~\cite{van2017neural} and finite scalar
quantization methods~\cite{mentzer2024finite}, offer promising tools
for constructing more efficient and data-adaptive state-space
partitions.

\subsection*{Data for benchmark cases}

The source-dependent and target-dependent benchmark cases comprise a
system of two variables $Q_1$ and $Q_2$ at discrete times $t_n =
n$. The system is initially set to $q_1(1) = q_2(1)= 0$. A stochastic
forcing, represented by $W_i$, acts on $Q_i$ and follows a Gaussian
distribution. The computation of causalities is performed for a time
lag of $\Delta T=1$ using 75 uniform bins per variable. The
integration of the system is carried out over $10^8$ time steps, with
the first 10,000 steps excluded from the analysis to avoid transient
effects. In the analysis presented in
Fig.~\ref{fig:example-contributions}, the causality leak for $q_2^+ =
q_2(n+1)$ is 53\% in the source-dependent case and 54\% in the
target-dependent case. This indicates that only 47\% and 46\% of the
causality to $Q_2^+$ can be accounted for using $Q_1$ and $Q_2$
alone. The remaining unexplained part is due to the influence of
$W_2$, which also affects $Q_2^+$ but is not included in the
analysis. An additional study with a system of three variables $Q_1$,
$Q_2$, and $Q_3$ with similar functional relationships is reported in
the Supplementary Materials, where we also discuss the structure of
the state-dependent non-causal components.

\subsection*{Data for turbulent boundary layer}

\begin{figure}
    \centering
    \begin{center}
    \scalebox{1.05}{
    \hspace{-0.6cm}
    \begin{minipage}{\textwidth}
    \begin{tikzpicture}
        \node[anchor=north west] (img) at (0, 0) {\includegraphics[height=0.215\linewidth]{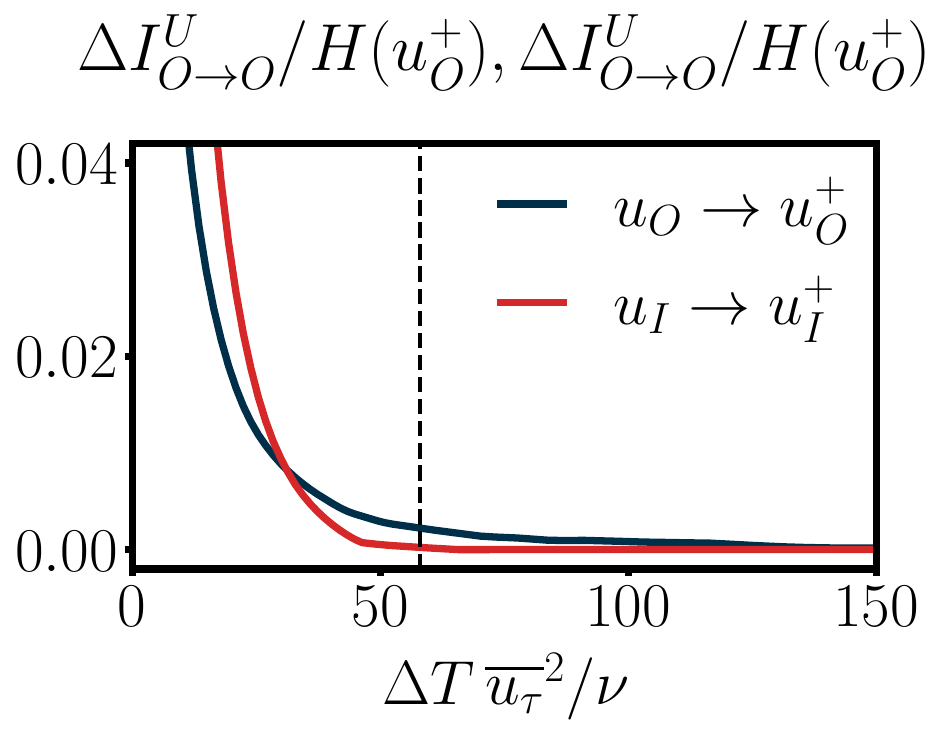}};
        \node[anchor=south, font=\normalsize, fill=white, text=black, inner sep=2pt, rounded corners, minimum width=0.2\textwidth,minimum height=6mm] at ([yshift=-6.25mm,xshift=0mm]img.north) {$\Delta I _ {O\to O} / H(u_O^+),\,\Delta I _ {I\to I} / H(u_I^+)$};
    \end{tikzpicture}
    \hspace{0.15cm}
    \begin{tikzpicture}
        \node[anchor=north west] (img) at (0, 0) {\includegraphics[height=0.215\linewidth]{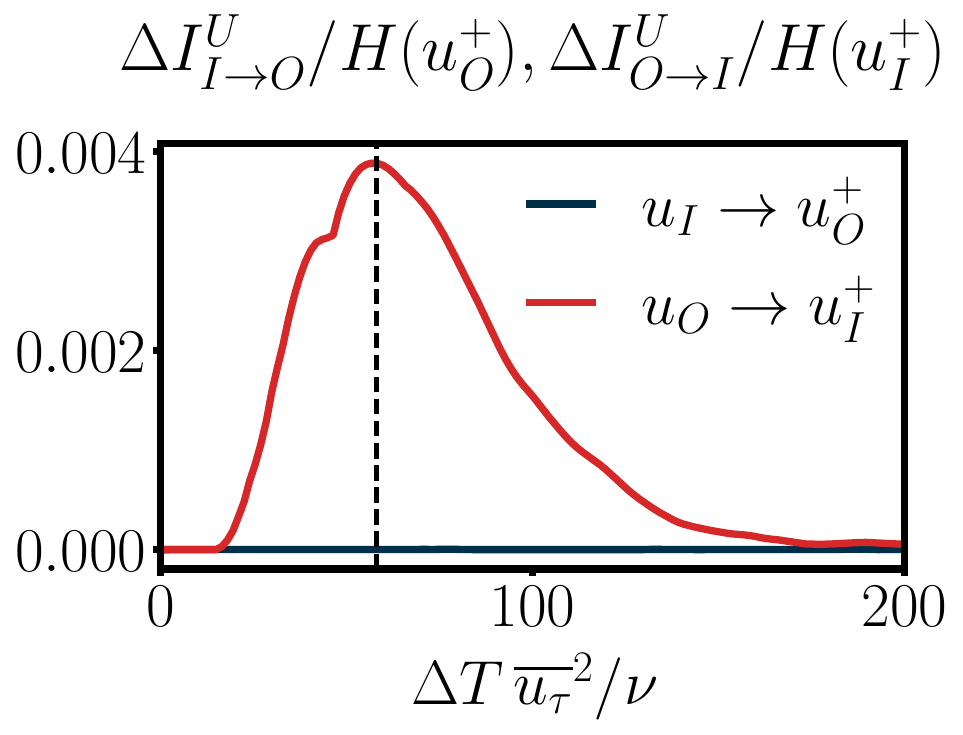}};
        \node[anchor=south, font=\normalsize, fill=white, text=black, inner sep=2pt, rounded corners, minimum width=0.2\textwidth,minimum height=6mm] at ([yshift=-6.25mm,xshift=0mm]img.north) {$\Delta I _ {I\to O} / H(u_O^+),\,\Delta I _ {O\to I} / H(u_I^+)$};
    \end{tikzpicture}
    \hspace{0.15cm}
    \begin{minipage}{0.33\textwidth}
    \vspace{-3.51cm}
    \includegraphics[width=\textwidth,trim=0 200 0 0, clip]{figures/inner_outer_self_surd.pdf}

    \includegraphics[width=\textwidth,trim=0 0 0 270, clip]{figures/inner_outer_self_surd.pdf}
    \end{minipage}
    \end{minipage}}
    \end{center}

    \vspace{-0.6cm}
    \caption{Left: Self-induced ($\Delta I_{O \to O}^U$, $\Delta I_{I
        \to I}^U$) and cross-induced ($\Delta I_{I \to O}^U$, $\Delta
      I_{O \to I}^U$) SURD causalities as a function of the time
      lag. Time is normalized using inner units based on the
      streamwise-averaged friction velocity $\overline{u_\tau}$ and
      viscosity $\nu$ and causalities with the Shannon information of
      the corresponding target variable, i.e. $H(u)$. The dashed
      vertical line denotes the time lag selected for the causal
      analysis. Right: redundant ($\Delta I^R_{OI \to I}$), unique
      ($\Delta I^U_{(\cdot) \to I}$), and synergistic ($\Delta I^S_{OI
        \to I}$) causalities from SURD to the outer-layer motions,
      $u_O^+ = u_O(t + \Delta T)$.}
    \label{fig:inner-outer-caus-time}
\end{figure}

The data used for analyzing inner/outer interactions in a turbulent
boundary layer were obtained for a dataset of a direct numerical
simulation of the Navier--Stokes equations
\cite{towne_lozano-duran_2022}, where all temporal and spatial scales
were resolved. The dataset covers the range of friction Reynolds
number from $Re_\tau \approx 292$ to $729$, with 10,000 flowfields
stored spanning 26 eddy-turnover times (after transients) based on
$\overline{\delta}/\overline{u_\tau}$, where the symbol
$\overline{(\cdot)}$ denotes the average along the streamwise
direction $x$. Further details about the numerical setup can be found
in Towne et al. \cite{Towne2023}. The data used in this study includes
the temporal evolution of the streamwise velocity at two different
wall-normal locations: $y_I = 4 \nu/u_\tau$ (for the inner layer) and
$y_O = 0.3\delta$ (for the outer layer). The analysis was conducted at
every streamwise location within the boundary layer, leveraging the
self-similarity of the flow, which results in $2\times10^7$ samples
for causal analysis. The time lag utilized to evaluate causality was
$\Delta T = 60 \nu/u_\tau^2$, which corresponds to the time lag for
maximum cross-induced unique causality. At this time lag, the
causality leak to the inner-layer velocity $u_I^+$ is 99\%. This high
value is expected, given that the vast majority of the degrees of
freedom of the system (associated with the full turbulent flow field)
are not accounted for~\cite{surd}.

The temporal evolution of the self-induced causalities ($\Delta I_{O
  \to O}^U$ and $\Delta I_{I \to I}^U$) and cross-induced causalities
($\Delta I_{I \to O}^U$ and $\Delta I_{O \to I}^U$) is presented in
Fig.~\ref{fig:inner-outer-caus-time}. The causalities directed toward
the outer-layer velocity $u_O^+$ at the same time lag are also shown
in Fig.~\ref{fig:inner-outer-caus-time}, revealing no significant
contribution from the inner-layer velocity $u_I$. Finally, the
probability distributions were discretized using 50 uniform bins per
variable. Additional tests conducted with half and double the number
of bins produced no significant differences in the results.

\subsection*{Data for climate science}

The climate time series data are regional averages from the reanalysis
for the period 1948-2012 with 780 months. WPAC denotes monthly surface
pressure anomalies in the West Pacific, and CPAC and EPAC surface air
temperature anomalies in the Central and East Pacific. Anomalies are
taken with respect to the whole period, which is publicly available in
\url{https://psl.noaa.gov/data/gridded/data.ncep.reanalysis.html}. The
time lag utilized to evaluate causality is $\Delta T = 4$ months,
which maximized the unique causality across different variables and is
within the range of lags analyzed in Runge \textit{et al.}
\cite{runge2019pcmci}. At this time lag, the causality leak to
WPAC$(t+\Delta T)$ and CPAC$(t+\Delta T)$ is 79\% and 65\%,
respectively, which quantifies the effect from variables not included
in the analysis.  The joint distribution of the variables is estimated
using a $k$-NN density estimator with $k = 7$ neighbors. Alternative
values of $k$ within the range [4,8] were tested, showing no
significant impact on the results. To quantify uncertainty in the
results, we performed 50 bootstrap resamplings on the available
data. Additionally, to assess statistical significance, we applied
1000 random permutations to compute the $p$-values of the redundant and
unique causalities reported in Fig.~\ref{fig:climate}, both of which
yielded $p$-values of zero. In contrast, when including synergistic
causalities, the $p$-values increased significantly, leading us to
conclude that the latter causalities lack sufficient statistical
support with the available data. More details about the statistical
significance of the results for this case are shown in the
Supplementary Materials, where we include the results for reduced
sample sizes.


\section*{Code availability}
The codes developed for this work are available at:
\url{https://github.com/ALD-Lab/SURD-states}.

\section*{Acknowledgements}

The authors would like to thank Jakob Runge for providing the data for
the climate science application and Gonzalo Arranz for his
contributions to this work.  This work was supported by the National
Science Foundation under Grant No. 2140775 and MISTI Global Seed Funds
and UPM. Á.~M.-S. received the support of a fellowship from the "la
Caixa" Foundation (ID 100010434). The fellowship code is
LCF/BQ/EU22/11930094. The authors acknowledge the MIT SuperCloud and
Lincoln Laboratory Supercomputing Center for providing HPC resources
that have contributed to the research results reported within this
paper.

\bibliography{main}

\section*{Author contributions statement}
Á.~M.-S.: Methodology, Software, Validation, Investigation, Data
Curation, Writing – Original Draft, Writing – Review \& Editing,
Visualization. A.~L.-D.: Ideation, Methodology, Writing – Review
\& Editing, Supervision, Resources, Funding acquisition.

\section*{Competing Interests}
The authors declare no competing interests.

\end{document}


\maketitle
\tableofcontents
\renewcommand{\thesection}{S\arabic{section}}
\renewcommand{\thesubsection}{\thesection.\arabic{subsection}}
\renewcommand{\thesubsubsection}{\thesubsection.\arabic{subsubsection}}
\newpage

\section{Method formulation}

\subsection{Fundamentals of information theory}

Consider $N$ quantities of interest at time $t$ represented by the
vector of observable variables $\bQ = [Q_1(t),Q_2(t), \ldots,Q_N(t)]$.
We treat $\bQ$ as a random variable and consider a finite partition of
the observable phase space $D =\{D_1, D_2, \dots, D_{N_D}\}$, where
$N_D$ is the number of cells, such that $D = \cup_{i=1}^{N_D} D_i$ and
$D_i \cap D_j = \emptyset$ for all $i \neq j$ (i.e., disjoint cover of
the observable phase space $D$).  We use upper case $Q$ to denote the
random variable itself; and lower case $q$ to denote a particular
state contained in one $D_i$ (also referred to as a value or an event)
of $Q$.  The probability of finding the system at state $D_i$ at time
$t$ is $p( \bQ(t) \in D_i )$, that in general depends on the partition
$D$.  For simplicity, we refer to the latter probability as $p(\bq)$.

The information contained in the variable $\bQ$ is given
by~\cite{shannon1948}:
\begin{equation}\label{eq:entropy}
  H(\bQ) = \sum_{\bq} -p(\bq) \log_2[p(\bq)]\geq 0,
\end{equation}
where the summation is over all the states of $\bQ$. The quantity $H$
is the Shannon information or entropy~\cite{shannon1948}. The units of
$H$ are set by the base chosen, in this case `bits' for base 2. For
example, consider a fair coin with
$Q\in\{\mathrm{heads},\mathrm{tails}\}$ such that
$p(\mathrm{heads})=p(\mathrm{tails})=0.5$. The information of the
system `tossing a fair coin $n$ times' is $H = -\sum 0.5^n
\log_2(0.5^n) = n$ bits, where the summation is carried out across all
possible outcomes (namely, $2^n$).  If the coin is completely biased
towards heads, $p(\mathrm{heads})=1$, then $H = 0$ bits (taking $0\log
0 = 0$), i.e., no information is gained as the outcome was
already known before tossing the coin.  The Shannon information can
also be interpreted in terms of uncertainty: $H(\bQ)$ is the average
number of bits required to unambiguously determine $\bQ$.  $H$ is
maximum when all the possible outcomes are equiprobable (indicating a
high level of uncertainty in the state of the system) and zero when
the process is completely deterministic (indicating no uncertainty in
the outcome).

The Shannon information of $\bQ$ conditioned on another variable
$\bQ'$ is defined as:
\begin{equation}\label{eq:entropy conditional}
H( \bQ | \bQ' ) = \sum_{\bq,\bq'} -p(\bq,\bq') \log_2[p(\bq|\bq')].
\end{equation}
where $p(\bq|\bq') = p(\bq,\bq')/p(\bq')$ with $p(\bq')\neq 0$ is the
conditional probability distribution, and $p(\bq') = \sum_{\bq}
p(\bq,\bq')$ is the marginal probability distribution of $\bq'$. It is
useful to interpret $H(\bQ|\bQ')$ as the uncertainty in the variable
$\bQ$ after conducting the `measurement' of $\bQ'$.  If $\bQ$ and
$\bQ'$ are independent random variables, then $H(\bQ|\bQ')=H(\bQ)$,
i.e., knowing $\bQ'$ does not reduce the uncertainty in
$\bQ$. Conversely, $H(\bQ|\bQ')=0$ if knowing $\bQ'$ implies that
$\bQ$ is completely determined. Finally, the mutual information
between the random variables $\bQ$ and $\bQ'$ is
\begin{equation}\label{eq:optimal}
   I(\bQ;\bQ') =  H(\bQ) - H(\bQ|\bQ') = H(\bQ') - H(\bQ'|\bQ),
\end{equation}
which is a symmetric measure $I(\bQ;\bQ')=I(\bQ';\bQ)$ representing
the information shared among the variables $\bQ$ and $\bQ'$. Figure
\ref{fig:basics} depicts the relationship between the Shannon
information, conditional Shannon information, and mutual information.

\begin{figure}[h]
\centering
  \begin{tikzpicture}[scale=2.4,>={Latex[length=.2cm]}]
    \colorlet{c1}{myc1}
    \colorlet{c2}{myc2}

    \colorlet{MI}{myc3}
    \pgfmathsetmacro{\dis}{.6}
    \begin{normalsize}
        \draw[c1!80!black,fill=c1!40] (-\dis,0) circle (1);
        \draw[c2!80!black,fill=c2!40] (\dis,0) circle (1);
        
        \begin{scope}
            \clip (-\dis,0) circle (1);
            \draw[draw=MI,fill=MI!40] (\dis,0) circle (1);
        \end{scope}
        \begin{scope}
            \clip (\dis,0) circle (1);
            \draw[MI!80!black] (-\dis,0) circle (1);
        \end{scope}

        \node[anchor=east] at (-\dis,0) {$H(\bQ |\bQ')$};
        \node[anchor=west] at (+\dis,0) {$H(\bQ'|\bQ)$};

        \node[anchor=center] at (0,0) {$I(\bQ;\bQ')$};

        \draw[<-] (+\dis,0) ++ (30:1)  --+ (30:.6)  node[fill=white,pos=1] {$H(\bQ')$};
        \draw[<-] (-\dis,0) ++ (150:1) --+ (150:.6) node[fill=white,pos=1] {$H(\bQ)$};

    \end{normalsize}
\end{tikzpicture} 
    \caption{Venn diagram of the Shannon information, conditional Shannon
      information and mutual information between two random variables
      $\bQ$ and $\bQ'$.}\label{fig:basics}
\end{figure}

\subsection{SURD: Synergistic-Unique-Redundant Decomposition of causality}
\label{sec:surd}

We provide a brief summary of the SURD formulation. For full
derivations and discussion, the reader is referred to
Ref.~\cite{surd}.  SURD quantifies causality as the increase in
information ($\Delta I$) about $Q_j^+$ obtained from observing
individual components or groups of components from $\bQ$. The
information in $Q_j^+$ is measured by the Shannon entropy.  Using the
principle of forward-in-time propagation of information (i.e.,
information only flows toward the future)~\cite{lozano2022},
$H(Q_j^+)$ can be decomposed as the sum of all causal contributions
from the past and present:
\begin{equation}
  \label{eq-sup:conservation_info_surd}
  H(Q_j^+) = \sum_{\mathbold{i} \in \mathcal{P}} \Delta I_{\bi
    \rightarrow j}^R + \sum_{i=1}^N \Delta I_{i\rightarrow j}^U +
  \sum_{\bi\in \mathcal{P}} \Delta I_{\bi \rightarrow j}^S + \Delta
  I_{\text{leak}\rightarrow j},
\end{equation}
where $\Delta I_{\bi \rightarrow j}^R$, $\Delta I_{i\rightarrow j}^U$,
and $\Delta I_{\bi \rightarrow j}^S$ are the redundant, unique, and
synergistic causalities, respectively, from the observed variables to
$Q_j^+$, and $\Delta I_{\text{leak}\rightarrow j}$ is the causality
from unobserved variables, referred to as the causality leak. Unique
causalities are associated with individual components of $\bQ$,
whereas redundant and synergistic causalities arise from groups of
variables from $\bQ$. Hence, the set $\mathcal{P}$ contains all
combinations involving more than one variable.  For instance,
Eq.~\eqref{eq:conservation_info} can be expanded for $N=3$ as
\begin{subequations}
\begin{align}
H(Q_j^+) &\equiv  \Delta I_{1\rightarrow j}^U +  \Delta I_{2\rightarrow j}^U  +  \Delta I_{3\rightarrow j}^U +  \Delta I_{12 \rightarrow j}^R + \Delta I_{13 \rightarrow j}^R +
\Delta I_{23 \rightarrow j}^R \\
&+  \Delta I_{12 \rightarrow j}^S + \Delta I_{13 \rightarrow j}^S +
\Delta I_{23 \rightarrow j}^S + \Delta I_{123 \rightarrow j}^R + \Delta I_{123 \rightarrow j}^S + \Delta
  I_{\text{leak}\rightarrow j}.
\end{align}
\end{subequations}

The SURD causalities are constructed using the specific mutual
information~\cite{DeWeese1999} from
$\bQ_{\bi}=[Q_{i_1},Q_{i_2},\ldots]$ to a particular event
$Q_j^+=q_j^+$, which is defined as
\begin{equation}\label{eq:sp_mi}
   \Is(Q_j^+ = q_j^+;\bQ_{\bi}) = \sum_{\bq_{\bi}} p(\bq_{\bi} | q_j^+)
  \log_2\left( \frac{p(q_j^+|\bq_{\bi})}{p(q_j^+)} \right) \geq 0.
\end{equation}
Note that the specific mutual information is a function of the random
variable $\bQ_{\bi}$ (which encompasses all its states) but only a
function of one particular state of the target variable (namely,
$q_j^+$). For the sake of simplicity, we will use the notation
$\Is_{\bi}(q_j^+) = \Is(Q_j^+ = q_j^+;\bQ_{\bi})$.  Similarly to
Eq.~\eqref{eq:optimal}, the specific mutual information quantifies the
dissimilarity between $p(q_j^+)$ and $p(q_j^+|\bq)$ but in this case
for the particular state $Q_j^+=q_j^+$. The mutual information between
$Q_j^+$and $\bQ_{\bi}$ is recovered by $I(Q_j^+;\bQ_{\bi}) =
\sum_{q_j^+} p(q_j^+) \Is_{\bi}(q_j^+)$.

\begin{figure}
    \centering
    \begin{tikzpicture}
        \node[anchor=north west] (fig) at (0,0) {\includegraphics[width=.75\textwidth]{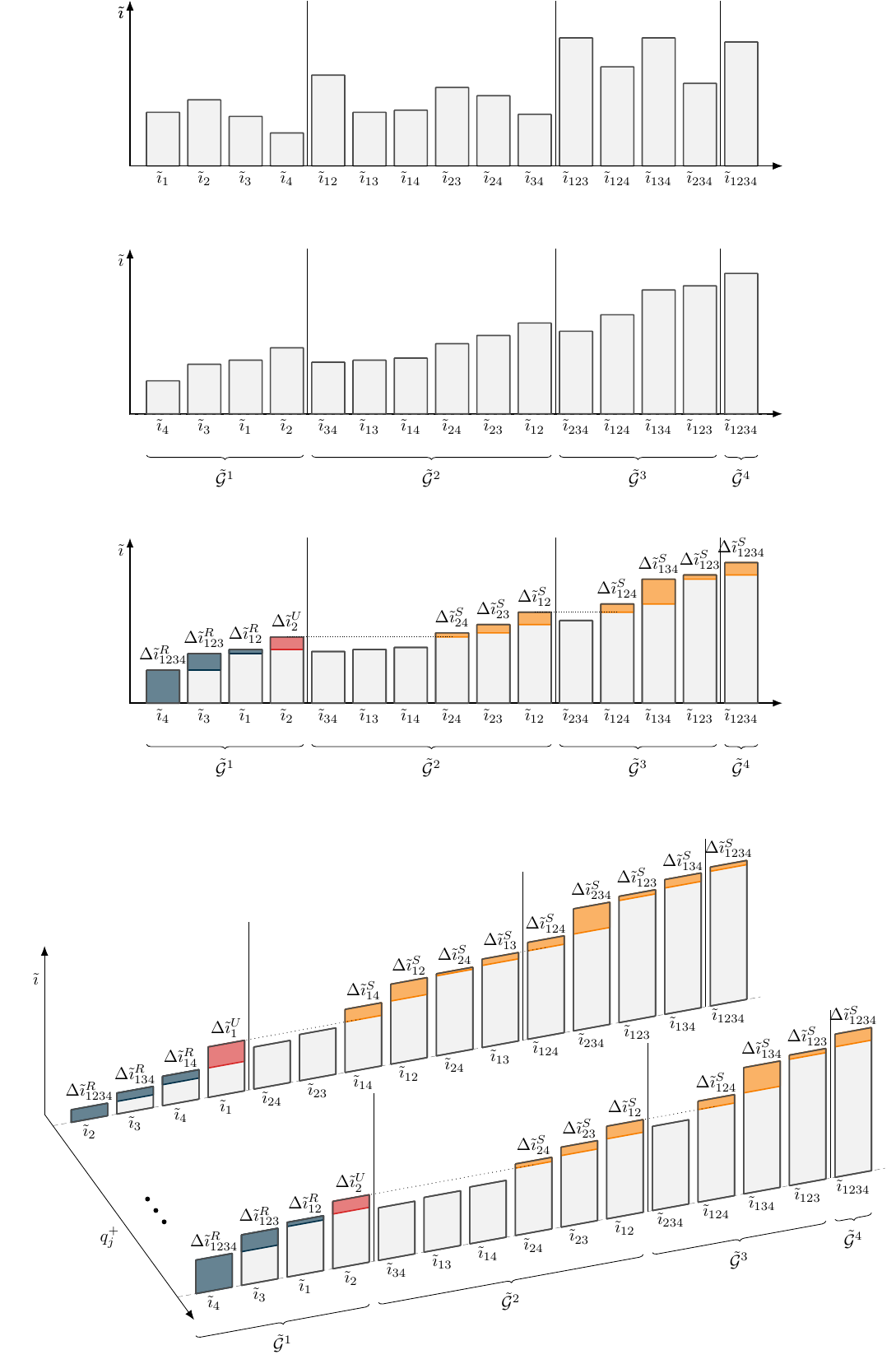}};
        \node at (0,0) {(a)};
        \node at (0,-3.5) {(b)};
        \node at (0,-7.5) {(c)};
        \node at (0,-12) {(d)};
    \end{tikzpicture}
\caption{ Schematic of the steps involved in the calculation of
  specific causalities. For a given state $Q_j^+=q_j^+$, the panels
  illustrate: (a) all possible specific mutual information values for
  a collection of four variables; (b) tuples of specific mutual
  information with the components organized in ascending order; (c)
  the increments corresponding to specific redundant (blue), unique
  (red), and synergistic (yellow) causalities; and (d) illustration of
  specific causalities for different states of $Q_j^+$.}
  \label{fig:sup_method}
\end{figure}

We now introduce the mathematical expressions for redundant, unique,
and synergistic causalities (Fig.~\ref{fig:sup_method}). These
definitions are guided by the following intuition:
\begin{itemize}
\item Redundant causality from $\bQ_{\bi}=[Q_{i_1},Q_{i_2},\ldots]$ to
  $Q_j^+$ is the common causality shared among all the components of
  $\bQ_{\bi}$, where $\bQ_{\bi}$ is a subset of $\bQ$ with two or more
  components.
\item Unique causality from $Q_i$ to $Q_j^+$ is the causality from
  $Q_i$ that cannot be obtained from any other individual variable
  $Q_k$ with $k\neq i$.
\item Synergistic causality from $\bQ_{\bi}=[Q_{i_1},Q_{i_2},\ldots]$
  to $Q^+_j$ is the causality arising from the joint effect of the
  variables in $\bQ_{\bi}$.

\item Redundant and unique causalities must depend only on probability
  distributions based on $Q_i$ and $Q_j^+$, that is,
  $p(q_i,q_j^+)$. On the other hand, synergistic causality must depend
  on the joint probability distribution of $\bQ_{\bi}$ and $Q^+_j$,
  i.e., $p(\bq_{\bi},q_j^+)$.
\end{itemize}

For a given value $Q_j^+=q_j^+$, the specific redundant, unique, and
synergistic causalities are calculated as follows:
\begin{enumerate}
\item The specific mutual information are computed for all possible
  combinations of variables in $\bQ$. This includes specific mutual
  information of order one ($\Is_1, \Is_2, \ldots$), order two
  ($\Is_{12}, \Is_{13}, \ldots$), order three ($\Is_{123}, \Is_{124},
  \ldots$), and so forth. One example is shown in
  Fig.~\ref{fig:sup_method}(a).
\item The tuples containing the specific mutual information of order
  $M$, denoted by $\GIs^M$, are constructed for $M=1,\ldots,N$.  The
  components of each $\GIs^M$ are organized in ascending order as
  shown in Fig.~\ref{fig:sup_method}(b).
\item
  The specific redundant causality is the increment in information
  gained about $q_j^+$ that is common to all the components of
  $\bQ_{\bj_k}$ (blue contributions in Fig.~\ref{fig:sup_method}c):
  \begin{equation}
  \label{eq:red_inc_spec}
    \Delta \Is^R_{\bj_k} =
    \begin{cases}
      \Is_{i_k}-\Is_{i_{k-1}},& \text{for} \ \Is_{i_k},\Is_{i_{k-1}} \in \GIs^1 \ \text{and} \ k\neq n_1\\
      0,              & \text{otherwise},
    \end{cases}
  \end{equation}
  where we take $\Is_{i_0}=0$, $\bj_k = [j_{k1}, j_{k2}, \ldots]$ is
  the vector of indices satisfying $\Is_{j_{kl}} \geq \Is_{i_k}$ for
  $\Is_{j_{kl}}, \Is_{i_k} \in \GIs^1$, and $n_1$ is the number of
  elements in $\GIs^1$. For instance, in Fig.~\ref{fig:sup_method}c,
  the specific redundant causality $\Delta \Is^R_{12}$ is calculated
  using $\Is_2$ and $\Is_3$. In this case, the indices $i_k$ and
  $i_{k-1}$ are 1 and 3, respectively, and $\bj_k = [1, 2]$ since
  $\Is_1\geq\Is_{i_k}$ and $\Is_2\geq\Is_{i_k}$.
\item
  The specific unique causality is the increment in information gained
  by $Q_{i_k}$ about $q_j^+$ that cannot be obtained by any other
  individual variable (red contribution in
  Fig.~\ref{fig:sup_method}c):
    \begin{equation}\label{eq:uni_inc_spec}
      \Delta \Is^U_{i_k} = 
      \begin{cases}
        \Is_{i_k}-\Is_{i_{k-1}}, & \text{for}\ i_k=n_1, \ \Is_{i_{k}},\Is_{i_{k-1}} \in \GIs^1\\
        0,                 & \text{otherwise}.
      \end{cases}
    \end{equation}
    %
  \item The specific synergistic causality is the increment in
    information gained by the combined effect of all the variables in
    $\bQ_{\bi_k}$ that cannot be gained by other combination of
    variables $\bQ_{\bj_k}$ (yellow contributions in
    Fig.~\ref{fig:sup_method}c) such that $\Is_{\bj_k} \leq
    \Is_{\bi_k}$ for $\Is_{\bi_k} \in \GIs^M$ and $\Is_{\bj_k} \in
    \{\GIs^1,\ldots,\GIs^{M}\}$ with $M>1$ (dotted line in
    Fig.~\ref{fig:sup_method}c):
    \begin{equation} \label{eq:syn_inc_spec}
      \Delta \Is^S_{\bi_k} = 
      \begin{cases}
        \Is_{\bi_k} - \Is_{\bi_{k-1}}, & \text{for} \ \Is_{\bi_{k-1}}\geq \max\{\GIs^{M-1}\}, \ \text{and} \ \Is_{\bi_{k}}, \Is_{\bi_{k-1}} \in \GIs^M \\
        \Is_{\bi_k} - \max\{\GIs^{M-1}\}, & \text{for} \ \Is_{\bi_{k}}>\max\{\GIs^{M-1}\}>\Is_{\bi_{k-1}}, \ \text{and} \ \Is_{\bi_{k}},  \Is_{\bi_{k-1}} \in \GIs^M\\
        0,              & \text{otherwise}.
      \end{cases}
    \end{equation}
    %
\item The specific redundant, unique and synergistic causalities that
  do not appear in the steps above are set to zero.
\item The steps (1) to (6) are repeated for all the states of $Q_j^+$
  (Figure~\ref{fig:sup_method}d).
\item Redundant, unique, and synergistic causalities are obtained as
  the expectation of their corresponding specific values with respect
  to $Q_j^+$,
  \begin{subequations} \label{eq:surd_causalities}
    \begin{align}  
      \Delta I^R_{\bi \rightarrow j} &= \sum_{q_j^+} p(q_j^+) \Delta \Is^R_{\bi}(q_j^+), \\
      \Delta I^U_{i\rightarrow j}   &= \sum_{q_j^+} p(q_j^+) \Delta \Is^U_{i}(q_j^+), \\
      \Delta I^S_{\bi\rightarrow j} &= \sum_{q_j^+} p(q_j^+) \Delta \Is^S_{\bi}(q_j^+). 
    \end{align}  
  \end{subequations}
\item 
  Finally, the average order of the specific causalities with respect
  to $Q_j^+$ is defined as
  \begin{equation}
    N^{\alpha}_{\bi \rightarrow j} = \sum_{q_j^+} p(q_j^+)
    n^{\alpha}_{\bi \rightarrow j}(q_j^+),
  \end{equation}
  where $\alpha$ denotes R, U, or S, $n^{\alpha}_{\bi \rightarrow
    j}(q_j^+)$ is the order of appearance of $\Delta
  \Is^{\alpha}_{\bi}(q_j^+)$ from left to right as in the example
  shown in Figure~\ref{fig:sup_method}.  The values of
  $N^{\alpha}_{\bi \rightarrow j}$ are used to plot $\Delta
  I^{\alpha}_{\bi \rightarrow j}$ following the expected order of
  appearance of $\Delta \Is^{\alpha}_{\bi \rightarrow j}$. All the
  causalities from SURD presented in this work are plotted in order
  from left to right, following $N^{\alpha}_{\bi \rightarrow j}$.
\end{enumerate}
%
%

\newpage
\subsection{State-dependent decomposition of causality}
\label{sec:intensities}

The derivation of our state-dependent causality builds on the SURD
formulation introduced above. The key distinction is that, rather than
quantifying the aggregate contribution of the entire sets $Q_j^+$ and
$\bQ$, we evaluate causality at the level of individual state pairs
$(q_j^+,\bq)$. Specifically, we aim to measure the influence of each
source state $\bq_{\bi}$ on each future state of the target variable,
denoted as $q_j^+ = q_j(t + \Delta T)$, where $\Delta T > 0$ is an
arbitrary time increment. To this end, we assess the information gain
about $q_j^+$ obtained by observing single or groups of states $\bq$
from the observable $\bQ$. The principle of forward-in-time
information propagation can be recast as:
\begin{equation}
  \label{eq-sup:conservation_info_2}
  H(Q_j^+) = \sum_{q_j^+\in Q_j^+} \sum_{\bq \in \bQ} \left[\Iss (q_j^+;\bq) + \Inn (q_j^+;\bq)\right]  + \Delta I_{\text{leak}\rightarrow j},
\end{equation}
where $\Iss(q_j^+,\bq)>0$ and $\Inn(q_j^+,\bq)\leq0$ denote the causal
and non-causal contributions of the source state $\bq$ to the target
state $q_j^+$, respectively, and $\Delta I_{\text{leak}\rightarrow j}$
again accounts for the causality leak.  The terms $\Iss(q_j^+, \bq)$
and $\Inn(q_j^+, \bq)$ can be further decomposed as:
\begin{equation} \label{eq:surd_causal}
    \Iss(q_j^+;\bq) = \sum_{\bi\in\mathcal{P}}\Iss_{\boldsymbol{i}\rightarrow j}^R + \sum_{i=1}^N\Iss_{\boldsymbol{i}\rightarrow j}^U + \sum_{\bi\in\mathcal{P}} \Iss_{\boldsymbol{i}\rightarrow j}^S,
\end{equation}
where $\Iss_{\bi \rightarrow j}^R$, $\Iss_{i\rightarrow j}^U$, and
$\Iss_{\bi \rightarrow j}^S$ are the redundant, unique, and
synergistic causalities, respectively, specific to each of the states
of the observed variables to the states of the target variable
$q_j^+$. As in SURD causality, unique causalities are associated with
individual components of $\bQ$, whereas redundant and synergistic
causalities arise from groups of variables from $\bQ$.  The set
$\mathcal{P}$ contains all combinations involving more than one
variable. For instance, for $N=2$, Eq.~\eqref{eq:surd} reduces to
$\Iss(q_j^+;\bq) = \Iss_{12\rightarrow j}^R + \Iss_{1\rightarrow j}^U
+ \Iss_{2\rightarrow j}^U + \Iss_{12 \rightarrow j}^S$. The same
expression can be derived for the non-causal component
$\Inn(q_j^+;\bq)$:
\begin{equation} \label{eq:surd_non_causal}
    \Inn(q_j^+;\bq) = \sum_{\bi\in\mathcal{P}}\Inn_{\boldsymbol{i}\rightarrow j}^R + \sum_{i=1}^N\Inn_{\boldsymbol{i}\rightarrow j}^U + \sum_{\bi\in\mathcal{P}} \Inn_{\boldsymbol{i}\rightarrow j}^S.
\end{equation}
Next, we detail how each component of $\Iss(q_j^+;\bq)$ and
$\Inn(q_j^+;\bq)$ can be further decomposed according to the
individual states of the target and source variables. We then
demonstrate that this state-dependent formulation is equivalent to the
total causality measures originally defined by SURD.

\subsubsection{State-dependent redundant causalities}

In our decomposition, causal and non-causal contributions are
accounted for in the sequence: redundant $\to$ unique $\to$
synergistic. We begin by deriving the state‐dependent redundant
terms. Denote by $\Iss^R_{\bi \rightarrow j}$ and $\Inn^R_{\bi
  \rightarrow j}$ the positive (causal) and negative (non‐causal)
contributions, respectively, from the joint state $(\bq_\bi,q_j^+)$ to
$\Delta I^R_{\bi \rightarrow j}$, defined as
\begin{equation}
  \label{eq:cau_per_int_red}
  \Delta I^R_{\bi \rightarrow j} = \sum_{q^+_j\in Q_j^+} \sum_{\bq_\bi
    \in \bQ} \left[\Iss^R_{\bi \rightarrow j} + \Inn^R_{\bi
      \rightarrow j}\right],
\end{equation}
where $\bi = [i_1,i_2,\dots]$ indexes the source variables involved in
the redundant causality, and $\bq_\bi = [q_{i_1},q_{i_2},\dots]$
denotes their particular states. The expressions for $\Iss^R_{\bi
  \rightarrow j}$ and $\Inn^{R}_{\bi \rightarrow j}$ follow from
Eqs.~\ref{eq:sp_mi}, \ref{eq:red_inc_spec}, and
\ref{eq:surd_causalities}:
\begin{equation}
    \Delta I^R_{\bi \rightarrow j} = \sum_{q_j^+} \sum_{q_{i_k}}
    \sum_{q_{i_{k-1}}} p(q_j^+,q_{i_k},q_{i_{k-1}})
    \log_2\left(\frac{p(q_j^+|q_{i_k})}{p(q_j^+|q_{i_{k-1}})} \right),
\end{equation}
for $\Is_{i_k},\Is_{i_{k-1}}\in\mathcal{G}_1$ with $k\neq n_1$, and
zero otherwise. The index vector $\bi=[i_1,i_2,\ldots]$ comprises
those $n$ satisfying $\Is_n\geq\Is_{i_k}$ (ordered so that $\Is_{i_k}$
increases). For example, if $\bi = [1,2]$ and
$\Is_4<\Is_3<\Is_1<\Is_2$ (see Fig.~\ref{fig:sup_method}b), then
$i_k=1$ and $i_{k-1}=3$. Therefore, the positive redundant causal
component is given by:
\begin{equation} \label{eq:redundant-causal}
  \Iss^R_{\bi \rightarrow j} = 
  \begin{cases}
  \sum_{q_{\cancel{\bi}}} p(q_j^+, \bq)
  \max\left(0,\log_2\left( \frac{p(q_j^+|q_{i_k})}{p(q_j^+|q_{i_{k-1}})} \right)\right), & \text{for} \ \Is_{i_k},\Is_{i_{k-1}} \in \GIs^1 \ \text{and} \ k\neq n_1,\\
   0,              & \text{otherwise},
\end{cases}
\end{equation}
where $\cancel{\bi}$ lists all indices not in $\bi$.

For example, in a system with two variables $Q_1$ and $Q_2$ targeting
$Q_j^+$ with joint distribution $p(q_j^+,q_1,q_2)$,
Eq.~\eqref{eq:redundant-causal} simplifies to
\begin{equation} \label{eq:contribution-redundant-example}
  \Iss^R_{12 \rightarrow j} = 
  \begin{cases}
  p(q^+_j, q_1, q_2)
  \max\left(0,\log_2\left( \frac{p(q_j^+|q_1)}{p(q^+_j)} \right)\right), & \text{for} \ \Is_{2}\geq \Is_{1}, \\
   0,              & \text{otherwise}.
\end{cases}
\end{equation}
Figure~\ref{fig:diagram-states-redundant} provides a schematic of the
state-dependent redundant causality $\Iss^R_{12\to j}$. Positive
values of $\Iss^R_{12\to j}$ indicate states where knowledge of $q_1$
increases the likelihood of $q_j^+$, i.e., $p(q_j^+ \mid q_1) >
p(q_j^+)$. 
%
\begin{figure}
    \centering
\pgfmathdeclarefunction{gauss}{3}{%
  \pgfmathparse{1/(#3*sqrt(2*pi))*exp(-((#1-#2)^2)/(2*#3^2))}%
}
\tikzmath{%
  function h1(\x, \lx) { return (9*\lx + 3*((\lx)^2) + ((\lx)^3)/3 + 9); };
  function h2(\x, \lx) { return (3*\lx - ((\lx)^3)/3 + 4); };
  function h3(\x, \lx) { return (9*\lx - 3*((\lx)^2) + ((\lx)^3)/3 + 7); };
  function skewnorm(\x, \l) {
    \x = (\l < 0) ? -\x : \x;
    \l = abs(\l);
    \e = exp(-(\x^2)/2);
    return (\l == 0) ? 1 / sqrt(2 * pi) * \e: (
      (\x < -3/\l) ? 0 : (
      (\x < -1/\l) ? \e / (8 * sqrt(2 * pi)) * h1(\x, \x*\l) : (
      (\x <  1/\l) ? \e / (4 * sqrt(2 * pi)) * h2(\x, \x*\l) : (
      (\x <  3/\l) ? \e / (8 * sqrt(2 * pi)) * h3(\x, \x*\l) : (
      sqrt(2/pi) * \e)))));
  };
}

\begin{tikzpicture}[scale=1.5]

    \def\B{13.3};   
    \def\S{16};   
    \def\T{7}; 

    \def\Bs{3.10};  
    \def\Ss{3.20};  
    \def\Ts{4.40};  

    \def\alphaT{-0.3}; 
    \def\alphaS{1}; 
    \def\xIntersect{9.9};

    \def\xmax{25};
    \def\ymin{{-0.15*gauss(\B,\B,\Bs)}};
 
    \begin{axis}[every axis plot post/.append style={
                 mark=none,domain={-0.5*(\xmax)}:{1.2*\xmax},samples=80,smooth},
                 xmin={-0.4*(\xmax)}, xmax=\xmax,
                 ymin=\ymin, ymax={1.3*gauss(\B,\B,\Bs)},
                 axis lines=middle,
                 axis line style=thick,
                 enlargelimits=upper, 
                 ticks=none,
                 xlabel=$q_j^+$,
                 y axis line style={opacity=0},
                 x label style={at={(axis description cs:0.69,0.065)},anchor=east},
                 width=9cm, height=5cm,
                ]

      \addplot[name path=B,thick,black!10!black] {0.95*gauss(x,\B,\Bs)};

      \addplot[name path=T, thick, black!10!myc1] {skewnorm((x-\T)/\Ts, \alphaT) / \Ts};
    
      \addplot [very thick, myc1!20, forget plot] fill between [
        of=T and B, soft clip={domain=-25:\xIntersect}];
      \addplot [very thick, black!8, forget plot] fill between [
        of=T and B, soft clip={domain=\xIntersect:50}];  
 
      \begin{huge}
      \node[above,black!20!black ] at (260,142) {$p(q_j^+)$};
      \node[above,black!20!myc1 ] at (185,112) {$p(q_j^+|q_1)$};
      \draw[-, align=center] (150,60) --++ (-40,25) node[anchor=east, align=center, yshift=10] {causal \\[2pt] $\Iss^R_{12\to j}>0$};

      
      \draw[-, align=center] (280,80) --++ (32,25) node[anchor=west, align=center, yshift=10] {non-causal \\[2pt] $\Inn^R_{12\to j}<0$};
      \end{huge}

    \end{axis}

\end{tikzpicture}
    \caption{State-dependent redundant causality $\Iss^R_{12\to j}$
      for a system with target variable $Q_j^+$ and source variables
      $Q_1$ and $Q_2$. In this example, the specific mutual
      information $\Is(Q_j^+=q_j^+;Q_2)$ is greater than
      $\Is(Q_j^+=q_j^+;Q_1)$ for all states $q_j^+\in Q_j^+$. Causal
      states ($\Iss^R_{12\to j}>0$) indicate causal states where
      knowledge of $q_1$ increases the likelihood of the reference
      state $q_j^+$, i.e., $p(q_j^+ \mid q_1) > p(q_j^+)$. Conversely,
      non-causal states ($\Inn^R_{12\to j}<0$) corresponds to negative
      values where knowledge of $q_1$ decreases the likelihood of
      $q_j^+$.}
    \label{fig:diagram-states-redundant}
\end{figure}

An analogous definition holds for the non-causal, state‐dependent
redundant causality, capturing those source states that increase
uncertainty about $q_j^+$ relative to the previous level. It is given
by:
\begin{equation} \label{eq:redundant-non-causal}
  \Inn^R_{\bi \rightarrow j} = 
  \begin{cases}
  \sum_{q_{\cancel{\bi}}} p(q_j^+, \bq)
  \min\left(0,\log_2\left( \frac{p(q_j^+|q_{i_k})}{p(q_j^+|q_{i_{k-1}})} \right)\right), & \text{for} \ \Is_{i_k},\Is_{i_{k-1}} \in \GIs^1 \ \text{and} \ k\neq n_1,\\
   0,              & \text{otherwise}.
\end{cases}
\end{equation}
In the example from Fig.~\ref{fig:diagram-states-redundant},
non-causal values indicate states where knowledge of $q_1$ decreases
the likelihood of $q_j^+$, i.e., $p(q_j^+ \mid q_1) < p(q_j^+)$.

\newpage
\subsubsection{State-dependent unique causalities}

\begin{figure}
    \centering
\pgfmathdeclarefunction{gauss}{3}{%
  \pgfmathparse{1/(#3*sqrt(2*pi))*exp(-((#1-#2)^2)/(2*#3^2))}%
}
\tikzmath{%
  function h1(\x, \lx) { return (9*\lx + 3*((\lx)^2) + ((\lx)^3)/3 + 9); };
  function h2(\x, \lx) { return (3*\lx - ((\lx)^3)/3 + 4); };
  function h3(\x, \lx) { return (9*\lx - 3*((\lx)^2) + ((\lx)^3)/3 + 7); };
  function skewnorm(\x, \l) {
    \x = (\l < 0) ? -\x : \x;
    \l = abs(\l);
    \e = exp(-(\x^2)/2);
    return (\l == 0) ? 1 / sqrt(2 * pi) * \e: (
      (\x < -3/\l) ? 0 : (
      (\x < -1/\l) ? \e / (8 * sqrt(2 * pi)) * h1(\x, \x*\l) : (
      (\x <  1/\l) ? \e / (4 * sqrt(2 * pi)) * h2(\x, \x*\l) : (
      (\x <  3/\l) ? \e / (8 * sqrt(2 * pi)) * h3(\x, \x*\l) : (
      sqrt(2/pi) * \e)))));
  };
}

\begin{tikzpicture}[scale=1.5]

    \def\B{13.3};   
    \def\S{16};   
    \def\T{7}; 

    \def\Bs{3.10};  
    \def\Ss{3.20};  
    \def\Ts{4.40};  

    \def\alphaT{-0.3}; 
    \def\alphaS{1}; 
    \def\xIntersect{13.25};

    \def\xmax{25.5};
    \def\ymin{{-0.15*gauss(\B,\B,\Bs)}};
 
    \begin{axis}[every axis plot post/.append style={
                 mark=none,domain={-0.5*(\xmax)}:{1.2*\xmax},samples=80,smooth},
                 xmin={-0.4*(\xmax)}, xmax=\xmax,
                 ymin=\ymin, ymax={1.3*gauss(\B,\B,\Bs)},
                 axis lines=middle,
                 axis line style=thick,
                 enlargelimits=upper, 
                 ticks=none,
                 xlabel=$q_j^+$,
                 y axis line style={opacity=0},
                 x label style={at={(axis description cs:0.69,0.065)},anchor=east},
                 width=9cm, height=5cm,
                ]

      \addplot[name path=B, thick,black!10!black] {0.95*gauss(x,\B,\Bs)};

      \addplot[name path=T, thick, black!10!myc2] {skewnorm((x-\T)/\Ts, \alphaT) / \Ts};
      \addplot[name path=S, thick, black!10!myc1] {0.5*skewnorm((x-\S)/\Ss, \alphaS) / \Ss};
    
      \addplot [very thick, myc2!20, forget plot] fill between [
        of=T and S, soft clip={domain=-25:\xIntersect}];
      \addplot [very thick, myc1!20, forget plot] fill between [
        of=T and S, soft clip={domain=\xIntersect:50}];  
 
      \begin{huge}
      \node[above,black!20!black ] at (260,142) {$p(q_j^+)$};
      \node[above,black!20!myc2 ] at (185,112) {$p(q_j^+|q_2)$};
      \node[above,black!20!myc1 ] at (315,95) {$p(q_j^+|q_1)$};
      \draw[-, align=center] (150,60) --++ (-40,25) node[anchor=east, align=center, yshift=10] {causal \\[2pt] $\Iss^U_{2\to j}>0$};

      
      \draw[-, align=center] (317,60) --++ (32,25) node[anchor=west, align=center, yshift=10,xshift=-2] {non-causal \\[2pt] $\Inn^U_{2\to j}<0$};
      \end{huge}

    \end{axis}


\end{tikzpicture}
    \caption{State‐dependent unique causality $\Iss^U_{2\to j}$ for a
      system with target variable $Q_j^+$ and source variables $Q_1$
      and $Q_2$, as given by Eq.~\eqref{eq:contribution example}. In
      this example, the specific mutual information $\Is(Q_j^+ =
      q_j^+; Q_2)$ exceeds $\Is(Q_j^+ = q_j^+; Q_1)$ for every state
      $q_j^+ \in Q_j^+$. Causal states ($\Iss^U_{2\to j}>0$) indicate
      states where knowing $q_2$ increases the likelihood of $q_j^+$
      relative to knowing $q_1$, i.e., $p(q_j^+ \mid q_2) > p(q_j^+
      \mid q_1)$. Conversely, non-causal states ($\Inn^U_{2\to j}<0$)
      correspond to states where knowledge of $q_2$ decreases that
      likelihood, $p(q_j^+ \mid q_2) < p(q_j^+ \mid q_1)$.}
    \label{fig:diagram-states-unique}
\end{figure}

We denote by $\Iss^U_{i \rightarrow j}$ and $\Inn^U_{i \rightarrow j}$
the positive (causal) and negative (non‐causal) state‐dependent
contributions from $(q_i,q_j^+)$ to the unique causality increment:
\begin{equation}
  \label{eq:cau_per_int}
  \Delta I^U_{i \rightarrow j} = \sum_{q^+_j\in Q_j^+} \sum_{q_i \in Q_i} \left[\Iss^U_{i \rightarrow j} + \Inn^{U}_{i \rightarrow j}\right].
\end{equation}
Comparing Eq.~\eqref{eq:cau_per_int} with the full expression for
$\Delta I^U_{i \rightarrow j}$ (Eqs.~\ref{eq:sp_mi},
\ref{eq:uni_inc_spec}, and \ref{eq:surd_causalities}) leads to
\begin{equation}
\label{eq:cau_unique_2}
  \Delta I^U_{i \rightarrow j} = \sum_{q_j^+} \sum_{q_{i}} \sum_{q_{k}} p(q_j^+, q_{i}, q_{k})
  \log_2\!\left( \frac{p(q_j^+|q_i)}{p(q_j^+|q_k)} \right),
\end{equation}
which holds for $i = n_1$ and $\Is_i,\Is_k \in \GIs^1$, and is zero
otherwise.  This represents the expected log‐ratio of the conditional
probabilities $p(q_j^+|q_i)$ and $p(q_j^+|q_k)$. From this, we obtain
the causal component:
\begin{equation} \label{eq:unique-causal}
  \Iss^U_{i \rightarrow j}(q_i,q_j^+) = 
  \begin{cases}
  \displaystyle
  \sum_{q_{k}} p(q_j^+, q_{i}, q_{k})
  \max\!\Bigl(0,\log_2\!\bigl( \tfrac{p(q_j^+|q_{i})}{p(q_j^+|q_{k})}\bigr)\Bigr), & \text{if } \Is_i \ge \Is_{k-1}, \\[1em]
   0,              & \text{otherwise},
\end{cases}
\end{equation}
where $\Is_{k}$ is the second-largest element of $\GIs^1$.

For instance, in a two-variable system $(Q_1,Q_2)$ targeting $Q_j^+$
with joint distribution $p(q_j^+,q_1,q_2)$, Eq.~\eqref{eq:unique-causal}
for the unique causality from $Q_2$ becomes
\begin{equation} \label{eq:contribution example}
  \Iss^U_{2 \rightarrow j} = 
  \begin{cases}
  \displaystyle
  \sum_{q_1} p(q^+_j, q_1, q_2)
  \max\!\Bigl(0,\log_2\!\bigl( \tfrac{p(q_j^+|q_2)}{p(q^+_j|q_1)}\bigr)\Bigr), & \text{if } \Is_2 \ge \Is_1, \\[0.5em]
   0,              & \text{otherwise}.
\end{cases}
\end{equation}

The non-causal component contains those source states that increase
uncertainty about $q_j^+$ compared to the largest redundant
contribution, and is always non‐positive:
\begin{equation} \label{eq:unique-non-causal}
  \Inn^U_{i \rightarrow j}(q_i,q_j^+) = 
  \begin{cases}
  \displaystyle
  \sum_{q_{k}} p(q_j^+, q_{i}, q_{k})
  \min\!\Bigl(0,\log_2\!\bigl( \tfrac{p(q_j^+|q_{i})}{p(q_j^+|q_{k})}\bigr)\Bigr), & \text{if } \Is_i \ge \Is_{k-1}, \\[1em]
   0,              & \text{otherwise}.
\end{cases}
\end{equation}

\newpage

\begin{figure}[t]
    \centering
\pgfmathdeclarefunction{gauss}{3}{%
  \pgfmathparse{1/(#3*sqrt(2*pi))*exp(-((#1-#2)^2)/(2*#3^2))}%
}
\tikzmath{%
  function h1(\x, \lx) { return (9*\lx + 3*((\lx)^2) + ((\lx)^3)/3 + 9); };
  function h2(\x, \lx) { return (3*\lx - ((\lx)^3)/3 + 4); };
  function h3(\x, \lx) { return (9*\lx - 3*((\lx)^2) + ((\lx)^3)/3 + 7); };
  function skewnorm(\x, \l) {
    \x = (\l < 0) ? -\x : \x;
    \l = abs(\l);
    \e = exp(-(\x^2)/2);
    return (\l == 0) ? 1 / sqrt(2 * pi) * \e: (
      (\x < -3/\l) ? 0 : (
      (\x < -1/\l) ? \e / (8 * sqrt(2 * pi)) * h1(\x, \x*\l) : (
      (\x <  1/\l) ? \e / (4 * sqrt(2 * pi)) * h2(\x, \x*\l) : (
      (\x <  3/\l) ? \e / (8 * sqrt(2 * pi)) * h3(\x, \x*\l) : (
      sqrt(2/pi) * \e)))));
  };
}

\begin{tikzpicture}[scale=1.5]

    \def\B{13.3};   
    \def\S{16};   
    \def\T{7}; 

    \def\Bs{3.10};  
    \def\Ss{3.20};  
    \def\Ts{4.40};  

    \def\alphaT{-0.3}; 
    \def\alphaS{1}; 
    \def\xIntersect{13.25};

    \def\xmax{25.5};
    \def\ymin{{-0.15*gauss(\B,\B,\Bs)}};
 
    \begin{axis}[every axis plot post/.append style={
                 mark=none,domain={-0.5*(\xmax)}:{1.2*\xmax},samples=80,smooth},
                 xmin={-0.4*(\xmax)}, xmax=\xmax,
                 ymin=\ymin, ymax={1.3*gauss(\B,\B,\Bs)},
                 axis lines=middle,
                 axis line style=thick,
                 enlargelimits=upper, 
                 ticks=none,
                 xlabel=$q_j^+$,
                 y axis line style={opacity=0},
                 x label style={at={(axis description cs:0.69,0.065)},anchor=east},
                 width=9cm, height=5cm,
                ]

      \addplot[name path=B,thick,black!10!black] {0.95*gauss(x,\B,\Bs)};

      \addplot[name path=T, thick, black!10!myc3] {skewnorm((x-\T)/\Ts, \alphaT) / \Ts};
      \addplot[name path=S, thick, black!10!myc2] {0.5*skewnorm((x-\S)/\Ss, \alphaS) / \Ss};
    
      \addplot [very thick, myc3!20, forget plot] fill between [
        of=T and S, soft clip={domain=-25:\xIntersect}];
      \addplot [very thick, myc2!20, forget plot] fill between [
        of=T and S, soft clip={domain=\xIntersect:50}];  
 
      \begin{huge}
      \node[above,black!20!black ] at (260,142) {$p(q_j^+)$};
      \node[above,black!20!myc3 ] at (185,112) {$p(q_j^+|q_1,q_2)$};
      \node[above,black!20!myc2 ] at (315,95) {$p(q_j^+|q_2)$};
      \draw[-, align=center] (150,60) --++ (-40,25) node[anchor=east, align=center, yshift=10] {causal \\[2pt] $\Iss^S_{12\to j}>0$};

      
      \draw[-, align=center] (314,60) --++ (32,25) node[anchor=west, align=center, yshift=10,xshift=-2] {non-causal \\[2pt] $\Inn^S_{12\to j}<0$};
      \end{huge}

    \end{axis}

\end{tikzpicture}
    \caption{ State-dependent synergistic causality $\Iss^S_{12\to j}$
      for a system with target variable $Q_j^+$ and source variables
      $Q_1$ and $Q_2$. In this example, the specific mutual
      information $\Is(Q_j^+ = q_j^+; Q_1, Q_2)$ is greater than
      $\Is(Q_j^+ = q_j^+; Q_2)$ for all states $q_j^+ \in
      Q_j^+$. Causal values ($\Iss^S_{12\to j}>0$) indicate states
      where knowledge of both $q_1$ and $q_2$ increases the likelihood
      of the reference state $q_j^+$ compared to knowing only $q_2$,
      i.e., $p(q_j^+ \mid q_1, q_2) > p(q_j^+ \mid q_2)$. Conversely,
      non-causal values ($\Inn^S_{12\to j}<0$) implies that knowledge
      of $q_1$ and $q_2$ reduces the likelihood of $q_j^+$ relative to
      knowing only $q_2$, i.e., $p(q_j^+ \mid q_1, q_2) < p(q_j^+ \mid
      q_2)$.}
    \label{fig:diagram-states-synergistic}
\end{figure}

\subsubsection{State-dependent synergistic causalities}

The derivation for the state-dependent synergistic causality parallels
those for the redundant and unique terms. We introduce the positive
(causal) and negative (non-causal) contributions $\Iss^S_{\bi
  \rightarrow j}$ and $\Inn^S_{\bi \rightarrow j}$ from the joint
state $(\bq_\bi,q_j^+)$ to the synergistic increment $\Delta I^S_{\bi
  \rightarrow j}$:
\begin{equation}
  \label{eq:cau_per_int_syn}
  \Delta I^S_{\bi \rightarrow j} = \sum_{q^+_j\in Q_j^+}
  \sum_{\bq_\bi\in \bQ_\bi} \left[\Iss^S_{\bi \rightarrow j} +
    \Inn^{S}_{\bi \rightarrow j} \right],
\end{equation}
where $\bi=[i_1,i_2,\dots]$ indexes the source variables in the
synergy and $\bq_\bi=[q_{i_1},q_{i_2},\dots]$ their particular
states. Hence, the dimensionality of $\Iss^S_{\bi\rightarrow j}$ grows
with $|\bi|$. For instance, if $\bq_\bi=[q_1,q_3,q_4]$, then
$\Iss^S_{134\rightarrow j}(q_1,q_3,q_4,q_j^+)$ depends on three source
states in addition to $q_j^+$.  Comparing Eq.~\eqref{eq:cau_per_int_syn}
with the full form of $\Delta I^S_{\bi\rightarrow j}$ (using
Eqs.~\ref{eq:sp_mi}, \ref{eq:syn_inc_spec}, and
\ref{eq:surd_causalities}) yields
\begin{equation}
  \Delta I^S_{\bi \rightarrow j} = \sum_{q_j^+} \sum_{\bq_{\bi}} \sum_{\bq_{\bk}} p(q_j^+, \bq_{\bi}, \bq_{\bk})
  \log_2\left( \frac{p(q_j^+|\bq_\bi)}{p(q_j^+|\bq_\bk)} \right),
\end{equation}
where the index vector $\bk$ selects the variables with the
next-highest specific mutual information after those in $\bi$. By
definition of synergistic causality (\S\ref{sec:surd}), $\bk$ may
either continue in the same order as $\bi$ or revert to the maximal entry of
the previous order.  This leads to the following decomposition of
the causal component:
\begin{equation} \label{eq:contribution-synergistic}
  \Iss^S_{\bi \rightarrow j}=
  \begin{cases}
    \sum_{\bq_{\bk}} p(q_j^+, \bq_{\bi}, \bq_{\bk})
    \max\left(0,\log_2\left( \frac{p(q_j^+|\bq_{\bi})}{p(q_j^+|\bq_{\bk})} \right)\right), & \Is_{\bi} \geq \Is_{\bk} \geq \max\{\GIs^{M-1}\},\\[0.5em]
    \sum_{\bq_{\boldm}} p(q_j^+, \bq_{\bi}, \bq_{\boldm})
    \max\left(0,\log_2\left( \frac{p(q_j^+|\bq_{\bi})}{p(q_j^+|\bq_{\boldm})} \right)\right), & \Is_{\bi} \geq \Is_{\boldm} = \max\{\GIs^{M-1}\}>\Is_{\bk},\\[0.5em]
    0,              & \text{otherwise},
  \end{cases}
\end{equation}
where $\boldm$ indexes the element with the maximum specific mutual
information in $\GIs^{M-1}$.

In the two-variable example $(Q_1,Q_2)$ targeting $Q_j^+$ with joint
distribution $p(q_j^+,q_1,q_2)$, this reduces to
\begin{equation} \label{eq:contribution-synergistic-example}
  \Iss^S_{12 \rightarrow j} =
  \begin{cases}
  p(q^+_j, q_1, q_2)
  \max\left(0,\log_2\left( \frac{p(q_j^+|q_1, q_2)}{p(q^+_j|q_2)} \right)\right), & \Is_{12}\geq \Is_{2}, \\
   0,              & \text{otherwise}.
\end{cases}
\end{equation}
Figure~\ref{fig:diagram-states-synergistic} illustrates $\Iss^S_{12\to
  j}$ for a system consisting of two variables $Q_1$ and $Q_2$ and a
target variable $Q_j^+$ with a joint probability distribution
$p(q_j^+,q_1,q_2)$. Positive values mark states where joint knowledge
of $q_1$ and $q_2$ adds information beyond the unique effect of $q_2$,
i.e.\ $p(q_j^+\mid q_1,q_2)>p(q_j^+\mid q_2)$. Zero values indicate no
synergy relative to the unique contribution from $q_2$.

Finally, the non-causal component follows analogously:
\begin{equation} \label{eq:contribution-synergistic-nc}
  \Inn^S_{\bi \rightarrow j}=
  \begin{cases}
    \sum_{\bq_{\bk}} p(q_j^+, \bq_{\bi}, \bq_{\bk})
    \min\left(0,\log_2\left( \frac{p(q_j^+|\bq_{\bi})}{p(q_j^+|\bq_{\bk})} \right)\right), & \Is_{\bi} \geq \Is_{\bk} \geq \max\{\GIs^{M-1}\},\\[0.5em]
    \sum_{\bq_{\boldm}} p(q_j^+, \bq_{\bi}, \bq_{\boldm})
    \min\left(0,\log_2\left( \frac{p(q_j^+|\bq_{\bi})}{p(q_j^+|\bq_{\boldm})} \right)\right), & \Is_{\bi}\geq \Is_{\boldm} =\max\{\GIs^{M-1}\}>\Is_{\bk},\\[0.5em]
    0,              & \text{otherwise}.
  \end{cases}
\end{equation}

\subsection{Alternative state-dependent decomposition of causality}

The definition of causality adopted here treats positive increments of
information as causal, while negative increments are deemed
non-causal. This choice is guided by the intuition that such a
distinction aligns better with how we interpret causality in the
context of understanding interactions among variables. Of course, this
is only one possible definition and alternative formulations may also be
considered plausible. Here, we present two such alternatives and
discuss the reasons why we did not adopt them in our framework.

First, we illustrate the inevitability of obtaining negative terms
when decomposing mutual information based on individual state pairs
$(q_j^+, \bq)$. Consider the decomposition
$$
I(Q_j^+;\bQ) = \sum_{q_j^+} \sum_{\bq} \left[\Iss(q_j^+,\bq) + \Inn(q_j^+,\bq) \right],
$$
which can also be written explicitly as:
\begin{equation}\label{eq:pmi}
    I(Q_j^+;\bQ) = \sum_{q_j^+ \in Q_j^+} \sum_{\bq \in \bQ} \Delta
    \mathcal{I} = \sum_{q_j^+ \in Q_j^+} \sum_{\bq \in \bQ} p(q_j^+,
    \boldsymbol{q}) \log_2 \frac{p(q_j^+, \boldsymbol{q})}{p(q_j^+)
      p(\boldsymbol{q})}.
\end{equation}
The term inside the sum is zero \emph{if and only if} $q_j^+$ and
$\bq$ are statistically independent, and it can be positive or
negative depending on whether the joint probability $p(q_j^+,
\boldsymbol{q})$ is greater or smaller than the product of the
marginals $p(q_j^+)p(\boldsymbol{q})$.

As a simple example, consider the task of forecasting whether it will
rain tomorrow, denoted by $Q_{\mathrm{rain}}^+$, where
$q_{\mathrm{rain}}^+ = 1$ indicates rain and $q_{\mathrm{rain}}^+ = 0$
indicates no rain. Suppose we use atmospheric pressure, $Q_p$, as an
observable, which can take only two values: low ($q_p = 0$) or high
($q_p = 1$). After collecting observations, we can construct a joint
probability table for all combinations of these variables.  For
example, consider the observations in Table \ref{tab:rain}.
\begin{table}[h!]
\centering
\begin{tabular}{|c|c|c|c|}
\hline
{pressure} & {rain} & $ p(\text{pressure}, \text{rain}) $ & $ \Delta\mathcal{I} (\text{pressure}; \text{rain}) $ \\
\hline
0 & 0 & 0.10 & $-0.10$ \\
0 & 1 & 0.70 & $0.16$ \\
1 & 0 & 0.15 & $0.24$ \\
1 & 1 & 0.05 & $-0.08$ \\
\hline
\end{tabular}
\caption{Joint probability distribution of pressure and rain, along
  with their contributions to the mutual information
  $\Delta\mathcal{I}$. The marginal probabilities are:
  $p(\text{pressure}=0) = 0.80$, $p(\text{pressure}=1) = 0.20$,
  $p(\text{rain}=0) = 0.25$, and $p(\text{rain}=1) = 0.75$. The total
  mutual information is given by the sum of all values in
  $\Delta\mathcal{I}$, i.e., $I(\text{pressure}; \text{rain})=0.22$.}
\label{tab:rain}
\end{table}
%

Given a table of probabilities, we can evaluate the expression in
Eq.~\eqref{eq:pmi} for all combinations of states. In the example above,
observing that the pressure is low (i.e., $q_p = 0$) yields a negative
increment of information for $q_{\mathrm{rain}}^+ = 0$ and a positive
increment for $q_{\mathrm{rain}}^+ = 1$.  Intuitively, this implies
that a particular state $\bq$ of the source variables may be
non-causal with respect to one target state $q_{j_1}^+$, while
simultaneously being causal with respect to a different target state
$q_{j_2}^+$. This behavior arises because, if for some pair $(\bq,
q_{j_1}^+)$ we have $p(\bq, q_{j_1}^+) < p(\bq)\, p(q_{j_1}^+)$, then
normalization of probabilities requires that this be offset by another
pair $(\bq, q_{j_2}^+)$ such that $p(\bq, q_{j_2}^+) > p(\bq)
p(q_{j_2}^+)$.

As a consequence, if there exists at least one pair of states $(q_j^+,
\bq)$ such that the contribution to the mutual information $\Delta\mathcal{I}(q_j^+, \bq)$ is negative, then any decomposition of
mutual information into terms that depend solely on specific pairs
$(q_j^+, \bq)$ must necessarily include at least one negative
component. In particular, the sum $\Iss(q_j^+, \bq) + \Inn(q_j^+,
\bq)$ must be negative for some state pair $(q_j^+, \bq)$.  If this
sum can take both positive and negative values, then each of the
state-dependent causal components $\Iss_{\boldsymbol{i} \rightarrow
  j}^\alpha$ and $\Inn_{\boldsymbol{i} \rightarrow j}^\alpha$, with
$\alpha \in \{R, U, S\}$, must also include negative values for at
least one such pair. 

\subsubsection{Alternative definition I: Causality as positive and negative information increments}

The first alternative definition considered treats the sum of both
positive and negative contributions as the effective measure of
causality at the state level, namely $\Iss_{\bi \rightarrow j}^\alpha
+ \Inn_{\bi \rightarrow j}^\alpha$. The main advantage of this
formulation is that the total SURD causality can be directly recovered
by summing this combined quantity over all states.  In contrast, in
our proposed approach, the total SURD causalities are obtained only
when the causal and non-causal components are explicitly
combined. While the alternative definition simplifies aggregation, it
may reduce interpretability at the state level, since the resulting
maps include both reinforcing and opposing contributions.
Nevertheless, this formulation still preserves the key qualitative
distinction between causal states—where $\Iss_{\bi \rightarrow
  j}^\alpha + \Inn_{\bi \rightarrow j}^\alpha > 0$—and non-causal
states—where $\Iss_{\bi \rightarrow j}^\alpha + \Inn_{\bi \rightarrow
  j}^\alpha < 0$---consistent with our proposed method.

\begin{figure}
    \centering
    \subfloat[Current]{
    \begin{minipage}{0.32\textwidth}
    \includegraphics[width=\linewidth]{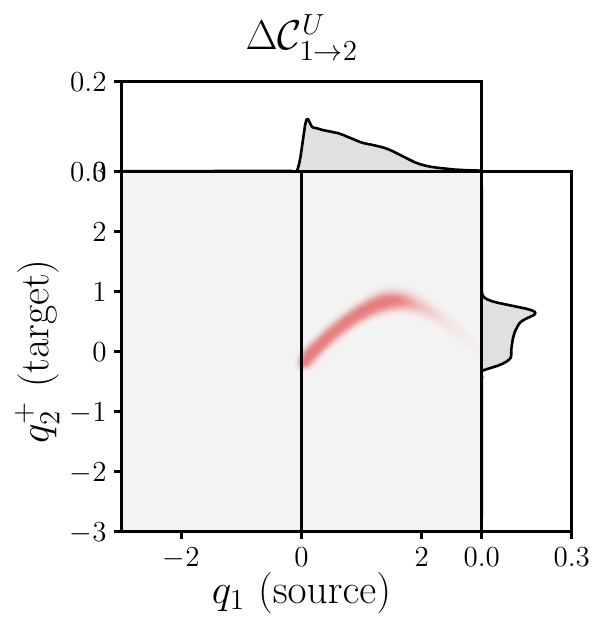}
    \end{minipage}}
    \hfill
    \subfloat[Alternative I]{
    \begin{minipage}{0.32\textwidth}
    \includegraphics[width=\linewidth]{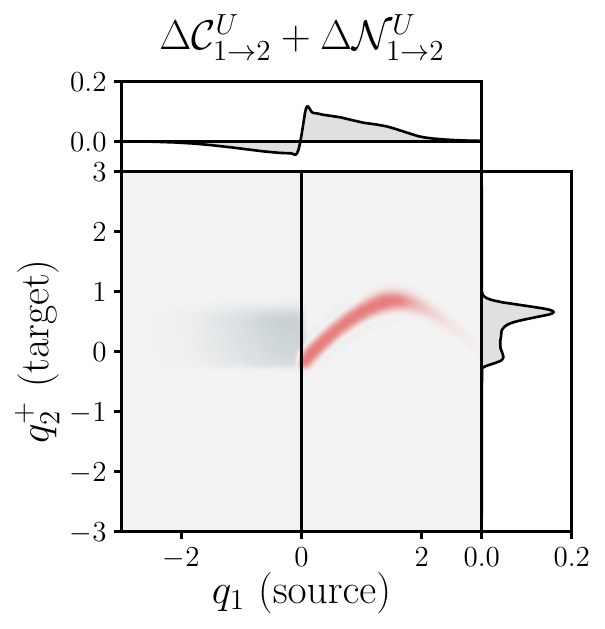}
    \end{minipage}}
    \hfill
    \subfloat[Alternative II]{
    \begin{minipage}{0.32\textwidth}
    \includegraphics[width=\linewidth]{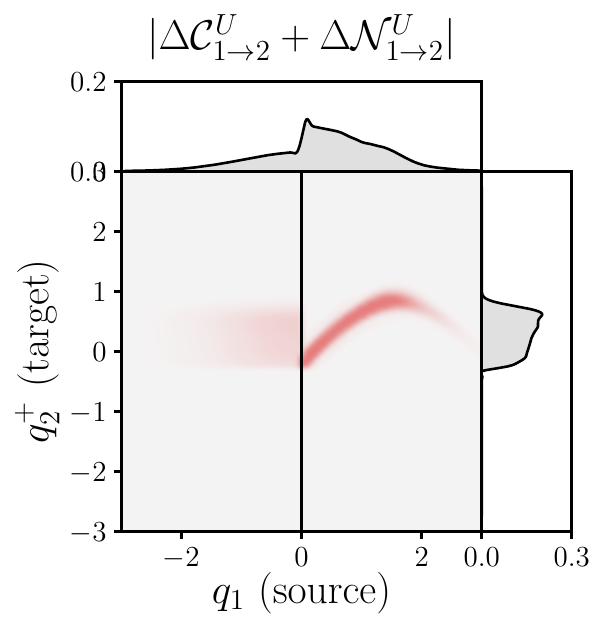}
    \end{minipage}}
    \caption{Comparison of different state-dependent causal
      decomposition methods: (a) Proposed method, which separates
      causal and non-causal contributions; (b) Alternative Method I,
      which combines causal and non-causal terms into a single
      quantity; and (c) Alternative Method II, which uses the absolute
      value of the combined contributions, discarding the sign
      information.}
    \label{fig:comparison-nn}
\end{figure}

To illustrate the distinction between our proposed method and the
alternative formulation discussed above, Fig.~\ref{fig:comparison-nn}
presents a comparison based on the source-dependent benchmark case
introduced in Fig.~\ref{fig:example-contributions}. In this example,
the interaction between variables is defined \textit{a priori} as $q_1
\to q_2^+$ if $q_1 > 0$, and $q_2 \to q_2^+$ if $q_1 \leq 0$.  We
focus on the unique causality from the source variable $q_1$ to the
target variable $q_2^+$, quantified by the term $\Iss^U_{1 \rightarrow
  2}$. Under our proposed method, only states with $q_1 > 0$ are
identified as uniquely causal, in agreement with the system functional
dependence. In contrast, the alternative formulation, which uses the
combined quantity $\Iss^U_{1 \rightarrow 2} + \Inn^U_{1 \rightarrow
  2}$, also identifies states with $q_1 < 0$ as contributing states to $\Delta I^U_{1 \rightarrow 2}$. We
argue that the latter is less useful for interpreting the functional
structure of the system.

\subsubsection{Alternative definition II: Causality as absolute value of information increments}

A second alternative approach defines state-dependent causality as the
absolute value of the combined positive and negative increments:
$|\Iss_{\bi \rightarrow j}^\alpha + \Inn_{\bi \rightarrow
  j}^\alpha|$. However, this formulation presents significant
drawbacks.  First, the total causality defined by SURD is no longer
preserved, as summing absolute values distorts the balance between
positive and negative contributions. More critically, taking the
absolute value eliminates the distinction between $\Iss_{\bi
  \rightarrow j}^\alpha$ and $\Inn_{\bi \rightarrow j}^\alpha$,
effectively interpreting all state pairs as causal regardless of
whether they reduce or increase uncertainty about the target.  This
conflation undermines interpretability: states in $\Iss_{\bi
  \rightarrow j}^\alpha$ provide useful, predictive information that
reduces uncertainty about the future, whereas states in $\Inn_{\bi
  \rightarrow j}^\alpha$ increase uncertainty and may reflect noise,
misleading associations, or indirect effects. While the latter may
carry some informational relevance, their role is fundamentally
different from genuinely causal states. As a result, treating both
types equivalently may obscure the intuitive notion of causality as
uncertainty reduction.

The results of this alternative formulation, applied to the
source-dependent benchmark case, are shown in
Fig.~\ref{fig:example-contributions}. We observe that this approach
assigns equal importance to both $q_1 > 0$ and $q_1 < 0$, which can be
misleading. While $q_1$ indeed plays a role in the regime where $q_1 <
0$—as its sign determines which variable governs the behavior of
$q_2^+$—the information from $q_2$ is still necessary to fully
determine the future state $q_2^+$. Consequently, in this regime,
$q_2^+$ cannot be uniquely determined by $q_1$ alone.

In conclusion, our proposed definition of state-dependent causality
offers clearer interpretability. Additionally, the sum of all causal
and non-causal components recovers the full mutual information between
the source variables $\bQ$ and the target $Q_j^+$, without requiring a
redefinition of the global redundant, unique, and synergistic
components in the SURD framework.

\newpage
\section{Verification and validation} \label{sec:validation}

\subsection{Multivariate correlated Gaussian variables}

We verify the implementation of the state-dependent causalities using
a Gaussian system, for which an analytical solution can be
obtained. We consider a system with two source variables, $Q_1$ and
$Q_2$, and a target variable, $Q_j^+$, following a joint Gaussian
distribution with arbitrary correlations between the variables. The
joint probability density function $p(q_j^+, q_1, q_2)$ takes the
form:
\begin{equation}
\label{eq: gaussian pdf}
    p(q_j^+, q_1, q_2) = \frac{1}{\sqrt{(2\pi)^3 \det \mathbf{\Sigma}}} \exp\left(-\frac{1}{2} {(\boldsymbol{z} - \boldsymbol{\mu})^\top} {\mathbf{\Sigma}}^{-1} (\boldsymbol{z} - \boldsymbol{\mu})\right),
\end{equation}
where $ \boldsymbol{z} $ represents the vector of variables $q_j^+$,
$q_1$, and $q_2$; $ \boldsymbol{\mu} = [\mu_{+}, \mu_{1},
  \mu_{2}]^\top $ is the mean vector of these variables, and
$\mathbf{\Sigma}$ is the covariance matrix which quantifies the degree
of dependence among the variables. Mathematically, the matrix is
represented as:
\begin{equation}
    \mathbf{\Sigma} = \begin{bmatrix}
    \sigma_+^2 & \rho_{1} \sigma_+ \sigma_1 & \rho_{2} \sigma_+ \sigma_2 \\
    \rho_{1} \sigma_+ \sigma_1 & \sigma_1^2 & \rho_{12} \sigma_1 \sigma_2 \\
    \rho_{2} \sigma_+ \sigma_2 & \rho_{12} \sigma_1 \sigma_2 & \sigma_2^2
    \end{bmatrix},
\end{equation}
where $\sigma_+$, $\sigma_1$, and $\sigma_2$ denote the standard
deviations of $q_j^+$, $q_1$, and $q_2$, respectively, and $\rho_{1}$,
$\rho_{2}$, and $\rho_{12}$ are the correlation coefficients that
measure the linear dependencies between the pairs of these
variables. In particular, $\rho_1$ denotes the correlation between
$Q_1$ and $Q_j^+$, $\rho_2$ the correlation between $Q_2$ and $Q_j^+$,
and $\rho_{12}$ the correlation between $Q_1$ and $Q_2$.

From Eq.~\eqref{eq: gaussian pdf}, we can directly obtain an analytical
equation for $\Iss^\alpha _ {\bi \rightarrow j}$. For example, for the
case in which $\rho_{1}=\rho_{12}=0$ and $\rho_{2}\neq 0$, the
analytical expression for $\Iss _ {2 \rightarrow j}$ is given by
Eq.~\eqref{eq:unique-causal}:
\begin{scriptsize}
\begin{equation} \label{eq:gaussian-unique}
    \Iss^U_{2 \rightarrow j} = \max\left(0,\frac{1}{\pi^{3/2} \sqrt{2(1 - \rho_{2}^2)}} \exp\left[-\frac{1}{1-\rho_{2}^2}\left(\frac{q_2^2}{2} + \frac{{q_j^+}^2}{2} + (1-\rho_{2})\frac{q_1^2}{2} - q_2 q_j^+ \rho_{2}\right) \right] \log\left(\frac{\exp\left(-\frac{{\rho_{2}} \left(-2 q_2 q_j^+ + q_2^2 \rho_{2} + {q_j^+}^2 \rho_{2}\right)}{2(1 - \rho_{2}^2)}\right)}{\sqrt{1 - \rho_{2}^2}}\right)^{1/2}\right),
\end{equation}
\end{scriptsize}
where we assume that the condition $\Is_2 \geq \Is_1 $ is satisfied
for all states of $Q_j^+$. A similar expression can be obtained for
the non-causal component $\Inn _ {2 \rightarrow j}$ using the function
$\min(\cdot)$.

\begin{figure}[h]
    \centering
    \includegraphics[height=0.365\linewidth]{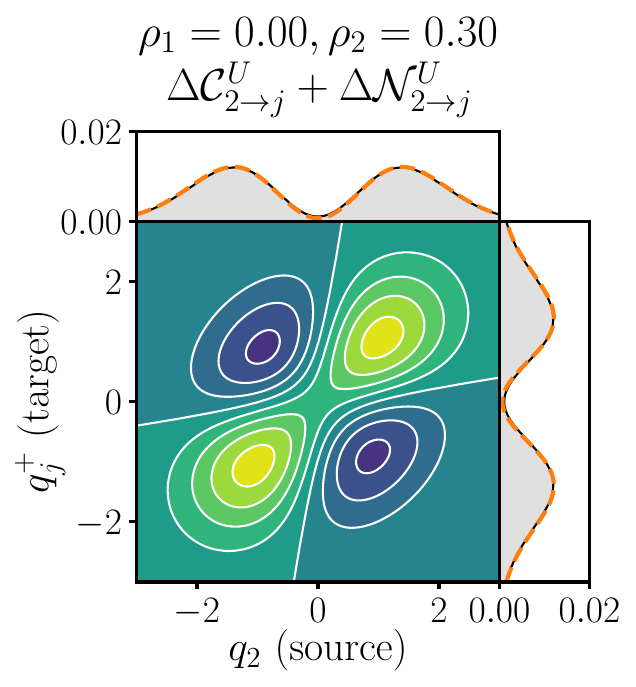}
    \includegraphics[height=0.365\linewidth]{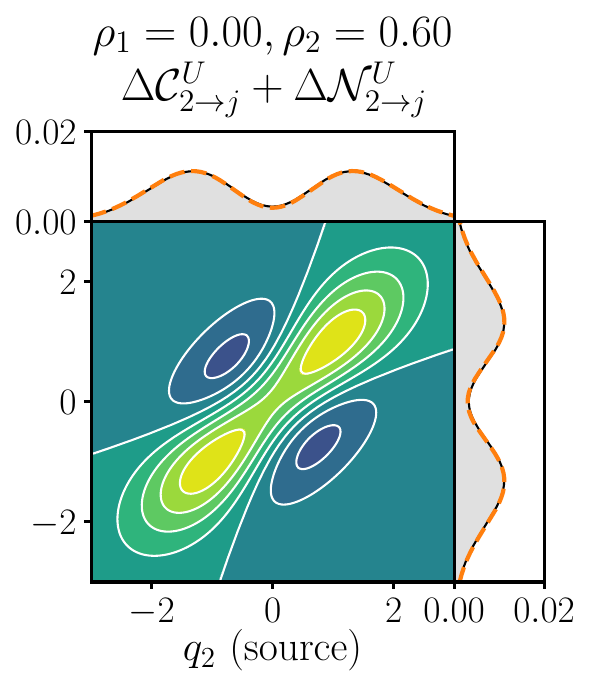}
    \includegraphics[height=0.365\linewidth]{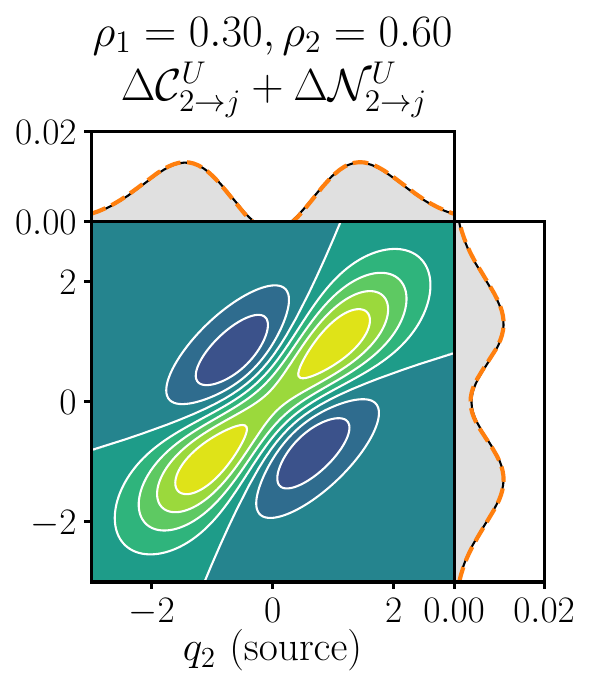}
    \caption{Causal $\Iss ^ U _ {2 \to j}$ and non-causal $\Inn ^ U _
      {2 \to j}$ contributions to the unique causality $\Delta I ^U _
      {2\to j}$ for a system with two source variables $Q_1$ and $Q_2$
      and a target variable $Q_j^+$ following a Gaussian distribution
      with three different combinations of correlation between $Q_1$
      and $Q_j^+$, i.e. $\rho_1$, and between $Q_2$ and $Q_j^+$,
      i.e. $\rho_2$. The analytical results are shown with filled
      contours and grey surfaces, whereas the numerical results from
      SURD are shown with white line contours and orange dashed lines
      for all cases.}
    \label{fig:gaussian-states}
\end{figure}

Figure~\ref{fig:gaussian-states} presents the causal and non-causal
contributions to the unique causality, given by $\Iss^U_{2 \to j} +
\Inn^U_{2 \to j}$, for a Gaussian system with three different
combinations of correlation parameters: $\rho_1$ (correlation between
$Q_1$ and $Q_j^+$) and $\rho_2$ (correlation between $Q_2$ and
$Q_j^+$). These analytical results are compared against the numerical
estimates obtained from our algorithmic implementation, which is used
throughout this study. This comparison provides a verification of the
numerical results for both the causal and non-causal components with
respect to the exact analytical solutions.

\subsection{Source-dependent unique causality}

\begin{figure}[h!]
    \centering
    \begin{minipage}{\textwidth}
    \begin{center}
        $
          q_3(n+1) =  
          \begin{cases}
            \cos\left[q_1(n)\right] + 0.1 w_3(n) & \text{if } q_1(n) < 0 \\
            0.9 q_3(n) + 0.1w_3(n) & \text{otherwise}
          \end{cases}
        $
    \end{center}
    \end{minipage}
    \vspace{0.0075\textwidth}
    
    \begin{minipage}{0.23\textwidth}
    \begin{tikzpicture}
        \node[anchor=north west] (img) at (0, 0) {\includegraphics[width=\linewidth]{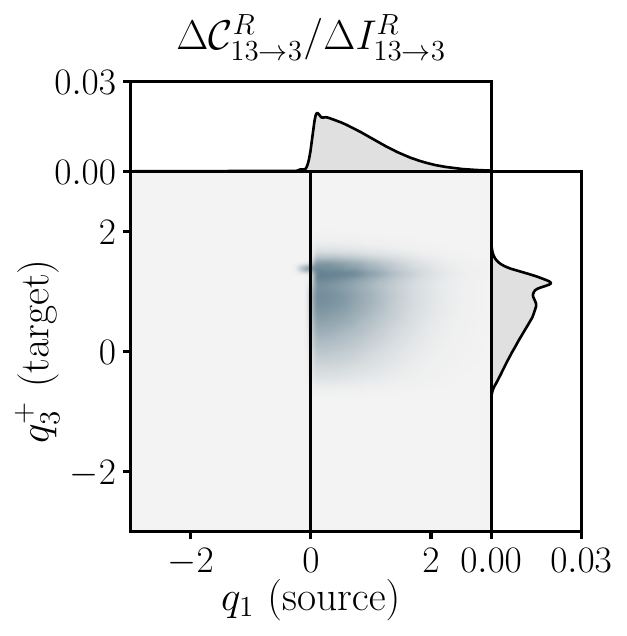}};
        \node[anchor=north east, align=center] at (2.3, -3.1) {\tiny{non-causal}};
        \node[anchor=north east, align=center] at (3.3, -3.1) {\tiny{causal}};
        \node[anchor=south, font=\small, fill=white, text=black, inner sep=2pt, rounded corners, minimum width=0.2\textwidth,minimum height=6mm] at ([yshift=-5.6mm,xshift=-1mm]img.north) {$\quad$Redundant R13$\quad$};
    \end{tikzpicture}
    \end{minipage}
    \begin{minipage}{0.21\textwidth}
    \begin{tikzpicture}
        \node[anchor=north west] (img) at (0, 0) {\includegraphics[width=\linewidth, trim=26 0 0 0, clip]{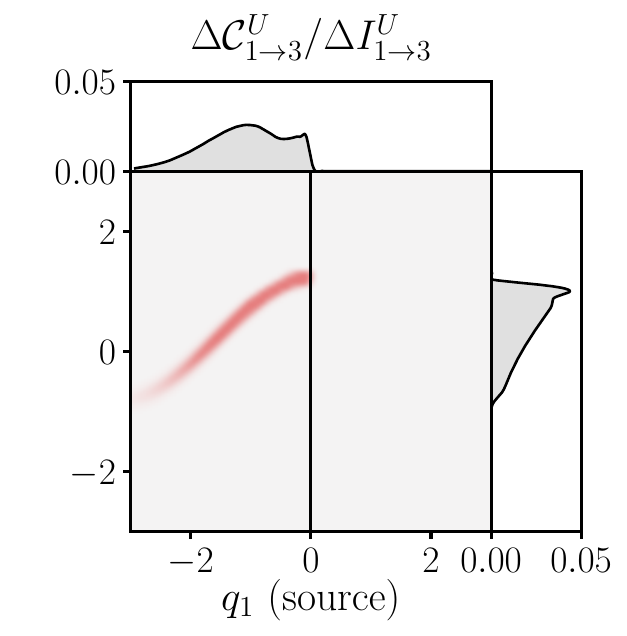}};
        \node[anchor=north east, align=center] at (1.65, -3.1) {\tiny{causal}};
        \node[anchor=north east, align=center] at (3.075, -3.1) {\tiny{non-causal}};
        \node[anchor=south, font=\small, fill=white, text=black, inner sep=2pt, rounded corners, minimum width=0.2\textwidth,minimum height=6mm] at ([yshift=-5.8mm,xshift=-3mm]img.north) {$\quad$Unique U1$\quad$};
    \end{tikzpicture}
    \end{minipage}
    \begin{minipage}{0.21\textwidth}
    \begin{tikzpicture}
        \node[anchor=north west] (img) at (0, 0) {\includegraphics[width=\linewidth, trim=26 0 0 0, clip]{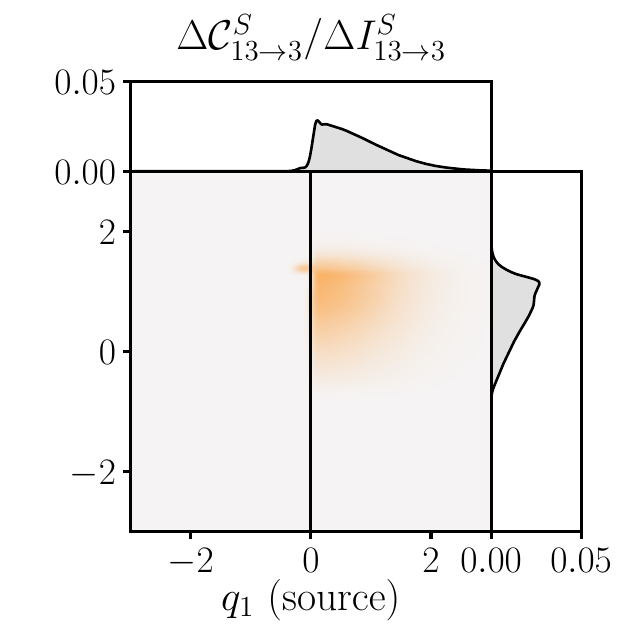}};
        \node[anchor=north east, align=center] at (1.85, -3.05) {\tiny{non-causal}};
        \node[anchor=north east, align=center] at (2.8, -3.05) {\tiny{causal}};
        \node[anchor=south, font=\small, fill=white, text=black, inner sep=2pt, rounded corners, minimum width=0.2\textwidth,minimum height=6mm] at ([yshift=-5.8mm,xshift=-2mm]img.north) {$\quad$Synergistic S13$\quad$};
    \end{tikzpicture}
    \end{minipage}
    \begin{minipage}{0.085\textwidth}
    \vspace{-0.25\textwidth}
    \begin{tikzpicture}
        \node[anchor=north west] (img) at (0, 0) {\includegraphics[width=\linewidth]{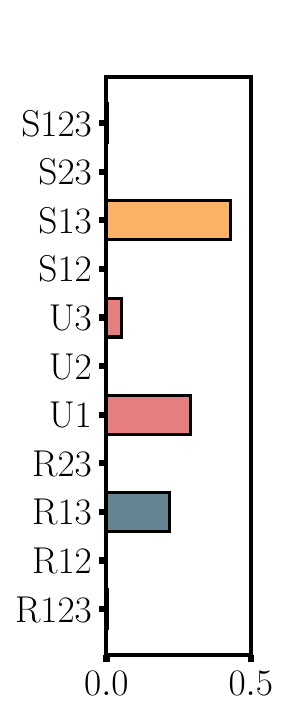}};
        \node[anchor=south, font=\small, fill=white, text=black, inner sep=2pt, rounded corners, minimum width=0.2\textwidth,minimum height=6mm] at ([yshift=-4.5mm,xshift=2mm]img.north) {SURD};
    \end{tikzpicture}
    \end{minipage}
    \begin{minipage}{0.21\textwidth}
    \begin{tikzpicture}
        \node[anchor=north west] (img) at (0, 0) {\includegraphics[width=\linewidth, trim=30 0 0 0, clip]{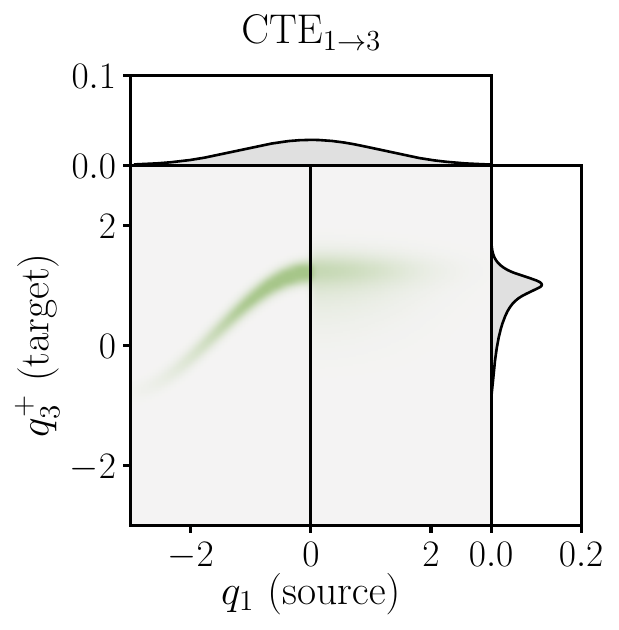}};
        \node[anchor=south, font=\small, fill=white, text=black, inner sep=2pt, rounded corners, minimum width=0.2\textwidth,minimum height=6mm] at ([yshift=-5.8mm,xshift=-2mm]img.north) {$\quad {\rm{CTE}}_{1\to 3} \quad$};
    \end{tikzpicture}
    \end{minipage}

    \vspace{0.0075\textwidth}
    \begin{minipage}{\textwidth}
    \centering
        $
          q_3(n+1) =  
          \begin{cases}
            q_1(n) \cos\left[q_1(n)\right] + 0.1 w_3(n) & \text{if } q_1(n) \cos\left[q_1(n)\right] + 0.1 w_3(n) < 0 \\
            q_3(n) + 0.1 w_3(n) & \text{otherwise}
          \end{cases}
        $
    \end{minipage}
    \vspace{0.0075\textwidth}

    \begin{minipage}{0.23\textwidth}
    \begin{tikzpicture}
        \node[anchor=north west] (img) at (0, 0) {\includegraphics[width=\linewidth]{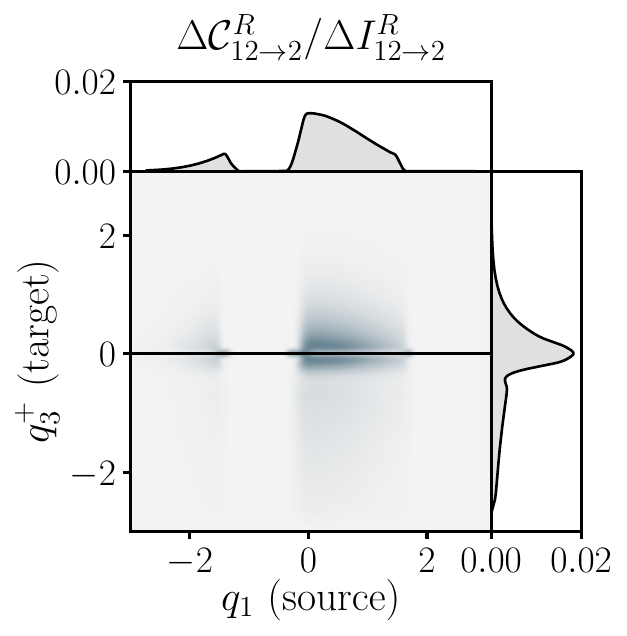}};
        \node[anchor=south, font=\small, fill=white, text=black, inner sep=2pt, rounded corners, minimum width=0.2\textwidth,minimum height=6mm] at ([yshift=-5.6mm,xshift=-1mm]img.north) {$\quad$Redundant R13$\quad$};
    \end{tikzpicture}
    \end{minipage}
    \begin{minipage}{0.21\textwidth}
    \begin{tikzpicture}
        \node[anchor=north west] (img) at (0, 0) {\includegraphics[width=\linewidth, trim=26 0 0 0, clip]{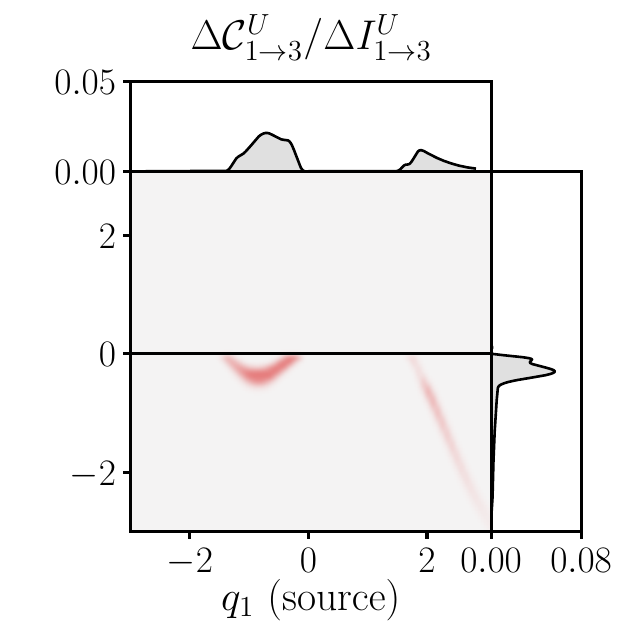}};
        \node[anchor=north east, align=center] at (2.3, -3.0) {\tiny{causal}};
        \node[anchor=north east, align=center] at (2.55, -1.49) {\tiny{non-causal}};
        \node[anchor=south, font=\small, fill=white, text=black, inner sep=2pt, rounded corners, minimum width=0.2\textwidth,minimum height=6mm] at ([yshift=-5.8mm,xshift=-3mm]img.north) {$\quad$Unique U1$\quad$};
    \end{tikzpicture}
    \end{minipage}
    \begin{minipage}{0.21\textwidth}
    \begin{tikzpicture}
        \node[anchor=north west] (img) at (0, 0) {\includegraphics[width=\linewidth, trim=26 0 0 0, clip]{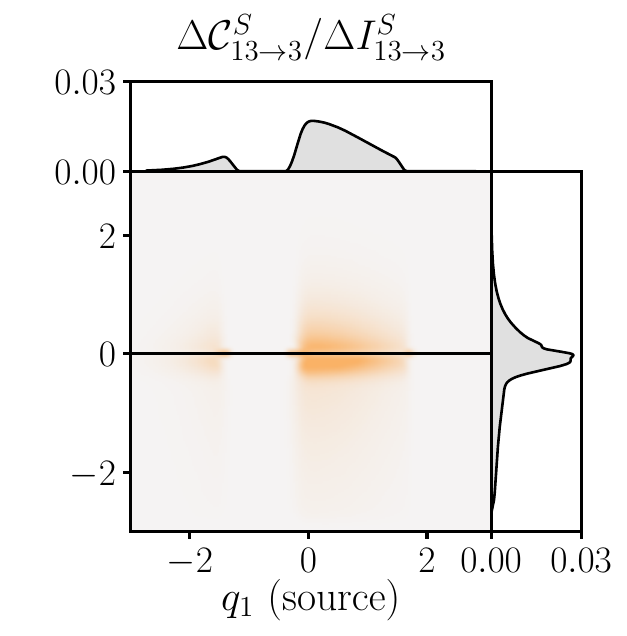}};
        \node[anchor=south, font=\small, fill=white, text=black, inner sep=2pt, rounded corners, minimum width=0.2\textwidth,minimum height=6mm] at ([yshift=-5.8mm,xshift=-2mm]img.north) {$\quad$Synergistic S13$\quad$};
        \node[anchor=south, font=\small, fill=white, text=black, inner sep=2pt, minimum width=0.2\textwidth,minimum height=2mm, rotate=90] at ([yshift=-23mm,xshift=-18mm]img.north) {$\quad\quad\quad\quad\quad$};
    \end{tikzpicture}
    \end{minipage}
    \begin{minipage}{0.08\textwidth}
    \vspace{-0.25\textwidth}
    \begin{tikzpicture}
        \node[anchor=north west] (img) at (0, 0) {\includegraphics[width=\linewidth]{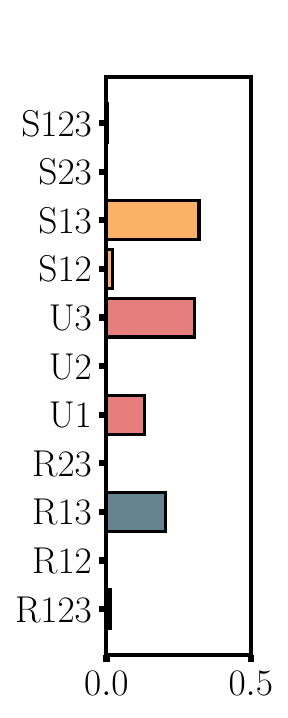}};
        \node[anchor=south, font=\small, fill=white, text=black, inner sep=2pt, rounded corners, minimum width=0.2\textwidth,minimum height=6mm] at ([yshift=-4.5mm,xshift=2mm]img.north) {SURD};
    \end{tikzpicture}
    \end{minipage}
    \begin{minipage}{0.21\textwidth}
    \begin{tikzpicture}
        \node[anchor=north west] (img) at (0, 0) {\includegraphics[width=\linewidth, trim=0 0 0 0, clip]{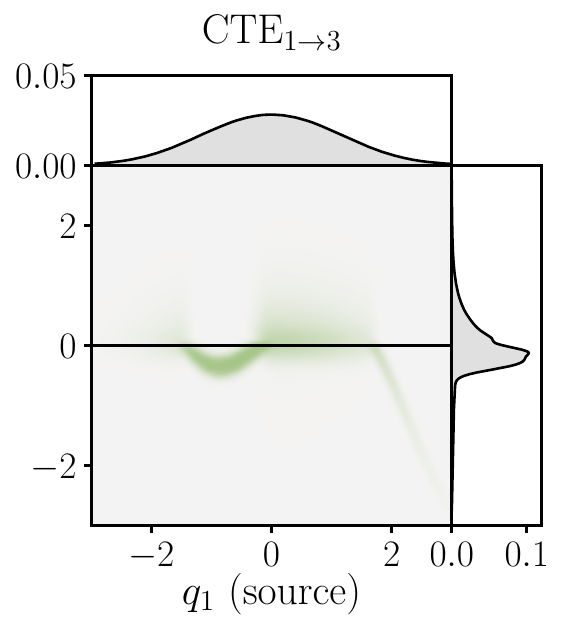}};
        \node[anchor=south, font=\small, fill=white, text=black, inner sep=2pt, rounded corners, minimum width=0.2\textwidth,minimum height=6mm] at ([yshift=-5.8mm,xshift=-2mm]img.north) {$\quad {\rm{CTE}}_{1\to 3} \quad$};
    \end{tikzpicture}
    \end{minipage}

    \vspace{0.0075\textwidth}
    \begin{minipage}{\textwidth}
    \centering
        $
          q_3(n+1) =  
          \begin{cases}
            \sin\left[q_1(n)q_2(n)\right] + 0.1 w_3(n) & \text{if } q_1(n) q_2(n) > 0 \\
            w_3(n) & \text{otherwise}
          \end{cases}
        $
    \end{minipage}
    \vspace{0.0075\textwidth}

    \begin{minipage}{0.23\textwidth}
    \begin{tikzpicture}
        \node[anchor=north west] (img) at (0, 0) {\includegraphics[width=\linewidth]{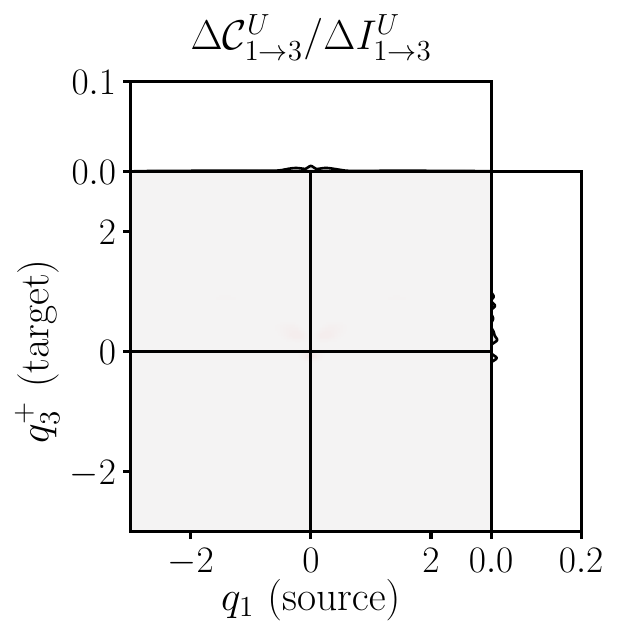}};
        \node[anchor=south, font=\small, fill=white, text=black, inner sep=2pt, rounded corners, minimum width=0.2\textwidth,minimum height=6mm] at ([yshift=-5.8mm,xshift=-1mm]img.north) {$\quad$Unique U1$\quad$};
    \end{tikzpicture}
    \end{minipage}
    \begin{minipage}{0.23\textwidth}
    \begin{tikzpicture}
        \node[anchor=north west] (img) at (0, 0) {\includegraphics[width=\linewidth, trim=0 0 0 0, clip]{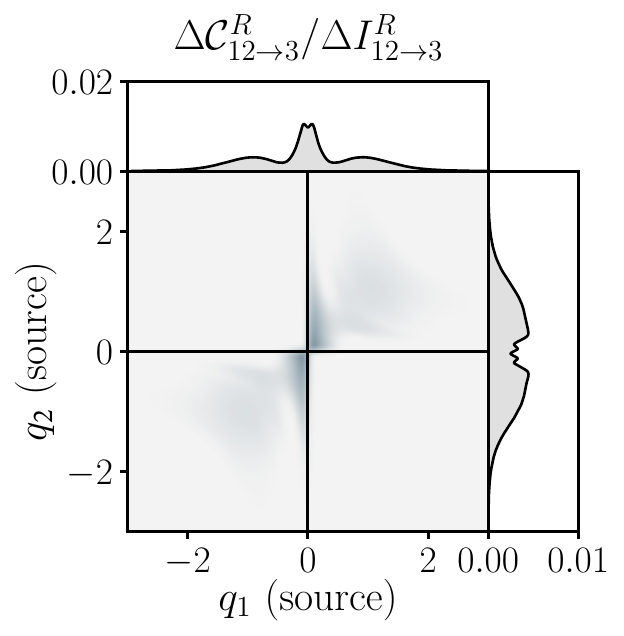}};
        \node[anchor=south, font=\small, fill=white, text=black, inner sep=2pt, rounded corners, minimum width=0.2\textwidth,minimum height=6mm] at ([yshift=-5.4mm,xshift=-1mm]img.north) {$\quad$Redundant R12$\quad$};
    \end{tikzpicture}
    \end{minipage}
    \begin{minipage}{0.21\textwidth}
    \begin{tikzpicture}
        \node[anchor=north west] (img) at (0, 0) {\includegraphics[width=\linewidth, trim=25 0 0 0, clip]{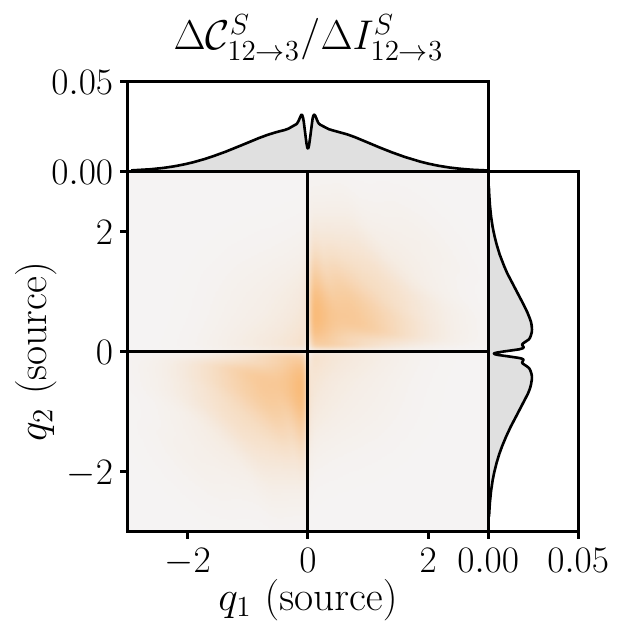}};
        \node[anchor=north east, align=center] at (1.65, -3.0) {\tiny{causal}};
        \node[anchor=north east, align=center] at (1.875, -1.49) {\tiny{non-causal}};
        \node[anchor=north east, align=center] at (3.05, -3.0) {\tiny{non-causal}};
        \node[anchor=north east, align=center] at (2.85, -1.49) {\tiny{causal}};
        \node[anchor=south, font=\small, fill=white, text=black, inner sep=2pt, rounded corners, minimum width=0.2\textwidth,minimum height=6mm] at ([yshift=-5.8mm,xshift=-2mm]img.north) {$\quad$Synergistic S12$\quad$};
        \node[anchor=south, font=\small, fill=white, text=black, inner sep=2pt, minimum width=0.2\textwidth,minimum height=2mm, rotate=90] at ([yshift=-23mm,xshift=-18mm]img.north) {$\quad\quad\quad\quad\quad$};
    \end{tikzpicture}
    \end{minipage}
    \begin{minipage}{0.08\textwidth}
    \vspace{-0.25\textwidth}
    \begin{tikzpicture}
        \node[anchor=north west] (img) at (0, 0) {\includegraphics[width=\linewidth]{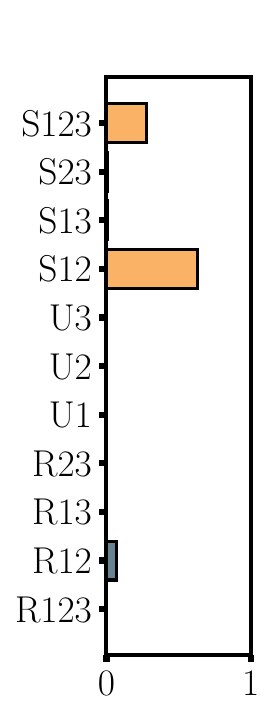}};
        \node[anchor=south, font=\small, fill=white, text=black, inner sep=2pt, rounded corners, minimum width=0.2\textwidth,minimum height=6mm] at ([yshift=-4.5mm,xshift=2mm]img.north) {SURD};
    \end{tikzpicture}
    \end{minipage}
    \begin{minipage}{0.21\textwidth}
    \begin{tikzpicture}
        \node[anchor=north west] (img) at (0, 0) {\includegraphics[width=\linewidth, trim=0 0 0 0, clip]{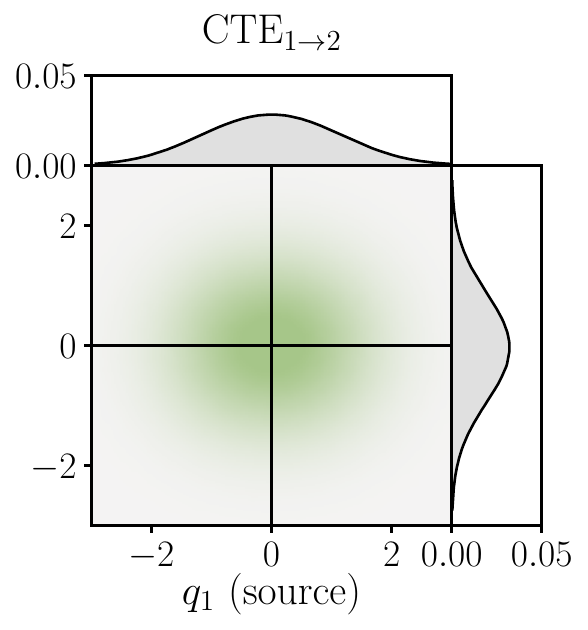}};
        \node[anchor=south, font=\small, fill=white, text=black, inner sep=2pt, rounded corners, minimum width=0.2\textwidth,minimum height=6mm] at ([yshift=-5.8mm,xshift=-2mm]img.north) {$\quad {\rm{CTE}}_{12\to 3} \quad$};
    \end{tikzpicture}
    \end{minipage}

    \caption{State-wise decomposition of causal states for (top)
      unique source-dependent, (middle) target-dependent cases, and
      (bottom) synergistic source-dependent cases. From left to right,
      panels show: state-dependent redundant (R), unique (U), and
      synergistic (S) causal contributions in blue, red, and yellow,
      respectively; the total causalities to the target variable
      $Q_3^+$ from SURD; and the state-dependent conditional transfer
      entropy (CTE). The notation used here is such that U1 represents
      the unique causality from $Q_1$ to $Q_3^+$, i.e., $\Delta
      I^U_{1\to 3}$, and similarly for other terms. The causal maps
      for redundant and synergistic causalities are average over all
      the states $q_3$.}
    \label{fig:example-contributions-sm}
\end{figure}

We investigate additional benchmark cases to further assess the
performance of our state-dependent causal inference method. To this
end, we extend the benchmark systems analyzed in the validation
section of the main text by introducing an additional target variable,
$Q_3^+$, whose dynamics mirror those of $Q_2^+$ but with an opposite
conditioning structure. The variables $Q_2^+$ and $Q_3^+$ are
constructed to be statistically independent of one another.  This
setup is designed to evaluate the ability of our method to distinguish
state-dependent confounding effects, where the same source variable,
$Q_1$, influences different targets in distinct regions of its state
space. Specifically, the interactions are defined such that $q_1
\rightarrow q_2^+$ for $q_1 > 0$, and $q_1 \rightarrow q_3^+$ for $q_1
< 0$.  The dynamics of $Q_2^+$ are identical to those presented in
Fig.~\ref{fig:example-contributions}, while the behavior of $Q_3^+$ is
shown in Fig.~\ref{fig:example-contributions-sm}. The source variable
$Q_1$ is sampled from a standard normal distribution with zero mean
and unit variance. The variables $Q_2$ and $Q_3$ are driven by
independent stochastic forcings $W_2$ and $W_3$, respectively, each
modeled as Gaussian random variables with means of $-2$ and $2$, and
unit standard deviation.

The state-dependent redundant, unique, and synergistic causalities for
the target variable $Q_2^+$ are identical to those shown in
Fig.~\ref{fig:example-contributions}, as the dynamics governing
$Q_2^+$ remain unchanged and the newly introduced variable $Q_3^+$
does not influence $Q_2^+$. Figure~\ref{fig:example-contributions-sm}
shows the corresponding results for the target variable $Q_3^+$.
First, for the redundant causality $\Iss^R_{13\to 3}$, we observe that
positive states $q_1 > 0$ are causal, whereas negative states $q_1 <
0$ are non-causal. This indicates that causality is shared between
$Q_1$ and $Q_3$ about $Q_3^+$ when $q_1 < 0$, as both $Q_1$ and $Q_3$
contribute information necessary to determine the future state of
$Q_3^+$.  For the unique causality $\Iss^U_{1\to 3}$, the results
clearly show that negative $q_1$ states ($q_1 < 0$) are causal
($\Iss^U_{1\to 3} > 0$). This implies that $Q_1$ uniquely drives
$Q_3^+$ in this regime without requiring information from $Q_3$.
Finally, for the synergistic causality $\Iss^S_{13\to 3}$, causal
states appear for $q_1 > 0$, where both $Q_1$ and $Q_3$ are jointly
required to predict $Q_3^+$. This synergistic behavior implies that
additional predictive information about $Q_3^+$ only emerges when
considering the combination of $Q_1$ and $Q_3$ together, beyond what
each variable can contribute individually through their unique
causalities.

Figure~\ref{fig:example-contributions-sm} also illustrates the SURD
causalities. The dominant causal influences originate from
combinations of the source variables $Q_1$ and $Q_3$, as reflected in
the redundant ($\Delta I^R_{13\to 3}$), unique ($\Delta I^U_{1\to 3}$
and $\Delta I^U_{3\to 3}$), and synergistic ($\Delta I^S_{13\to 3}$)
contributions. These results are consistent with the functional
dependencies of the system, where $Q_1$ and $Q_3$ govern the future
evolution of $Q_3^+$ depending on the specific state of $Q_1$.  The
analysis further shows that the role of $Q_2$ in determining $Q_3^+$
is negligible, indicating that the method correctly identifies the
relevant causal variables. However, as in the benchmark shown in the
main text, SURD causalities do not reveal the specific system states
in which each type of causality becomes active.
 
Additionally, we compare the results with those obtained using
state-dependent CTE, shown on the right-hand side of
Fig.~\ref{fig:example-contributions-sm}. As in the case of $Q_2^+$,
the CTE framework aggregates all causal contributions into a single
measure, which obscures the distinctions between redundant, unique,
and synergistic interactions.  In particular, CTE fails to reveal that
when $q_1 < 0$, $Q_1$ uniquely determines the future of $Q_3^+$,
whereas when $q_1 > 0$, information from both $Q_1$ and $Q_3$ is
required.

\subsection{Target-dependent unique causality}

The second case illustrates the decomposition of causality in a
target-dependent system, extending the validation example from the
main text by introducing an additional variable, $Q_3^+$, which also
exhibits target-dependent dynamics. The objective is to evaluate
whether the method can detect a different type of state-dependent
behavior and effectively handle confounding effects, given that the
source variable $Q_1$ influences both $Q_2^+$ and $Q_3^+$.  In this
system, the relationship is defined such that $f(q_1) \rightarrow
q_3^+$ when $f(q_1) < 0$, and $q_3 \rightarrow q_3^+$ when $f(q_1)
\geq 0$, where $f(\cdot)$ is a nonlinear function mapping the past of
the source variable to the future state of the target. The dynamics of
$Q_2^+$ are identical to those presented in the main text, ensuring
that any observed differences can be attributed solely to the
introduction of $Q_3^+$ and its interaction with $Q_1$.  The variable
$Q_1$ is drawn from a standard normal distribution (mean zero, unit
variance), while $Q_3$ is influenced by stochastic forcing $W_3$,
modeled as a Gaussian random variable with mean 1 and unit standard
deviation. The dynamics of the target variable $Q_3^+$ are illustrated
in Fig.~\ref{fig:example-contributions-sm}.

First, it was assessed that the results for $Q_2^+$ remain unchanged
after introducing the additional variable $Q_3$, as the relevant
causalities are identical to those reported in the main text. This
demonstrates that the proposed method is robust to confounding
effects, particularly in scenarios where a single variable
simultaneously influences multiple targets.

The results for $Q_3^+$ are presented in
Fig.~\ref{fig:example-contributions-sm}. Similarly to the findings for
$Q_2^+$, we observe that the total causality maps obtained from SURD
for $Q_3^+$---both in the source-dependent and target-dependent
regimes---appear to be remarkably similar, despite the underlying differences
in the system interactions. This highlights the limitation of
aggregate causal measures and underscores the value of a
state-resolved decomposition, which provides a more faithful
representation of the underlying causal structure.
Figure~\ref{fig:example-contributions-sm} illustrates the
state-dependent decomposition of redundant ($\Iss^R_{13 \rightarrow
  3}$), unique ($\Iss^U_{1 \rightarrow 3}$), and synergistic
($\Iss^S_{13 \rightarrow 3}$) causalities as functions of the source
state $q_1$ and target state $q_3^+$. The analysis clearly shows that
the unique causality $\Iss^U_{1 \rightarrow 3}$ varies primarily with
the target state, rather than the source. In particular, negative
values of the target variable ($q_3^+ < 0$) correspond to the
strongest unique influence from $Q_1$, reflecting the regime where
$Q_1$ alone drives $Q_3^+$.  In contrast, both the synergistic
$\Iss^S_{13 \rightarrow 3}$ and redundant $\Iss^R_{13 \rightarrow 3}$
causalities exhibit nonzero contributions across the full range of
$q_3^+$, indicating that $Q_1$ and $Q_3$ jointly provide information
about $Q_3^+$ in both positive and negative target states.

Figure \ref{fig:example-contributions-sm} also compares the
state-dependent causalities with the results obtained from
state-dependent CTE for $Q_3^+$. As in the source-dependent case,
state-dependent CTE is unable to disentangle redundant, unique, and
synergistic causal interactions. For instance, it does not capture the
unique causality from $Q_1$ to $Q_3^+$ when $q_3^+ < 0$, nor does it
reveal that the states for which causality becomes redundant or
synergistic.

\subsection{Source-dependent synergistic causality}

This benchmark case involves a system in which the direction of
causality varies synergistically depending on the joint intensity of
two source variables. Specifically, we consider a system with three
variables, where $[q_1, q_2] \rightarrow q_3^+$ when $q_1 q_2 > 0$,
and $w_3 \rightarrow q_3^+$ otherwise. The source variables $Q_1$ and
$Q_2$ are independent and sampled from a standard normal distribution
(mean zero, unit variance). The target variable $Q_3$ is also
influenced by a stochastic forcing term $W_3$, which is independent,
normally distributed with zero mean and unit variance, and not
included in the set of observed variables.

The results for the state-dependent causalities are shown in
Fig.~\ref{fig:example-contributions-sm}, alongside the corresponding
SURD causalities. The key insight from this analysis arises from the
synergistic term $\Iss^S_{12 \rightarrow 3}$, which clearly highlights
strong synergistic effects in regions where $q_1 q_2 > 0$—i.e., in the
first and third quadrants of the $q_1$–$q_2$ space. In contrast, no
causality is detected in the regime where $q_1 q_2 \leq 0$, as the
dynamics of $Q_3^+$ are entirely driven by the unobserved stochastic
forcing $W_3$ in this region. The regime $q_1 q_2 > 0$ also exhibits a small
amount of redundant causality, as captured by the SURD decomposition,
though its magnitude remains relatively low. Moreover, the unique
causalities from $Q_1$ and $Q_2$ are nearly zero throughout,
reflecting the fact that neither source variable alone is sufficient
to drive $Q_3^+$—causal influence emerges only when both are
considered jointly.

Finally, Fig.~\ref{fig:example-contributions-sm} also compares the
state-dependent causalities with those obtained from
CTE. Specifically, the figure presents the results for the combination
$\mathrm{CTE}_{12 \rightarrow 3}$, where no discernible difference is
observed between the cases $ q_1 q_2 > 0 $ and $ q_1 q_2 < 0 $. Hence,
CTE is unable to capture state-dependent relationships involving
synergistic effects.

\subsection{Coupled Rössler--Lorenz system}

We study a coupled version of the Lorenz system and the R\"ossler
system. The former was developed by Lorenz as a simplified model of a
viscous fluid flow. R\"ossler proposed a simpler version of the
Lorenz's equations in order to facilitate the study of its chaotic
properties. The governing equations are:
\begin{subequations}
\begin{align}
  \label{eq:RL}
 \frac{\dd q_1}{\dd t} &= -6[ q_2 + q_3 ], \\
 \frac{\dd q_2}{\dd t} &= 6[q_1 + 0.2q_2 ], \\
 \frac{\dd q_3}{\dd t} &= 6\left[ 0.2+q_3[q_1-5.7] \right], \\
 \frac{\dd q_4}{\dd t} &= 10[ q_5 - q_4 ], \\
 \frac{\dd q_5}{\dd t} &= q_4 [28 - q_6 ] - q_5 + c(q_2) q_2^2, \\
 \frac{\dd q_6}{\dd t} &= q_4 q_5 - \frac{8}{3}q_6,
\end{align}
\end{subequations}
where $[Q_1,Q_2,Q_3]$ correspond to the R\"ossler system and
$[Q_4,Q_5,Q_6]$ to the Lorenz system. The coupling between the systems
is unidirectional from the R\"ossler system to the Lorenz system from
$ Q_2 \rightarrow Q_5$ via the parameter $c(q_2)$ depending on the
intensity of $q_2$. The aim is to infer the states of the R\"ossler
system $[Q_1,Q_2,Q_3]$ that are the most causal to $Q_5$ from the
Lorenz system. Therefore, the observable source variables are $\bQ =
[Q_1, Q_2, Q_3]$. The system was integrated for $10^6 t_{\text{ref}}$
where $t_{\text{ref}}$ is the time for which $I(Q_1^+ ; Q_1)/I(Q_1 ;
Q_1) = 0.5$. The time-lag selected for causal inference is the one
that maximizes the cross-induced unique causality for $Q_5$ and 25
bins per variable were used to partition the observed phase space.
\begin{figure}[t!]
    \centering
    \subfloat[Rössler system]{\includegraphics[width=0.29\linewidth]{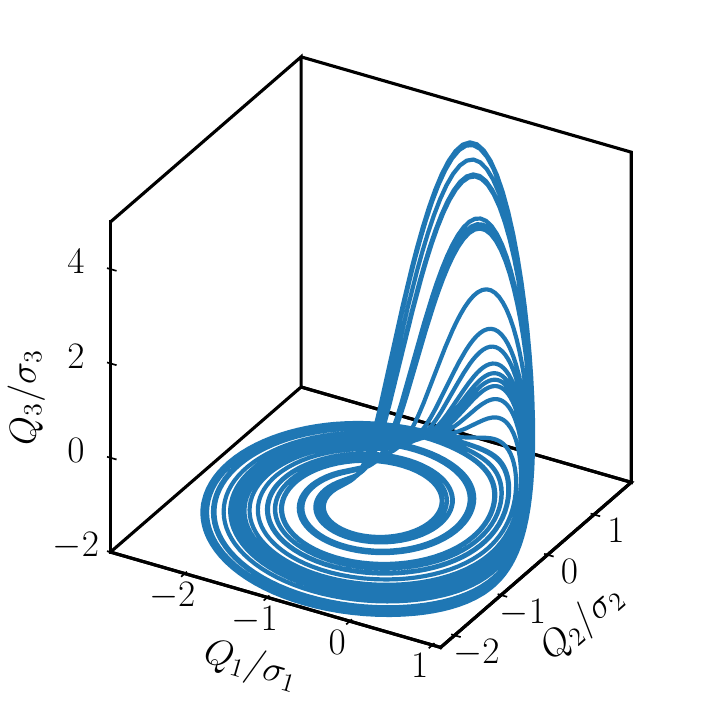}}
    \subfloat[$c=2$ if $q_2<0$ ]{\includegraphics[width=0.29\linewidth]{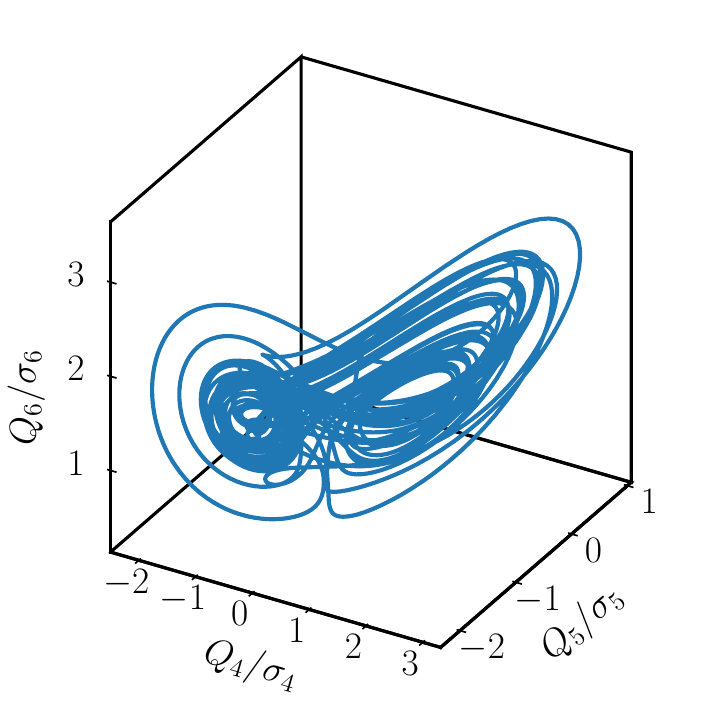}}
    \subfloat[$c=2$ if $q_2>0$]{\includegraphics[width=0.29\linewidth]{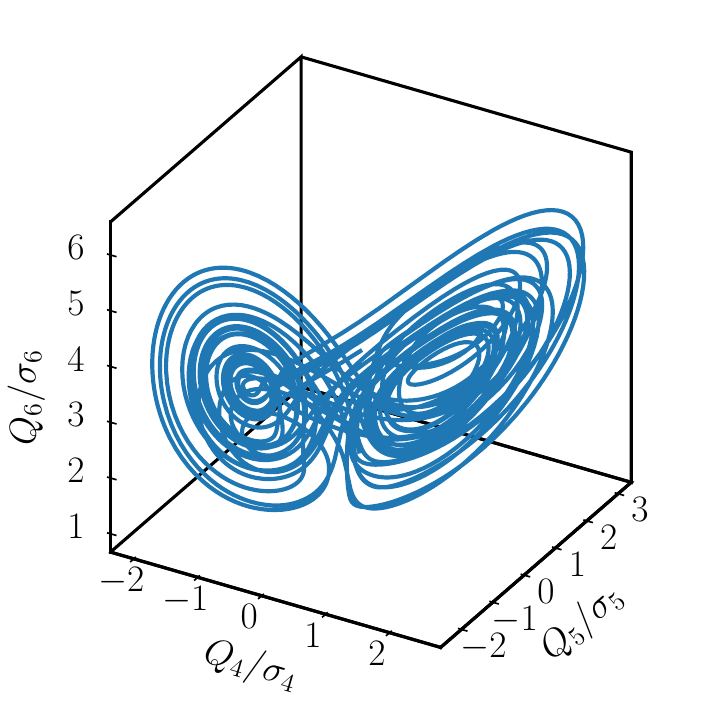}}

    \hspace{-0.045\linewidth}
    \subfloat[$c=2$ if $q_2<0$]{\includegraphics[width=0.46\linewidth, trim=0 0 70 0, clip]{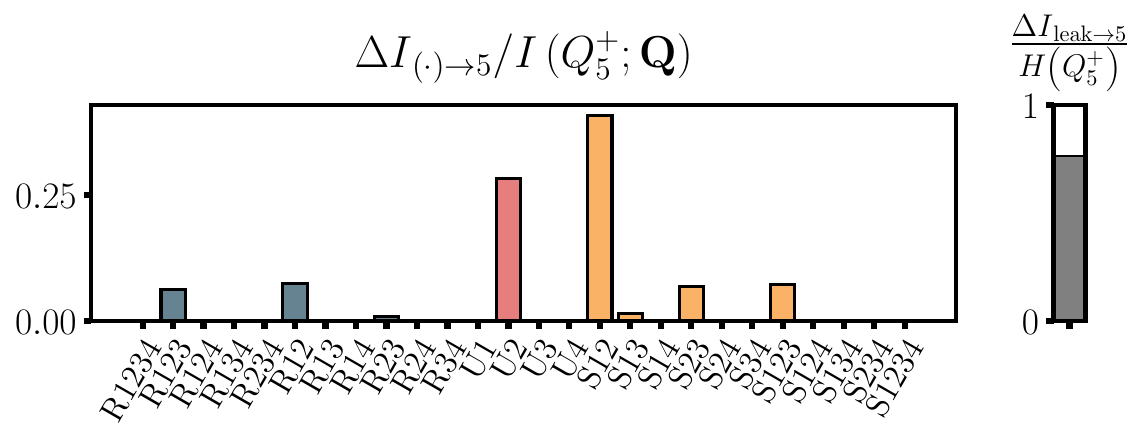}}
    \hspace{0.06\linewidth}
    \subfloat[$c=2$ if $q_2>0$]{\includegraphics[width=0.45\linewidth, trim=0 0 70 0, clip]{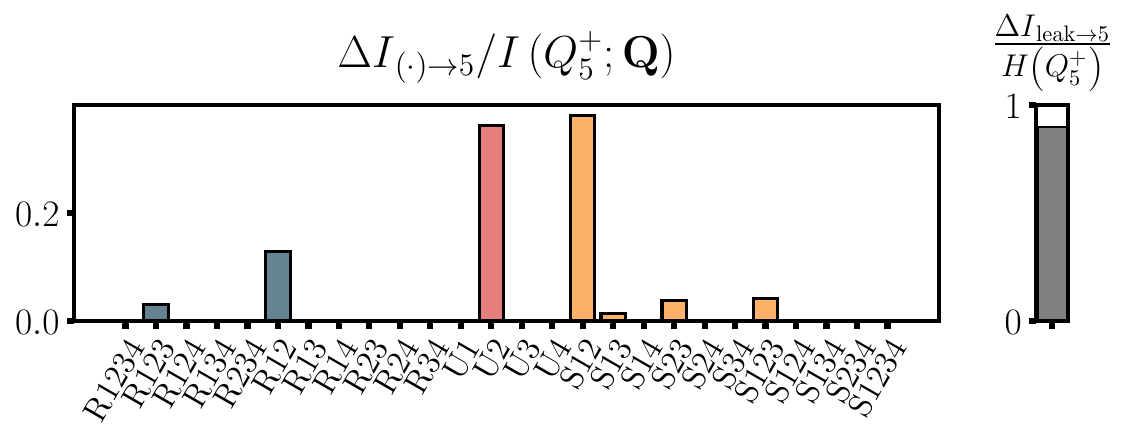}}

    \hspace{-0.05\linewidth}
    \subfloat[$c=2$ if $q_2<0$]{\includegraphics[width=0.45\linewidth]{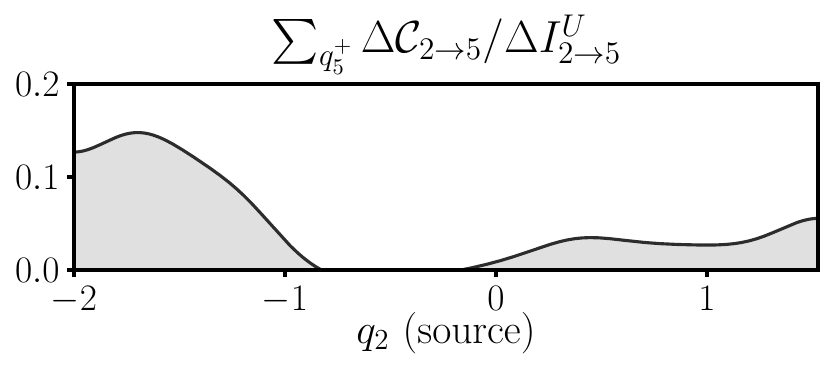}}
    \hspace{0.06\linewidth}
    \subfloat[$c=2$ if $q_2>0$]{\includegraphics[width=0.45\linewidth]{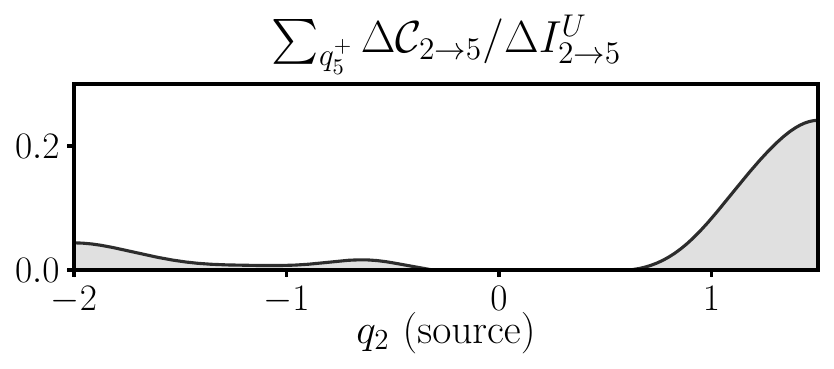}}
    
    \caption{Coupled R\"ossler--Lorenz system. The upper panels show
      excerpts of the trajectories pertaining to (a) the R\"ossler
      system $[Q_1,Q_2,Q_3]$ and the Lorenz system $[Q_4,Q_5,Q_6]$
      with (b) $c=2$ if $q_2<0$ and (c) $c=2$ if $q_2>0$. Redundant
      (R), unique (U), and synergistic (S) causalities from SURD among
      $[Q_1,Q_2,Q_3]$ for (d) $c=2$ if $q_2<0$ and (e) $c=2$ if
      $q_2>0$. The causality leak for each variable is also shown in
      the right-hand side bar. Panels (f) and (g) show the
      state-dependent decomposition of the unique causality
      $\Iss_{2\to 5} ^ U$ for both couplings.}
    \label{fig:rossler-lorenz}
\end{figure}

This section explores two coupling scenarios between the R\"ossler and
Lorenz systems. In the first scenario, the coupling from $q_2$
(R\"ossler system) to $q_5$ (Lorenz system) is activated only when the
states of $q_2$ are negative; in the second, the coupling is active
only for positive $q_2$ values.  When $q_2 > 0$, the R\"ossler
attractor tends to occupy regions where $q_3$ reaches large values,
whereas for $q_2 < 0$, $q_3$ remains negative and relatively
stable. This state-dependent coupling results in distinct dynamic
behaviors in the Lorenz system, giving rise to two different
attractors depending on the coupling regime.  The corresponding
attractors are shown in Fig.~\ref{fig:rossler-lorenz}(b) for the case
where coupling is active for $q_2 < 0$, and in
Fig.~\ref{fig:rossler-lorenz}(c) for the case where coupling is active
for $q_2 > 0$.

The results from the SURD decomposition for both coupling scenarios
are presented in Figs.~\ref{fig:rossler-lorenz}(d) and
\ref{fig:rossler-lorenz}(e), respectively. In both cases, the most
prominent unique causality originates from variable $q_2$, quantified
by $\Delta I^U_{2 \rightarrow 5}$. This indicates that $q_2$ has a
direct and distinctive influence on the target variable $q_5$.  In
addition, a strong synergistic causality from the pair $(q_1, q_2)$ to
$q_5$, denoted by $\Delta I^S_{12 \rightarrow 5}$, is observed. This
suggests that $q_1$ also contributes to the evolution of $q_5$,
despite not being directly coupled in the Lorenz equations. This
influence arises due to the interaction between $q_1$ and $q_2$ within
the original R\"ossler system, which becomes manifest over the
integration time scale used in the analysis---ultimately causing all
variables to appear in the right-hand side of the system equations.

As in the other cases discussed above, the Lorenz attractor exhibits
visibly different structures under the two coupling
scenarios. However, the total causal influence measured by SURD
remains largely similar across both cases. This highlights once again
that SURD alone may not fully capture the nuanced, state-dependent
dynamics of the system. In such instances, decomposing causality into
state-specific components offers additional insights.  This
decomposition is illustrated in Figs.~\ref{fig:rossler-lorenz}(f) and
\ref{fig:rossler-lorenz}(g), where it becomes clear that negative
$q_2$ states are the most influential when the coupling is active only
for $q_2 < 0$, while the opposite holds when the coupling is
restricted to $q_2 > 0$. Moreover, the decomposition reveals that
positive $q_2$ states still contribute relevant information about
$q_5$ in the $q_2 < 0$ coupling regime, provided that $c \neq 0$,
within the time scale considered for causal inference.  This effect
arises from the tendency of the system to transition into the $q_2 <
0$ regime shortly after experiencing large positive values of $q_2$,
thus allowing past positive states to carry predictive information
about the future. A similar phenomenon occurs in the $q_2 > 0$
coupling scenario, where large negative $q_2$ states can retain
residual influence on $q_5^+$ over the time-scale considered.

\newpage
\section{Analysis of non-causal contributions}

We examine the non-causal contributions corresponding to the cases
analyzed in the main text. These contributions arise in instances
where knowledge of the source states leads to a loss of
information---i.e., an increase in uncertainty---about the target
variable, relative to the previous level of SURD causality. An
alternative interpretation is that non-causal contributions indicate
causal influence that has already been accounted for at a lower level
in the SURD hierarchy.  For example, consider a system with one
redundant, one unique, and one synergistic S-causality component:
\begin{itemize}
    \item Non-causal redundant contributions reflect a loss of
      information relative to the baseline information already
      contained in the target $q_j^+$.
    \item Non-causal unique contributions represent a loss of
      information relative to what is already explained by the
      redundant component.
    \item Non-causal synergistic contributions represent a loss
      relative to the unique component.
\end{itemize}
In all cases, these non-causal terms are negative by
construction. Importantly, the average over both causal and non-causal
contributions recovers the total SURD causality at each level.

\subsection{Source-dependent and target-dependent systems}

\begin{figure}[t]
    \centering
    \begin{minipage}{\textwidth}
    \begin{center}
        $
          q_2(n+1) =  
          \begin{cases}
            \sin\left[q_1(n)\right] + 0.1 w_2(n) & \text{if } q_1(n) > 0 \\
            0.9 q_2(n) + 0.1 w_2(n) & \text{otherwise}
          \end{cases}
        $
    \end{center}
    \end{minipage}
    \vspace{0.0075\textwidth}
    
    \begin{minipage}{\textwidth}
    \begin{center}
    \begin{minipage}{0.31\textwidth}
    \begin{tikzpicture}
        \node[anchor=north west] (img) at (0, 0) {\includegraphics[width=\linewidth]{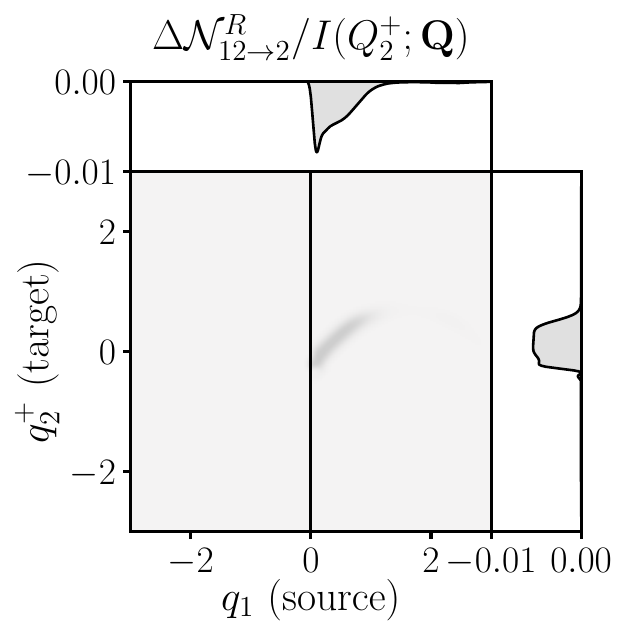}};
    \end{tikzpicture}
    \end{minipage}
    \begin{minipage}{0.31\textwidth}
    \begin{tikzpicture}
        \node[anchor=north west] (img) at (0, 0) {\includegraphics[width=\linewidth, trim=0 0 0 0, clip]{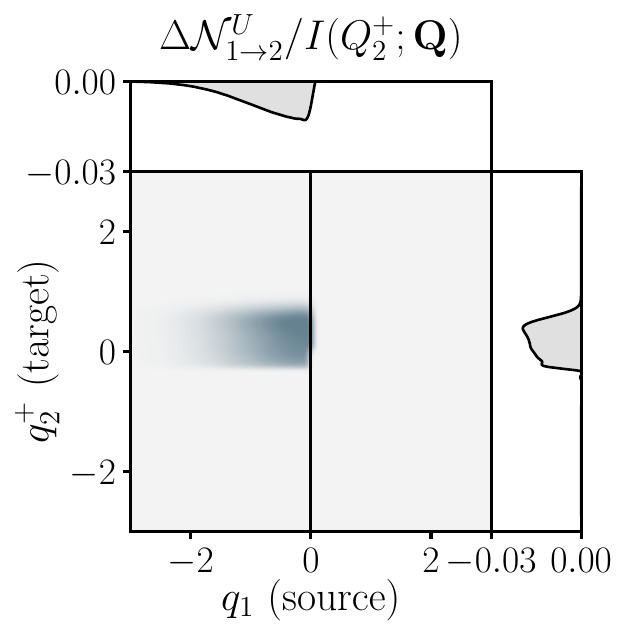}};
    \end{tikzpicture}
    \end{minipage}
    \begin{minipage}{0.31\textwidth}
    \begin{tikzpicture}
        \node[anchor=north west] (img) at (0, 0) {\includegraphics[width=\linewidth, trim=0 0 0 0, clip]{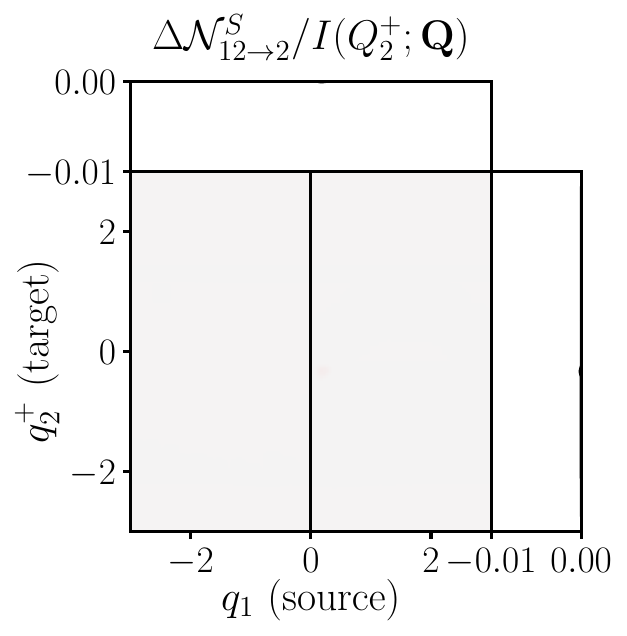}};
    \end{tikzpicture}
    \end{minipage}
    \end{center}
    \end{minipage}

    \begin{minipage}{\textwidth}
    \begin{center}
        $
          q_2(n+1) =  
          \begin{cases}
            q_1(n) \sin\left[q_1(n)\right] + 0.1 w_2(n) & \text{if } q_1(n) \sin\left[q_1(n)\right] + 0.1 w_2(n) > 0 \\
            0.9 q_2(n) + 0.1 w_2(n) & \text{otherwise}
          \end{cases}
        $
    \end{center}
    \end{minipage}
    \vspace{0.0075\textwidth}

    \begin{minipage}{\textwidth}
    \begin{center}
    \begin{minipage}{0.31\textwidth}
    \begin{tikzpicture}
        \node[anchor=north west] (img) at (0, 0) {\includegraphics[width=\linewidth]{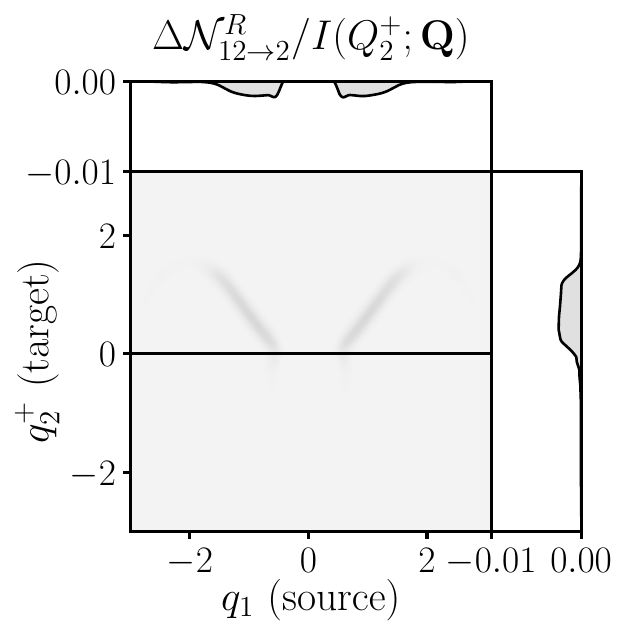}};
    \end{tikzpicture}
    \end{minipage}
    \begin{minipage}{0.31\textwidth}
    \begin{tikzpicture}
        \node[anchor=north west] (img) at (0, 0) {\includegraphics[width=\linewidth, trim=0 0 0 0, clip]{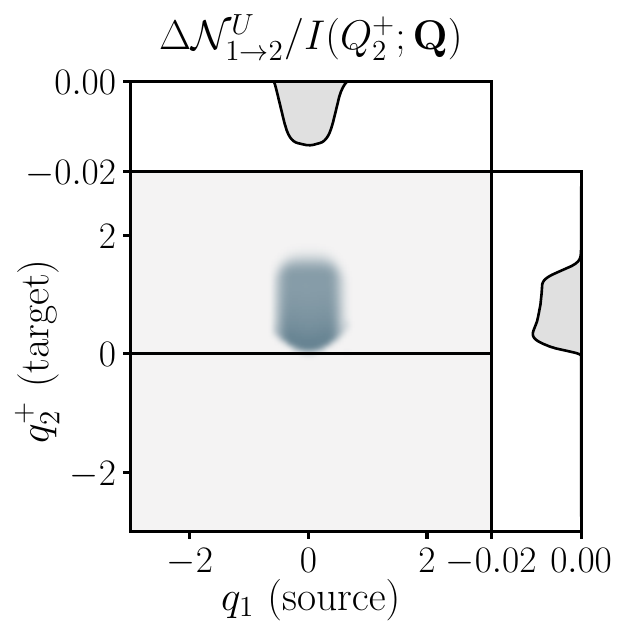}};
    \end{tikzpicture}
    \end{minipage}
    \begin{minipage}{0.31\textwidth}
    \begin{tikzpicture}
        \node[anchor=north west] (img) at (0, 0) {\includegraphics[width=\linewidth, trim=0 0 0 0, clip]{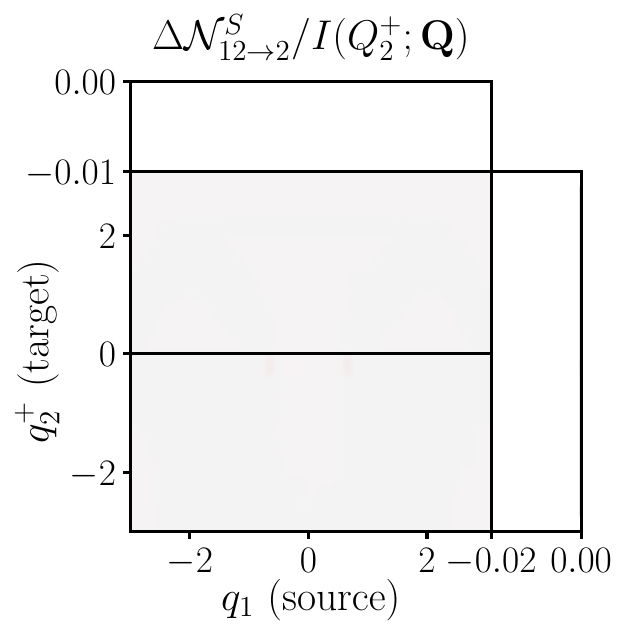}};
    \end{tikzpicture}
    \end{minipage}
    \end{center}
    \end{minipage}    
    
    \caption{Non-causal redundant (R), unique (U), and synergistic (S)
      states for (top) source-dependent and (bottom) target-dependent
      benchmark cases. The maps for redundant and synergistic
      components illustrate (top) the sum over all states $q_2$ and
      (bottom) the specific state $q_2$ that maximizes the total sum
      of individual contributions.}
    \label{fig:benchmark-nc}
\end{figure}

First, we examine the non-causal components for the two benchmark
systems (source-dependent and target-dependent) discussed in the
validation section of the main text. Figure~\ref{fig:benchmark-nc}
illustrates the decomposition of non-causal states into their
redundant, unique, and synergistic components for the target variable
$Q_2^+$ in both cases.

Starting with the source-dependent system, recall that the causal
states exhibited a strong unique contribution for $q_1 > 0$, where
$Q_1$ uniquely drives $Q_2^+$, as well as redundant and synergistic
causalities for $q_1 < 0$, where $Q_1$ and $Q_2$ jointly influence
$Q_2^+$.  For the non-causal contributions, the redundant component
$\Inn^R_{12 \rightarrow 2}$ is primarily located in the region $q_1 >
0$, where no additional information is gained beyond what is already
contained in the target variable $Q_2^+$. This is consistent with the
fact that, in this regime, no redundancy arises between the $q_1$ and
$q_2$ states.  For the unique component $\Inn^U_{1 \rightarrow 2}$,
the non-causal states are concentrated around $q_1 < 0$, indicating
that the causal influence of $Q_1$ in this region is already captured
by the total redundant contribution $\Delta I^R_{12 \rightarrow 2}$.
Finally, the non-causal synergistic component $\Inn^S_{12 \rightarrow
  2}$ is negligible across the state space.

In the target-dependent system, causal states were previously
identified in the region $q_2^+ > 0$ for unique causality, and in the
region $q_2^+ < 0$ for redundant and synergistic causalities.
Figure~\ref{fig:benchmark-nc} shows the corresponding non-causal
contributions. The redundant non-causal component displays the
opposite trend of its causal counterpart: non-causal states are now
concentrated in the region $q_2^+ > 0$, where no additional
information is gained beyond what is already contained in the target
variable $H(Q_2^+)$. This behavior is consistent with the
functional dependencies of the system, where no overlap of information between $q_1$
and $q_2$ was imposed for $q_2^+ > 0$.  Additionally, non-causal
contributions to the unique component appear in the same region
($q_2^+ > 0$), particularly around $q_1 = 0$, where no causal
influence had been detected in the earlier analysis. This further
confirms that these states contribute little to the predictability of
$Q_2^+$ from $Q_1$ alone.  Finally, the non-causal synergistic
component $\Inn^S_{12 \rightarrow 2}$ remains negligible across the
entire state space.

\subsection{Turbulent boundary layer and Walker circulation}

\begin{figure}[t!]
    \centering
    \begin{minipage}{\textwidth}
    \begin{minipage}{0.32\textwidth}
    \begin{tikzpicture}
        \node[anchor=north west] (img) at (0, 0) {\includegraphics[width=\linewidth]{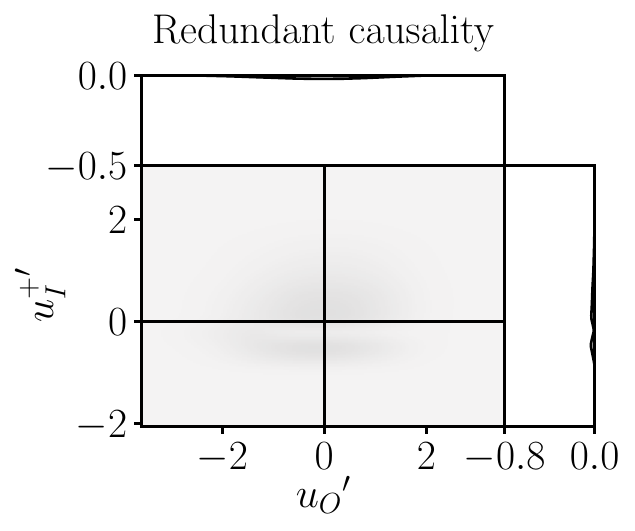}};
        \node[anchor=south, font=\normalsize, fill=white, text=black, inner sep=2pt, rounded corners, minimum width=0.2\textwidth,minimum height=6mm] at ([yshift=-6.5mm,xshift=-0.5mm]img.north) {Redundant non-causality};
    \end{tikzpicture}
    \end{minipage}
    \begin{minipage}{0.32\textwidth}
    \begin{tikzpicture}
        \node[anchor=north west] (img) at (0, 0) {\includegraphics[width=\linewidth, trim=0 0 0 0, clip]{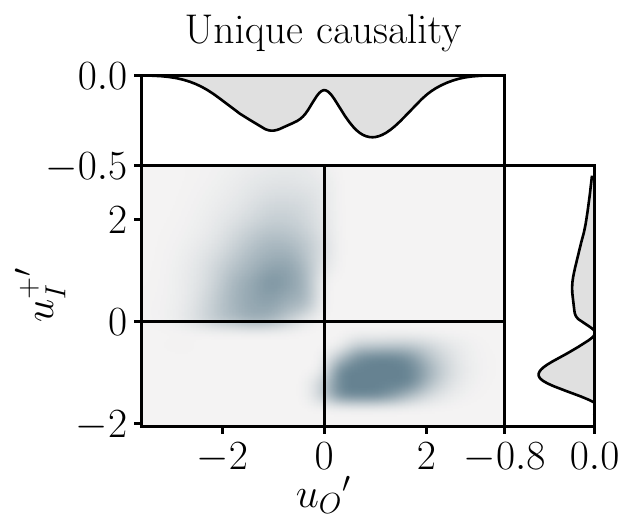}};
        \node[anchor=south, font=\normalsize, fill=white, text=black, inner sep=2pt, rounded corners, minimum width=0.2\textwidth,minimum height=6mm] at ([yshift=-6.5mm,xshift=-0.5mm]img.north) {Unique non-causality};
    \end{tikzpicture}
    \end{minipage}
    \begin{minipage}{0.32\textwidth}
    \begin{tikzpicture}
        \node[anchor=north west] (img) at (0, 0) {\includegraphics[width=\linewidth, trim=0 0 0 0, clip]{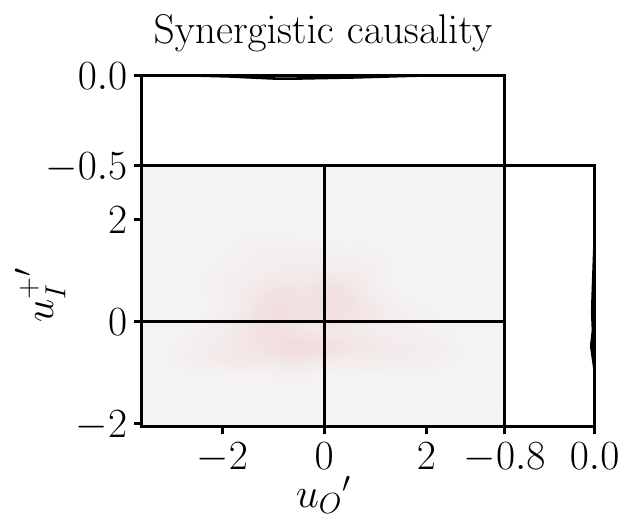}};
        \node[anchor=south, font=\normalsize, fill=white, text=black, inner sep=2pt, rounded corners, minimum width=0.2\textwidth,minimum height=6mm] at ([yshift=-6.5mm,xshift=0mm]img.north) {Synergistic non-causality};
    \end{tikzpicture}
    \end{minipage}
    \end{minipage}
    
    \begin{minipage}{\textwidth}
    \centering
    \hspace{-0.01\textwidth}
    \begin{minipage}{0.24\textwidth}
    \begin{tikzpicture}
        \node[anchor=north west] (img) at (0, 0) {\includegraphics[width=\textwidth]{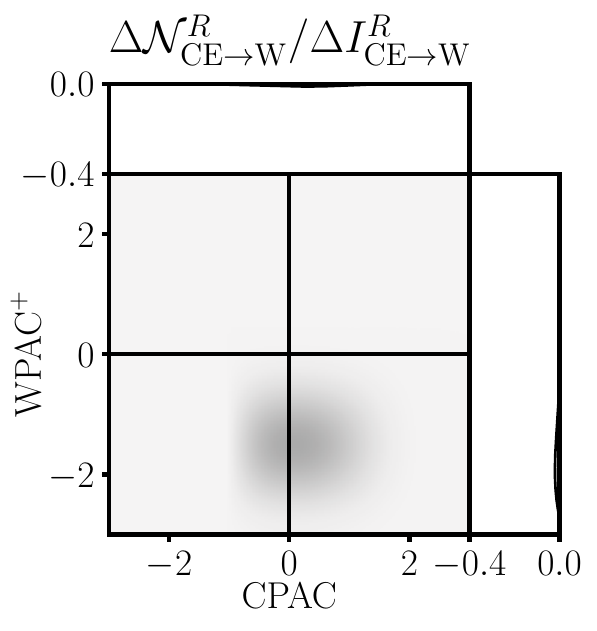}};
        \node[anchor=south, font=\small, fill=white, text=black, inner sep=2pt, rounded corners, minimum width=0.2\textwidth,minimum height=6mm] at ([yshift=-6.25mm,xshift=-1mm]img.north) {$\quad$[CPAC,EPAC]$\to$WPAC$^+\quad$};
        \node[anchor=south, font=\small, fill=white, text=black, inner sep=2pt, rounded corners, minimum width=0.2\textwidth,minimum height=6mm] at ([yshift=-2mm,xshift=-1mm]img.north) {Redundant non-causality};
    \end{tikzpicture}
    \end{minipage}
    \begin{minipage}{0.24\textwidth}
    \begin{tikzpicture}
        \node[anchor=north west] (img) at (0, 0) {\includegraphics[width=\textwidth]{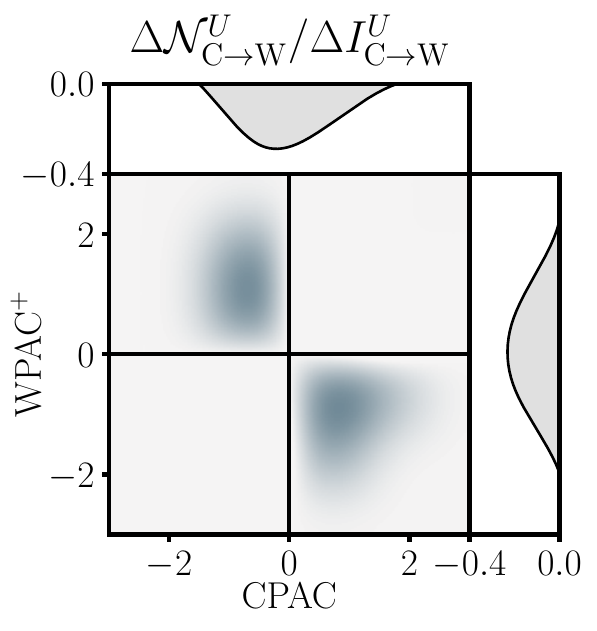}};
        \node[anchor=south, font=\small, fill=white, text=black, inner sep=2pt, rounded corners, minimum width=0.2\textwidth,minimum height=6mm] at ([yshift=-6.25mm,xshift=-1mm]img.north) {$\quad$CPAC$\to$WPAC$^+\quad$};
        \node[anchor=south, font=\small, fill=white, text=black, inner sep=2pt, rounded corners, minimum width=0.2\textwidth,minimum height=6mm] at ([yshift=-2mm,xshift=-1mm]img.north) {Unique non-causality};
    \end{tikzpicture}
    \end{minipage}
    \begin{minipage}{0.24\textwidth}
    \begin{tikzpicture}
        \node[anchor=north west] (img) at (0, 0) {\includegraphics[width=\textwidth]{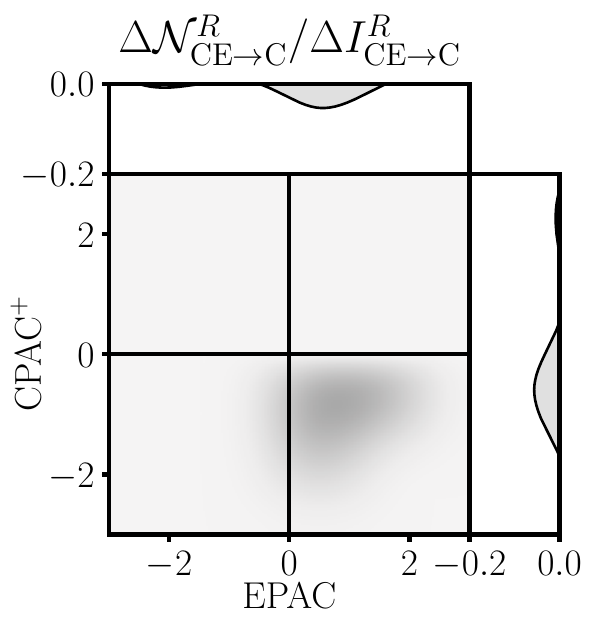}};
        \node[anchor=south, font=\small, fill=white, text=black, inner sep=2pt, rounded corners, minimum width=0.2\textwidth,minimum height=6mm] at ([yshift=-6.25mm,xshift=-1mm]img.north) {$\quad$[CPAC,EPAC]$\to$CPAC$^+\quad$};
        \node[anchor=south, font=\small, fill=white, text=black, inner sep=2pt, rounded corners, minimum width=0.2\textwidth,minimum height=6mm] at ([yshift=-2mm,xshift=-1mm]img.north) {Redundant non-causality};
    \end{tikzpicture}
    \end{minipage}
    \begin{minipage}{0.24\textwidth}
    \begin{tikzpicture}
        \node[anchor=north west] (img) at (0, 0) {\includegraphics[width=\textwidth]{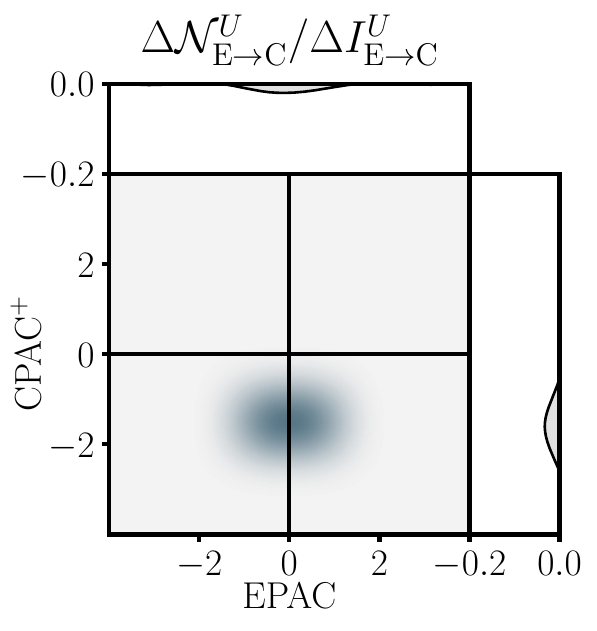}};
        \node[anchor=south, font=\small, fill=white, text=black, inner sep=2pt, rounded corners, minimum width=0.2\textwidth,minimum height=6mm] at ([yshift=-6.25mm,xshift=-1mm]img.north) {EPAC$\to$CPAC$^+$};
        \node[anchor=south, font=\small, fill=white, text=black, inner sep=2pt, rounded corners, minimum width=0.2\textwidth,minimum height=6mm] at ([yshift=-2mm,xshift=-1mm]img.north) {Unique non-causality};
    \end{tikzpicture}
    \end{minipage}
    \end{minipage}
    
    \caption{Non-causal redundant (R), unique (U), and synergistic (S)
      states for the two applications discussed in the main text:
      (top) inner/outer interactions in a turbulent boundary layer and
      (bottom) Walker circulation in the tropical Pacific. Note that
      for some of the panels, the color scale has been intentionally
      saturated to enhance visual contrast; however, the causalities
      are of small magnitude, as evidenced by the horizontal and
      vertical projections.  }
    \label{fig:applications-nc}
\end{figure}

We analyze the decomposition of non-causal contributions to the
causalities in the turbulent boundary layer and Walker circulation
applications, as shown in Fig.~\ref{fig:applications-nc}.

For the turbulent boundary layer, we observe that the redundant and
synergistic non-causal components are essentially negligible. The
dominant contribution arises from the unique non-causal component,
which is primarily concentrated in the second ($u_O^\prime > 0$,
${u_I^+}^\prime < 0$) and fourth ($u_O^\prime < 0$, ${u_I^+}^\prime >
0$) quadrants of the $u_O^\prime$–${u_I^+}^\prime$ space. These
quadrants correspond to states in which the outer- and inner-layer
motions deviate from the mean velocity in opposite
directions---indicating an asynchronous relationship between the two
layers.  Notably, the non-causal states in the fourth quadrant
contribute more strongly to $\Inn_{O\to I}^U$ than those in the second
quadrant. This asymmetry aligns with the flow structures shown in
Fig.~\ref{fig:inn-out-intensities}, where non-causal states are
associated with a lack of structural coherence across the boundary
layer. In these cases, no coherent flow structure spans both regions;
instead, the velocity fields linked to second- and fourth-quadrant
states appear fragmented and detached from the wall.  This
fragmentation reflects a breakdown of spatial coherence and a
corresponding reduction in causal influence from the outer to the
inner layer, consistent with the presence of non-causal states
identified in Fig.~\ref{fig:applications-nc}.

For the Walker circulation cases, the decomposition of non-causal
contributions is also shown in Fig.~\ref{fig:applications-nc}.  For
CPAC to WPAC$^+$, the redundant non-causal component is negligible,
while the unique non-causal contributions dominate the overall
structure.  The unique non-causal contribution from CPAC to WPAC$^+$
is primarily concentrated in the second (CPAC $> 0$, WPAC$^+$ $< 0$)
and fourth (CPAC $< 0$, WPAC$^+$ $> 0$) quadrants of the CPAC–WPAC$^+$
state space. These quadrants correspond to states where the CPAC and
WPAC$^+$ anomalies differ in sign—that is, one variable deviates above
its mean while the other falls below it. This misalignment suggests
that, in these states, the causal influence from CPAC to WPAC$^+$ is
already captured by redundant information.

For EPAC to CPAC$^+$, both redundant and unique non-causal
contribution are relevant.  We observe dominant non-causal
contributions to the redundant interaction from [CPAC, EPAC] to
CPAC$^+$ for positive EPAC anomalies (EPAC$>0$) and negative CPAC$^+$
anomalies (CPAC$^+<0$). The unique non-causal contribution
from EPAC to CPAC$^+$ is primarily associated with states where EPAC
fluctuates around its mean value and CPAC$^+ < 0$. These results
highlight the assymetry between El Niño and La Niña events in the
Walker circulation in the Pacific Region.

\newpage
\section{Other methods for quantifying state-dependent influence among variables}

We review four additional methods that provide state-dependent and/or
time-varying measures of influence among variables: local transfer
entropy (LTE)~\cite{Lizier2014}, specific transfer entropy
(STE)~\cite{Darmon2017specific}, local Granger causality
(LGC)~\cite{Stramaglia2021local}, and time-varying Liang--Kleeman
(TvLK) information flow~\cite{tvlk}. LTE and STE are designed to
quantify interactions among variables in dynamical systems without
assuming an underlying causal structure, whereas LGC and TvLK are
explicitly formulated as causal inference methods. For each approach,
we provide a brief description and, where applicable, compare its
performance against benchmark cases from the original studies to
demonstrate the correct implementation and use of the method.

\subsection{State-dependent conditional transfer entropy}

Transfer entropy (TE) is a non-parametric measure of directional
information transfer between two time series~\cite{schreiber2000}. The
TE between two time signals quantifies how much past values of one
variable help predict the present value of another. Here, we are
describe its multivariate extension, Conditional Transfer Entropy
(CTE)~\cite{verdes2005, lozano2022}, which generalizes this concept by
quantifying the information that a subset of variables $\bQ_\bi$
provides about a target $Q_j^+$, conditioned on the remaining
variables $\bQ_{\cancel{\bi}}$:
\begin{equation}
\label{eq: cte}
    \overline{\mathrm{CTE}}_{\bi\rightarrow j} = H(Q_j^+ |
    \bQ_{\cancel{\bi}}) - H(Q_j^+ | \bQ),
\end{equation}
where $\bQ = [\bQ_{\bi},  \bQ_{\cancel{\bi}}]$. The expression above can
be decomposed into state-dependent contributions following the same
approach proposed in this study:
\begin{equation}
\label{eq:contribution cte}
    {\mathrm{CTE}} _{\bi\rightarrow j} = \sum_{\bq_{\cancel{\bi}}} p(q_j^+,\bq) \log_2 \frac{p(q_j^+ \vert \bq)}{p(q_j^+ \vert \bq_{\cancel{\bi}} )},
\end{equation}
which allows one to identify the states of the source and target
variables that contribute to the information transfer from $\bQ_\bi$
to $Q_j^+$.

For example, in a system with the variables $Q_1$, $Q_2$, and target
$Q_j^+$, CTE simplifies to:
\begin{equation}
  {\mathrm{CTE}} _{2 \rightarrow j} = 
  \sum_{q_1} p(q_j^+, q_1, q_2)
  \log_2\left( \frac{p(q_j^+|q_1,q_2)}{p(q_j^+|q_1)} \right),
\end{equation}
which denotes the ratio between the conditional probabilities $p(q_j^+
| q_1,q_2)$ and $p(q_j^+ | q_1)$ weighted by the joint probability $
p(q_j^+, q_1, q_2)$. Comparing this result with the specific unique
causality defined in Eq.~\eqref{eq:Iss-decomp}, we observe that the
key difference lies in the conditional probabilities: our method uses
$p(q_j^+ \mid q_2)$ and $p(q_j^+ \mid q_1)$, rather than $p(q_j^+ \mid
q_1, q_2)$ and $p(q_j^+ \mid q_1)$. This distinction allows our
approach to more effectively isolate the influence of individual
variable states by conditioning out the effects of other variables.

\subsection{Local transfer entropy}

The localized definition of transfer entropy proposed by Lizier
\textit{et al.}~\cite{lizier2008} follows a similar derivation to the
state-dependent CTE discussed above but focuses on the pointwise
(state-specific) contributions to the expected information
transfer. Accordingly, the local transfer entropy (LTE) from $Q_i$ to
$Q_j^+$ is defined as:

\begin{equation}
\label{eq:lte}
    \mathrm{LTE} _{\bi\rightarrow j} = \log_2 \frac{p(q_j^+ \vert \bq)}{p(q_j^+ \vert \bq_{\cancel{\bi}} )}.
\end{equation}
The approach exhibits several desirable properties: it is identically
zero if and only if the total transfer entropy is zero, and its
average over all source and target states recovers the total transfer
entropy. Individual values can be either positive or negative—although
the overall average is non-negative---with negative values interpreted
as misleading or misinformative contributions.

We employ nearest-neighbor estimators for local transfer entropy based
on the approach outlined in Refs.~\cite{jidt,
  Darmon2017specific}. Specifically, we adopt the
Kraskov-Stögbauer-Grassberger (KSG) estimator for mutual
information~\cite{ksg2004} in the context of TE computation. For a
fixed nearest-neighbor parameter $k$ and a single past time lag, the
estimator of the local transfer entropy is defined as:
\begin{equation}
    \tilde{\rm{LTE}}_{i \to j} =
    \psi(k) + \psi\left(N_{Q_j}(t; \rho_{t,k}) + 1\right) 
    - \psi\left(N_{Q_j^+}(t; \rho_{t,k}) + 1\right) 
    - \psi\left(N_{Q_i \times Q_j}(t; \rho_{t,k}) + 1\right),
\end{equation}
where $\psi(\cdot)$ denotes the digamma function, and $\rho_{t,k}$
represents the distance to the $k$th-nearest neighbor of the state
vector $(Q_i, Q_j, Q_j^+)$ under the infinity norm. The function
$N_\mathcal{S}(t; \rho_{t,k})$ counts the number of sample points
within a radius $\rho_{t,k}$ in the space $\mathcal{S}$. The TE can be
recovered by averaging the LTE over all available samples.  The
parameter $k$ controls the trade-off between bias and variance in the
estimator $\tilde{\mathrm{LTE}}_{i \to j}$. While the KSG-based
estimator is asymptotically unbiased for any fixed $k$, finite sample
sizes introduce a nonzero bias that decreases with smaller values of
$k$. However, smaller $k$ values also lead to increased estimator
variance. Following standard practice, we fix $k = 4$ in all our
analyses.  To project the results onto a selected variable, we adopt
the approach proposed in Ref.~\cite{Darmon2017specific} for
$Q_j^+$. In this setting, the projection is carried out with respect
to a selected variable, enabling the estimation of
$\tilde{\mathrm{LTE}}_{i \to j}$ as a function of both $q_j^+$ and
$q_i$ when projecting onto $q_j$.

\subsection{Specific transfer entropy}

Darmon and Rapp~\cite{Darmon2017specific} introduced specific transfer
entropy (STE), an alternative to LTE that retains many of its
desirable properties while introducing others, such as the non-negativity of individual terms
and a clearer interpretation in terms of the state-dependent predictive impact
of source states on target states. The STE formulation is derived by
applying the law of iterated expectations to TE:
\begin{equation}
\label{eq:ste}
    \mathrm{TE} _{\bi\rightarrow j} = \sum_{q_j^+} \sum_{\bq}
    p(q_j^+,\bq)\log_2 \frac{p(q_j^+ \vert \bq)}{p(q_j^+ \vert
      \bq_{\cancel{\bi}} )} = \sum_{\bq} p(\bq) \sum_{q_j^+}
    p(q_j^+\vert \bq)\log_2 \frac{p(q_j^+ \vert \bq)}{p(q_j^+ \vert
      \bq_{\cancel{\bi}} )},
\end{equation}
Darmon and Rapp~\cite{Darmon2017specific} define STE using the inner
conditional expectation from the decomposition of total transfer
entropy. This quantity corresponds to the Kullback--Leibler divergence
($D_{KL}$) between $p(q_j^+ \mid \bq)$ and $p(q_j^+ \mid
\bq_{\cancel{\bi}})$, and is therefore guaranteed to be non-negative:
\begin{equation}
    \mathrm{STE} _{\bi\rightarrow j} = \sum_{q_j^+} p(q_j^+\vert
    \bq)\log_2 \frac{p(q_j^+ \vert \bq)}{p(q_j^+ \vert
      \bq_{\cancel{\bi}} )} = D_{KL}\left[ p(q_j^+\vert \bq)
      \vert\vert p(q_j^+\vert \bq_{\cancel{\bi}})\right].
\end{equation}

An estimator of STE can be obtained from the LTE by performing a
regression of the LTE values onto the past input-output states. To
achieve this, we follow the approach in Ref.~\cite{Darmon2017specific}
and use a $k_{\text{reg}}$-nearest-neighbor regression as a
nonparametric smoother. Given a nearest-neighbor parameter
$k_{\text{reg}}$, the estimator of the STE is given by:
\begin{equation}
    \tilde{\rm{STE}}_{i \to j} =
    \sum_{t=p+1}^{T} W_{t,k_{\text{reg}}} \cdot \tilde{\rm{LTE}}_{i \to j} (q_j^+, q_i, q_j),
\end{equation}
where the weights $W_{t,k_{\text{reg}}}$ are defined as
${1}/{k_{\text{reg}}}$ if the pair $q_i$-$q_j$ is among the
$k_{\text{reg}}$ nearest neighborr, and zero otherwise. In this
formulation, $k_{\text{reg}}$ controls the degree of smoothing and
balances the bias-variance trade-off in the estimation. Following
standard practice, we set $k_{\text{reg}} = \lfloor \sqrt{T} \rfloor$,
where $T$ is the total number of samples. Notably, unlike kernel
density-based estimators of the specific entropy rate, the
$k$th-nearest-neighbor-based estimator can yield negative STE values.
\begin{figure}
    \centering
    \begin{minipage}{0.32\textwidth}
    \vspace{-4.75cm}
    \begin{tikzpicture}
        \node[anchor=north west] (img) at (0, 0) {\includegraphics[width=\linewidth]{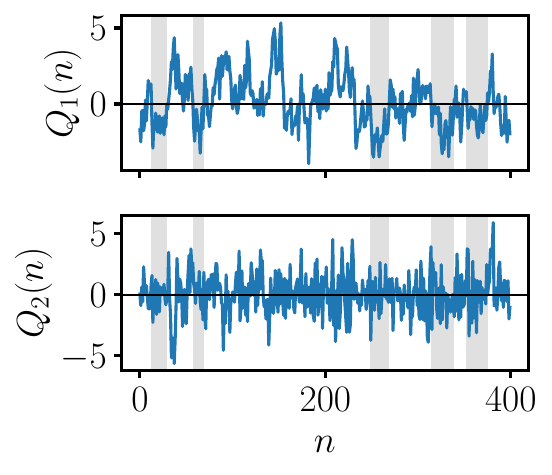}};
        \node[anchor=south, font=\normalsize, fill=white, text=black, inner sep=2pt, rounded corners, minimum width=0.2\textwidth,minimum height=6mm,rotate=90] at ([yshift=15mm,xshift=10mm]img.west) {$\quad\quad\quad$};
        \node[anchor=south, font=\normalsize, fill=white, text=black, inner sep=2pt, rounded corners, minimum width=0.2\textwidth,minimum height=6mm,rotate=90] at ([yshift=15mm,xshift=7.5mm]img.west) {$q_1(n)$};
        \node[anchor=south, font=\normalsize, fill=white, text=black, inner sep=2pt, rounded corners, minimum width=0.2\textwidth,minimum height=6mm,rotate=90] at ([yshift=-5mm,xshift=7.5mm]img.west) {$q_2(n)$};
    \end{tikzpicture}
    \end{minipage}
    \hspace{0.1cm}
    \includegraphics[width=0.32\linewidth]{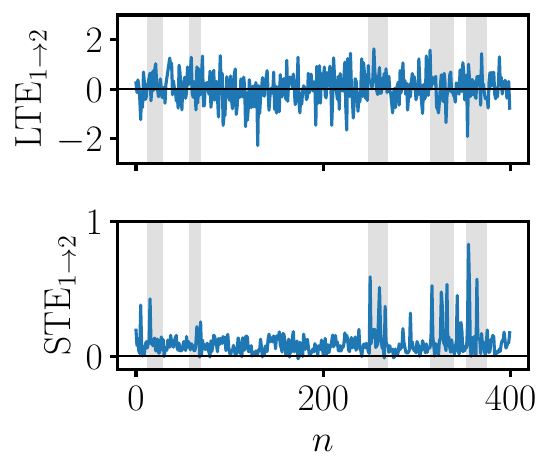}
    \begin{minipage}{0.33\linewidth}
        \vspace{-5.3cm}
        \includegraphics[width=\linewidth]{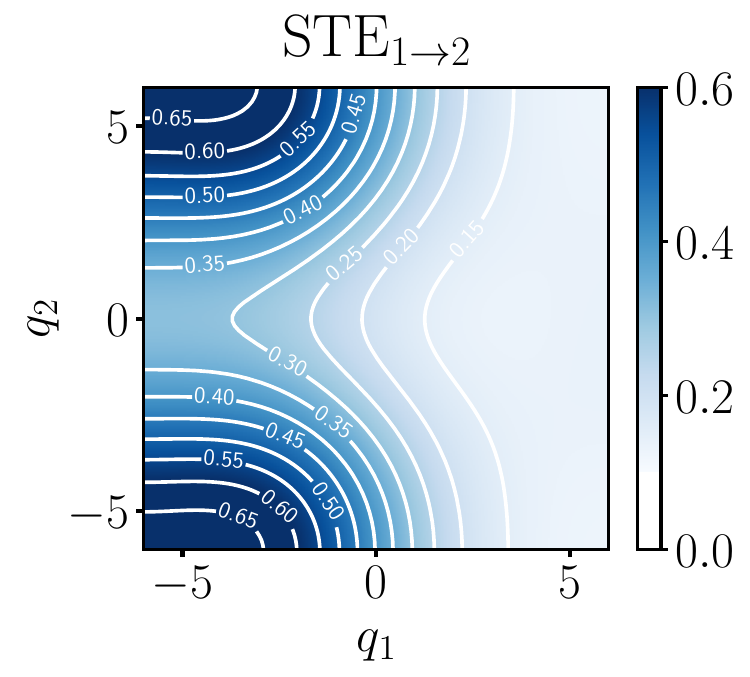}
    \end{minipage}
    \caption{Verification example for local and specific transfer
      entropy from Darmon \& Rapp \cite{Darmon2017specific}. The left
      panels show the temporal evolution of $q_1(n)$ and $q_2(n)$. The
      middle panels show the temporal evolution of local transfer
      entropy (LTE) as given by Eq.~\eqref{eq:lte} and specific
      transfer entropy (STE) as given by Eq.~\eqref{eq:ste}. The
      gray boxes show the regions in which $q_1(n) < 0$ for at least
      10 consecutive steps. The right panel depicts STE as a function
      of the values of source variables $q_1$ and $q_2$.}
    \label{fig:starx}
\end{figure}

We use the implementation and examples from Darmon \& Rapp
\cite{Darmon2017specific} to evaluate the performance of LTE and STE
on the benchmark cases discussed in this study. To verify the
implementation, we first consider the smooth threshold autorregresive
model with an exogenous driver (STARX). The system consists of a
nonlinear autorregresive model that incorporates and exogenous input
($Q_1$) whose value induces a smooth thresholding between two linear
autorregressive models for the output ($Q_2$). Mathematically, the
model can be expressed as:
\begin{align}
    q_1(n+1) &= 0.8 q_1(n) + w_1(n) \\
    q_2(n+1) &= \phi\left[q_1(n)\right] \left[ 0.5 q_2(n) + 2 w_2(n)\right] + \{1 - \phi\left[q_1(n)\right] \} \left[-0.5 q_2(n) + w_2(n)\right],
\end{align}
where $W_1$ and $W_2$ are mutually independent, identically
distributed standard normal random variables. The function
$\phi\left[q_1(n)\right]$ controls the switching between two
autoregressive models and corresponds to a sigmoidal function obtained
by rescaling the cumulative distribution function.
Figure~\ref{fig:starx} displays a segment of the time evolution of
$Q_1$ and $Q_2$, where gray boxes highlight intervals during which
$Q_1$ remains negative for more than 10 consecutive time steps. We
observe that when $q_1 < 0$, the output $Q_2$ exhibits low-amplitude
oscillations, whereas for $q_1 > 0$, the oscillations in $Q_2$ are
amplified.  Additionally, Fig.~\ref{fig:starx} shows the results for
LTE and STE computed for this system. As expected, LTE values
fluctuate between positive and negative, while STE remains strictly
non-negative. Peaks in STE align with the shaded regions (i.e.,
intervals where $q_1 < 0$). When examining STE as a function of both
$q_1$ and $q_2$ states, we find that the most significant influence
from $q_1$ to $q_2$ occurs when $q_1 < 0$ and $\lvert q_2 \rvert >
2.5$.  These results are consistent with the results reported by
Darmon and Rapp~\cite{Darmon2017specific}, which verifies our
implementation.

\subsection{Local Granger causality}

Local Granger causality (LGC), proposed by Stramaglia \textit{et
  al.}~\cite{Stramaglia2021local}, aims at providing a linear
counterpart of LTE. In this case, the method is obtained from a vector
autoregressive (VAR) model. Consider a system with $n$ zero-mean
variables $\bQ=[Q_1,Q_2,\ldots,Q_n]$, which is stationary and ergodic
with time-invariant variances and covariances. The distribution of
$\bQ$ is assumed to be a multivariate Gaussian in the stationary
regime. Given these conditions, the LGC from a vector of source
variables $\bQ_\bi$ to a target variable $Q_j^+$ is defined as:
\begin{equation}
\label{eq:lgc}
\mathrm{LGC}_{\bi \rightarrow j}(q_j^+, \bq_\bi,\bq_{\cancel{\bi}}) = {\rm{GC}}_{\bi \rightarrow j} + \mathcal{L}(q_j^+,\bq_\bi, \bq_{\cancel{\bi}}),
\end{equation}
where $\bq_{\cancel{\bi}}$ represents the states of the vector of
variables in $\bQ$ that are not $\bQ_\bi$. In the previous equation,
the first term is independent of the states $q_j^+$ and $\bq_\bi$, and
corresponds to the classical Granger causality. In contrast, the term
$\mathcal{L}(q_j^+, \bq_\bi)$ captures the instantaneous,
state-specific fluctuations around this baseline
value~\cite{Stramaglia2021local}. Specifically, these terms can be
mathematically expressed as:
\begin{equation}
    {\rm{GC}}_{\bi \rightarrow j} = \log \frac{|\mathbf{\Sigma}_{{\cancel{\bi}}\bi}| |\mathbf{\Sigma}_{j\cancel{\bi}}|}{|\mathbf{\Sigma}_{j{\cancel{\bi}}\bi}| |\mathbf{\Sigma}_{\cancel{\bi}}|},
\end{equation}
where $|\cdot|$ stands for the determinant of a matrix, and
\begin{equation}
    \mathcal{L}(q_j^+,\bq_\bi, \bq_{\cancel{\bi}}) = \mathbf{Z}_{{\cancel{\bi}}\bi}^\top \mathbf{\Sigma}_{{\cancel{\bi}}\bi}^{-1} \mathbf{Z}_{{\cancel{\bi}}\bi} + \mathbf{Z}_{j{\cancel{\bi}}}^\top \mathbf{\Sigma}_{j{\cancel{\bi}}}^{-1} \mathbf{Z}_{j{\cancel{\bi}}} - \mathbf{Z}_{j{\cancel{\bi}}\bi}^\top \mathbf{\Sigma}_{j{\cancel{\bi}}\bi}^{-1} \mathbf{Z}_{j{\cancel{\bi}}\bi} - \bq_{\cancel{\bi}}^\top \mathbf{\Sigma}_{\cancel{\bi}}^{-1} \bq_{\cancel{\bi}},
\end{equation}
where $\mathbf{Z}_{{\cancel{\bi}}\bi} = [\bq_{\cancel{\bi}}^\top
  \bq_\bi^\top]^\top$, $\mathbf{Z}_{j{\cancel{\bi}}} = [q_j^+
  \bq_{\cancel{\bi}}^\top]^\top$, and $\mathbf{Z}_{j{\cancel{\bi}}\bi}
= [q_j^+\bq_{\cancel{\bi}}^\top \bq_\bi^\top]^\top$ are the
observations of the present and past states of the variables, and
$\mathbf{\Sigma}_{{\cancel{\bi}}\bi}=\mathbb{E}[\mathbf{Z}_{{\cancel{\bi}}\bi}\mathbf{Z}_{{\cancel{\bi}}\bi}^\top]$,
$\mathbf{\Sigma}_{j{\cancel{\bi}}}=\mathbb{E}[\mathbf{Z}_{j{\cancel{\bi}}}\mathbf{Z}_{j{\cancel{\bi}}}^\top]$,
$\mathbf{\Sigma}_{j{\cancel{\bi}}\bi}=\mathbb{E}[\mathbf{Z}_{j{\cancel{\bi}}\bi}\mathbf{Z}_{j{\cancel{\bi}}\bi}^\top]$,
and $\mathbf{\Sigma}_{\cancel{\bi}}=\mathbb{E}[\bq_{\cancel{\bi}}
  \bq_{\cancel{\bi}}^\top]$ are the covariance matrices. Importantly,
$\mathrm{LGC}_{\bi \rightarrow j}$ retains the property that its
time-average recovers the standard Granger causality, since the term
$\mathcal{L}(q_j^+,\bq_\bi, \bq_{\cancel{\bi}})$ has a vanishing
expected value. However, individual local values can be either
positive or negative. According to Stramaglia \textit{et al.}
\cite{Stramaglia2021local}, negative LGC values indicate instances
where the knowledge of the source variables misleads the prediction of
the target.

\begin{figure}
    \centering
    \begin{minipage}{0.7\textwidth}
    \hspace{-0.15cm}
    \begin{tikzpicture}
        \node[anchor=north west] (img) at (0, 0) {\includegraphics[width=\linewidth]{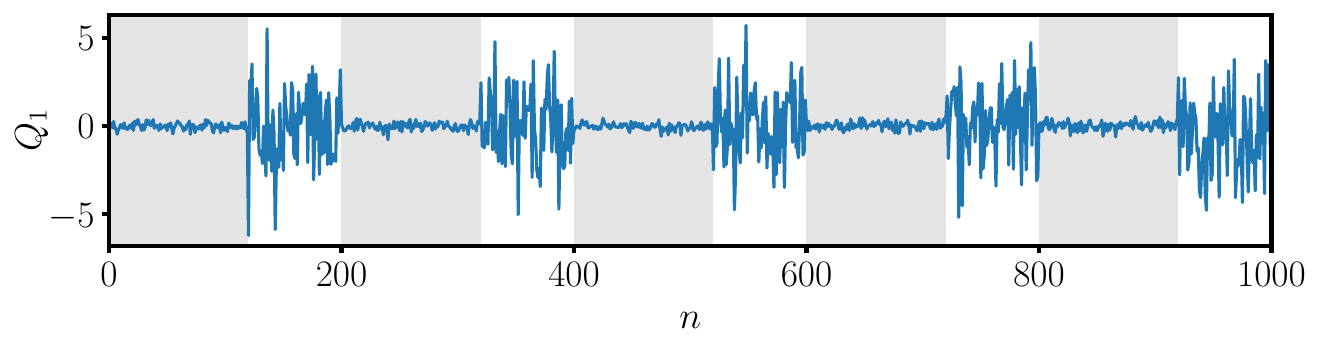}};
        \node[anchor=south, font=\normalsize, fill=white, text=black, inner sep=2pt, rounded corners, minimum width=0.1\textwidth,minimum height=0mm,rotate=90] at ([yshift=4mm,xshift=6mm]img.west) {$q_1(n)$};
    \end{tikzpicture}
    \end{minipage}
    \includegraphics[width=0.71\linewidth]{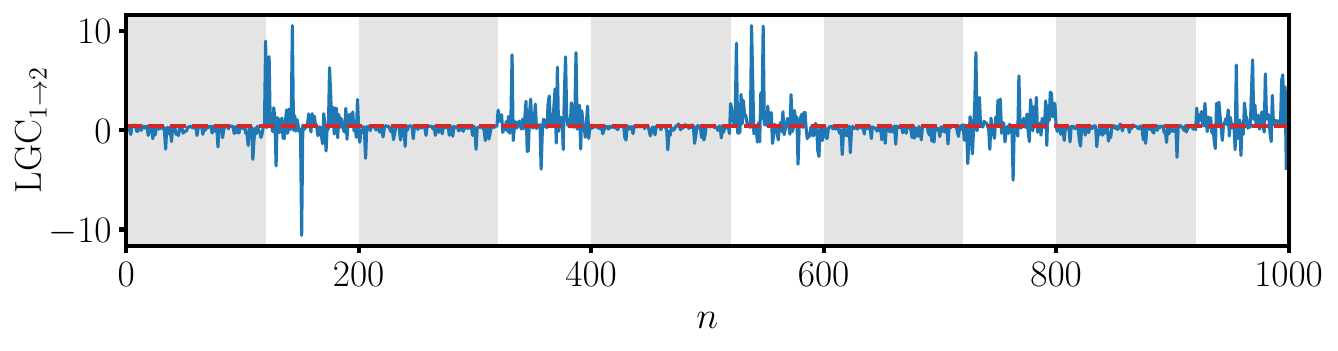}
    \caption{Time evolution of (top) $q_1(n)$ and (bottom) local
      Granger causality (LGC) from $Q_1$ to $Q_2^+$ for the toy model
      described in Eq.~\eqref{eq:lgc_toy}. The dashed line in the
      bottom panel denotes the total Granger causality GC$_{1\\to
        2}$. Light gray boxes denote periods where $\sigma_1 = 0.2$,
      while white intervals correspond to $\sigma_1 = 2$.}
\label{fig:lgc-toy}
\end{figure}

To estimate LGC in practice, we first fit a VAR model globally to the
data, selecting the model order via the Akaike information
criterion. We then compute the covariance matrices required for the
evaluation of Eq.~\eqref{eq:lgc}. To verify the implementation of the
method, we consider the following toy model with two variables $\bQ =
[Q_1, Q_2]$ reported by Stramaglia \textit{et
  al.}~\cite{Stramaglia2021local}:
\begin{align} \label{eq:lgc_toy}
q_1(n) &= w_1(n), \\
q_2(n) &= 0.2\, q_2(n-1) + 0.4\, q_1(n-1) + w_2(n),
\end{align}
where $w_1(n)$ and $w_2(n)$ are independent white noise terms with
standard deviation of $\sigma_{1}$ and $\sigma_{2}$, respectively. In
particular, $\sigma_1$ alternates between 0.2 and 2 at the switching
times $n \in \{120, 200, 320, 400, 520, 600, 720, 800, 920\}$, while
$\sigma_2 = 0.8$ remains constant throughout. An excerpt of the
evolution of $Q_1$ is shown in Fig.~\ref{fig:lgc-toy}. This system
establishes a unidirectional causal relationship from $q_1(n)$ to
$q_2^+=q_2(n+1)$, which is modulated by the strength of the noise
acting on $Q_1$.

The LGC analysis for this system is presented in
Fig.~\ref{fig:lgc-toy}. The results exhibit pronounced fluctuations
that span both positive and negative values, particularly during
intervals where $\sigma_{1}$ is large. Notably, large negative values
of local Granger causality arise when the noise term $w_2(n)$ drives
the system in opposition to the causal influence of
$q_1(n-1)$. According to the authors, such instances render the
driver's contribution misinformative with respect to the target.
Conversely, large positive fluctuations occur when the noise and the
source variable act in the same direction, reinforcing the causal
relationship. As discussed by Stramaglia \textit{et
  al.}~\cite{Stramaglia2021local}, these fluctuations reflect not
merely variations in the noise amplitude, but rather the interplay
between stochastic forcing and the underlying driver dynamics. The
results reported here are consistent with those previously described
by Stramaglia \textit{et al.}~\cite{Stramaglia2021local}, supporting
the correct implementation of LGC.

\subsection{Time-varying Liang--Kleeman information flow}

The last causal inference method considered in this study that
supports time-resolved analysis is based on the concept of
Liang--Kleeman (LK) information flow~\cite{liang2016if}.  Consider a
$n$-dimensional stochastic system governed by:
\begin{equation}
\label{eq:stochastic_system}
\rm{d}\bq = \mathbf{F}(\bq, t)\,\rm{d}t + \mathbf{B}(\bq, t)\,\rm{d}\mathbf{w},
\end{equation}
where $\mathbf{F}=[F_1,F_2,\cdots]$ is the drift vector representing
the deterministic dynamics, $\mathbf{B}$ is a matrix describing the
stochastic perturbations, and $\mathbf{w}$ is a vector of independent
Wiener processes. Within this framework, the rate of information flow
from a source variable $Q_2$ to a target variable $Q_1$ quantifies the
amount of entropy contributed by $Q_2$ to the marginal entropy
evolution of $Q_1$. Mathematically, the information flow from $Q_2$ to
$Q_1$ is given by:
\begin{equation}
\label{eq:lk_info_flow}
T_{2 \rightarrow 1} = -\mathbb{E}\left[\frac{1}{p_1} \frac{\partial (F_1 p_1)}{\partial q_1}\right] + \frac{1}{2}\mathbb{E}\left[\frac{1}{p_1} \frac{\partial^2\left[(b_{11}^2+b_{12}^2) p_1\right]}{\partial q_1^2}\right],
\end{equation}
where $p_1$ is the marginal probability density function of $Q_1$,
$F_1$ is the drift component associated with $Q_1$ and $\mathbb{E}$ is
the expection operator. The information flow is asymmetric; $T_{2
  \rightarrow 1}\neq T_{1 \rightarrow 2}$, and it guarantees that if
the evolution of $Q_1$ is independent of $Q_2$, then $T_{2 \rightarrow
  1}=0$. In the special case where the system is linear, i.e.,
$\mathbf{F}(\bq) = \bs{f} + A\mathbf{q}$ and $\mathbf{B}$ is constant,
and the joint distribution of $[Q_1, Q_2]$ follows a bivariate
Gaussian distribution, Eq.~\eqref{eq:lk_info_flow} simplifies to:
\begin{equation}
\label{eq:linear_lk_flow}
T_{2 \rightarrow 1} = \frac{\sigma_{12}}{\sigma_{11}} a_{12},
\end{equation}
where $\sigma_{11}$ and $\sigma_{12}$ are elements of the covariance
matrix of $[Q_1, Q_2]$, and $a_{12}$ is the coefficient representing
the direct effect of $Q_2$ on $Q_1$ in the drift term. In practice, we
do not know the parameters of the system, but an estimate can be
obtained from time series:
\begin{equation}
\label{eq:empirical_lk_flow}
T_{2 \rightarrow 1} =  \frac{\mathbf{\Sigma}_{12}\mathbf{\Sigma}_{11}\mathbf{\Sigma}_{2,d1} - \mathbf{\Sigma}_{12}^2\mathbf{\Sigma}_{1,d1} }{\mathbf{\Sigma}_{11}^2\mathbf{\Sigma}_{22} - \mathbf{\Sigma}_{11}\mathbf{\Sigma}_{12}^2},
\end{equation}
where $\mathbf{\Sigma}_{ij}$ denotes the sample covariance between
$Q_i$ and $Q_j$, and $\mathbf{\Sigma}_{i,dj}$ is the sample covariance
between $Q_i$ and the time derivative of $Q_j$ estimated via finite
differences.

The time-varying extension of the LK formalism, known as
TvLK~\cite{tvlk}, was developed by dynamically estimating covariances
at each time step using a Kalman filter.  Specifically, let
$\mathbf{P}_{ij}(t)$ denote the covariance between variables $Q_i$ and
$Q_j$ at time $t$, estimated via a square-root Kalman filter. The
time-varying causality from $Q_2$ to $Q_1$ is then given by:
\begin{equation}
\label{eq:tvlk_21}
T_{2 \rightarrow 1}(t) =  \frac{\mathbf{P}_{12}\mathbf{P}_{11}\mathbf{P}_{2,d1} - \mathbf{P}_{12}^2\mathbf{P}_{1,d1} }{\mathbf{P}_{11}^2\mathbf{P}_{22} - \mathbf{P }_{11}\mathbf{P}_{12}^2},
\end{equation}
Here, $\mathbf{P}_{i,dj}(t)$ represents the covariance between $Q_i$
and the derivative of $Q_j$ at time $t$.

We adopt the implementation described in Ref.~\cite{tvlk} to estimate
the TvLK information flow, applying a uniformly weighted moving
average (UWMA) filter with a window length of 10 points and a sliding
estimation window of 100 points. We verified that varying the
estimation window length between 30 and 150 points had no significant
impact on the results or the conclusions drawn from the analysis.

\subsection{Comparison of methods}

\begin{figure}[t!]
    \centering
    \begin{minipage}{\textwidth}
    \centering
        $
          q_2(n+1) =  
          \begin{cases}
            \sin\left[q_1(n)\right] + 0.1 w_2(n) & \text{if } q_1(n) > 0 \\
            0.9 q_2(n) + 0.1 w_2(n) & \text{otherwise}
          \end{cases}
        $
    \end{minipage}
    \vspace{0.0075\textwidth}
    \includegraphics[width=0.335\linewidth]{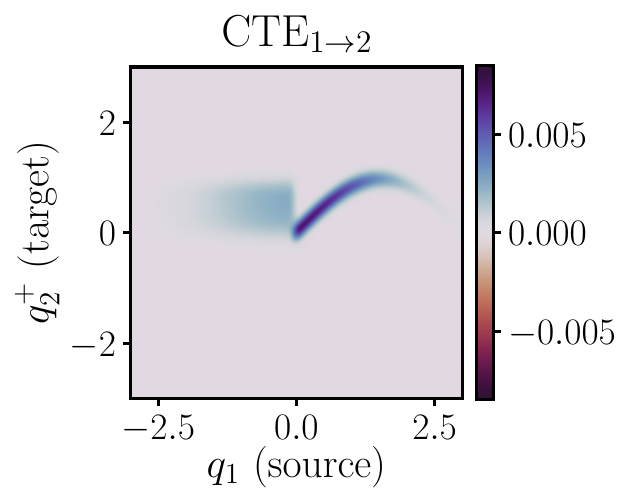}
    \includegraphics[width=0.31\linewidth]{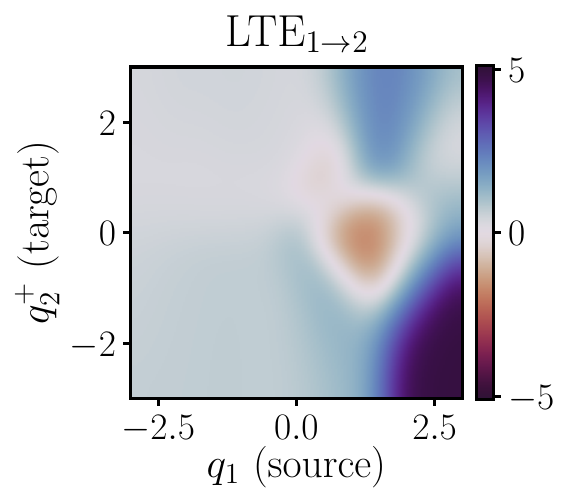}
    \includegraphics[width=0.31\linewidth]{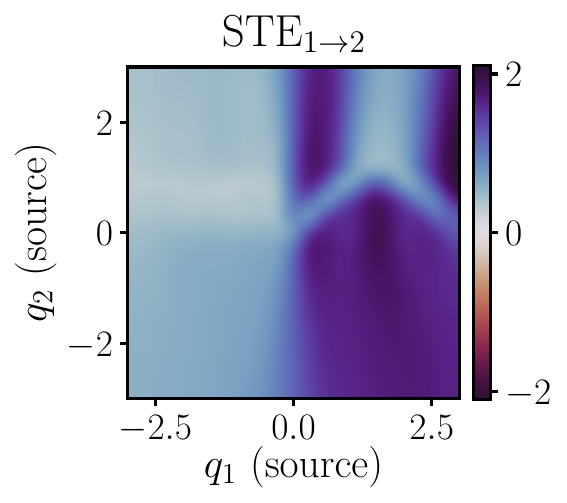}
    \vspace{0.0075\textwidth}

    \begin{minipage}{\textwidth}
    \centering
        $
          q_3(n+1) =  
          \begin{cases}
            \cos\left[q_1(n)\right] + 0.1 w_3(n) & \text{if } q_1(n) < 0 \\
            0.9 q_3(n) + 0.1 w_3(n) & \text{otherwise}
          \end{cases}
        $
    \end{minipage}
    \vspace{0.0075\textwidth}
    \includegraphics[width=0.335\linewidth]{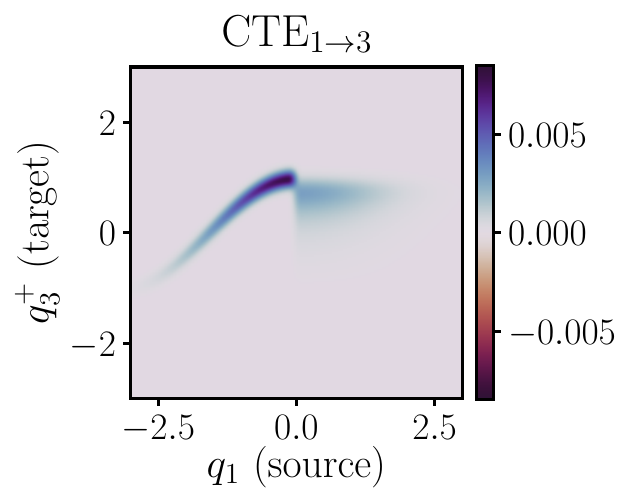}
    \includegraphics[width=0.31\linewidth]{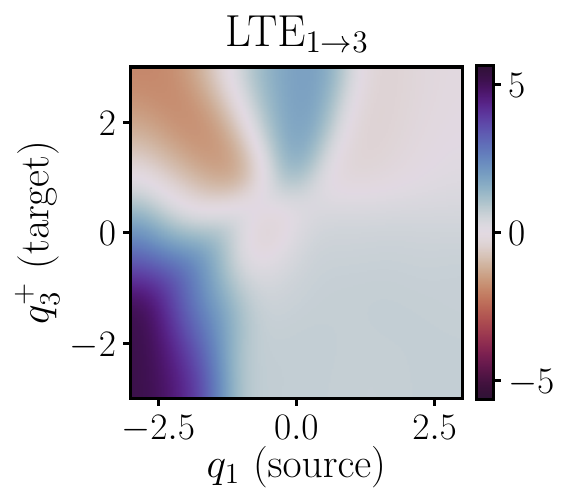}
    \includegraphics[width=0.31\linewidth]{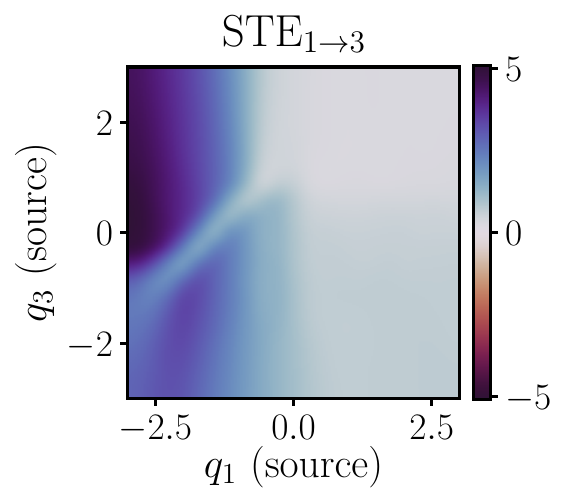}
    \caption{Results for the source-dependent system using conditional
      transfer entropy (CTE), local transfer entropy (LTE), and
      specific transfer entropy (STE).}
    \label{fig:source-dependent-unique-te}
\end{figure}

A summary of the time-varying results obtained from existing methods
for the benchmark cases was already provided in the main text. Here,
we offer an additional comparison limited to those cases where the
methods allow for a representation of causal influence on a
source-target state map, similar to that used in our approach. These
methods are CTE, LTE, and STE. For LGC and TvLK, such a map could in
principle be constructed by averaging the time-resolved estimates over
state bins. However, since these methods do not explicitly incorporate
the same notion of state-dependent causality, we exclude them from
this comparison. As discussed in the main text, we hypothesize that
LGC would exhibit behavior similar to LTE in linear systems, while the
estimates from TvLK information flow may lack the sensitivity required
to capture the abrupt, short-time-scale causal transitions present in
our benchmark cases.

We present the results for the source- and target-dependent benchmark
cases analyzed in Fig.~\ref{fig:example-contributions} and
\ref{fig:example-contributions-target} using CTE, LTE, and STE
methods. We also include in the comparison the results for the
additional target variable $Q_3^+$, whose results using our proposed
method were discussed in
\S\ref{sec:validation}. Figure~\ref{fig:source-dependent-unique-te}
shows the results for the source-dependent case. We observe that CTE
successfully detects the sinusoidal relationship between $q_1$ and
$q_2^+$ for $q_1 > 0$, while also identifying that $q_1 < 0$ provides
some information about $q_2^+$. However, CTE cannot clearly
distinguish between the different types of causality in each case. In
particular, CTE cannot detect that $q_1$ is the sole driver of $q_2^+$
for $q_1>0$ (unique causality) and that $q_1$ is required together
with $q_2$ to drive $q_2^+$ for $q_1<0$ (synergistic causality). Our
method accurately differentiates these dependencies, distinguishing
between the unique dependency for $q_1 > 0$ and the synergistic
dependency for $q_1 < 0$.

The maps for LTE and STE differ substantially from that of CTE. This
discrepancy arises because LTE and STE are not weighted by the
probability of each state, making them more sensitive to
low-probability regions. As a result, their maps can be heavily
influenced by rare states.  Despite this difference, the highest
values of $\text{LTE}_{1 \to 2}$ and $\text{STE}_{1 \to 2}$ occur for
$q_1 > 0$, which is consistent with the relationship $q_1 \to q_2^+$
for $q_1 < 0$. Furthermore, both LTE and STE exhibit non-zero values
across a range of states, reflecting the fact that $q_1$ is also
required to drive the future state $q_2^+$ when $q_1 < 0$.  A similar
conclusion holds for $q_3^+$ across all three methods: the most
significant contributions are observed in the region where $q_1 < 0$,
again aligning with the underlying functional dependencies.

\begin{figure}
    \centering
    \begin{minipage}{\textwidth}
    \centering
        $
          q_2(n+1) =  
          \begin{cases}
            q_1(n) \sin\left[q_1(n)\right] + 0.1 w_2(n) & \text{if } q_1(n) \sin\left[q_1(n)\right] + 0.1 w_2(n) > 0 \\
            q_2(n) + 0.1 w_2(n) & \text{otherwise}
          \end{cases}
        $
    \end{minipage}
    \vspace{0.0075\textwidth}
    \includegraphics[width=0.335\linewidth]{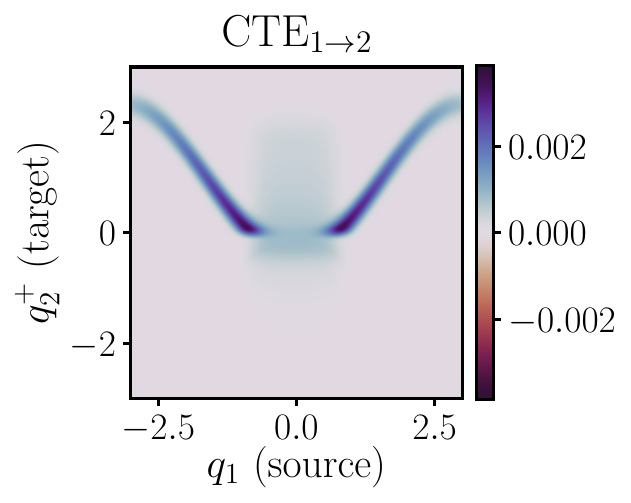}
    \includegraphics[width=0.31\linewidth]{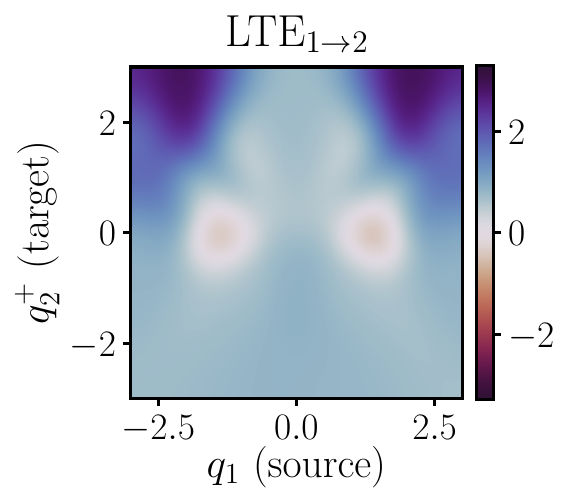}
    \includegraphics[width=0.31\linewidth]{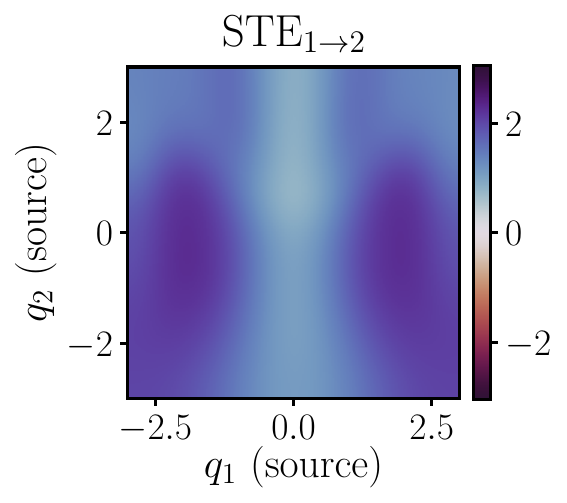}
    \vspace{0.0075\textwidth}

    \begin{minipage}{\textwidth}
    \centering
        $
          q_3(n+1) =  
          \begin{cases}
            q_1(n) \cos\left[q_1(n)\right] + 0.1 w_3(n) & \text{if } q_1(n) \cos\left[q_1(n)\right] + 0.1 w_3(n) < 0 \\
            q_3(n) + 0.1 w_3(n) & \text{otherwise}
          \end{cases}
        $
    \end{minipage}
    \vspace{0.0075\textwidth}
    \includegraphics[width=0.335\linewidth]{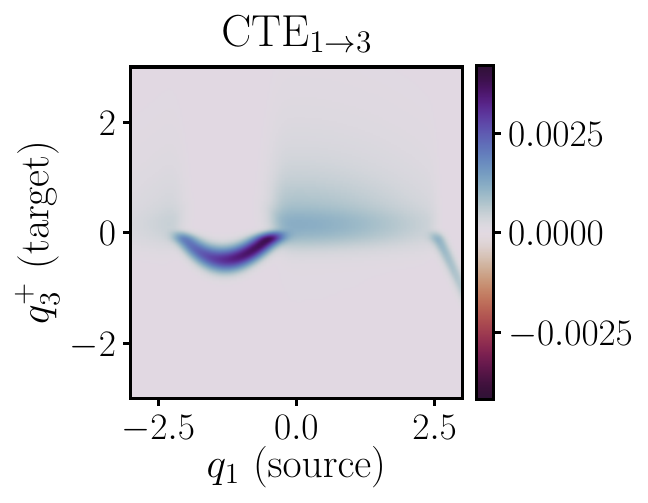}
    \includegraphics[width=0.31\linewidth]{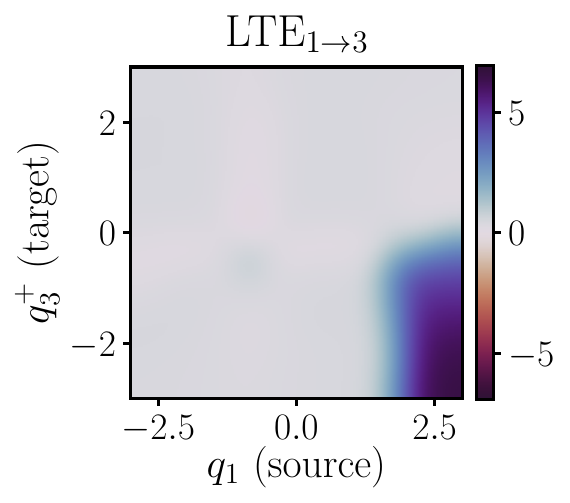}
    \includegraphics[width=0.31\linewidth]{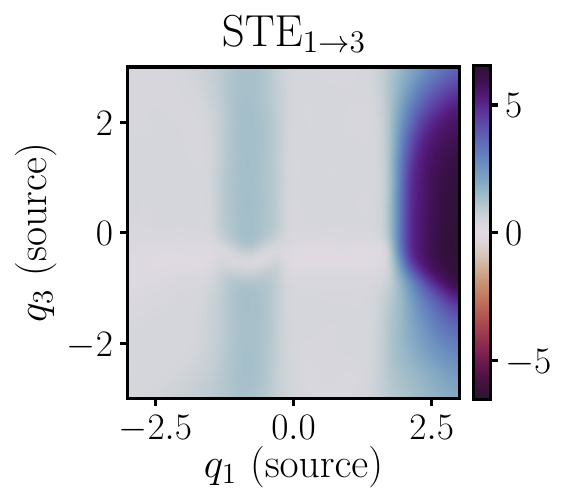}
    \caption{Results for the target-dependent system using conditional transfer entropy (CTE), local transfer entropy (LTE), and specific transfer entropy (STE).}
    \label{fig:target-dependent-te}
\end{figure}

Next, we analyze the results for the target-dependent system. The
results from CTE, LTE and STE are shown in
Fig.~\ref{fig:target-dependent-te}. In this case, CTE successfully
detects the non-linear relationship that occurs for $q_2^+ >
0$. However, it cannot fully differentiate between all types of
causality, which leads to the identification of some information
transfer even for $q_2^+ < 0$. According to the results from our
method, this information transfer corresponds to redundant and
synergistic causalities, while the unique causality from $q_1$ to
$q_2^+$ occurs exclusively for $q_2^+ > 0$. This highlights the
importance of disentangling causal relationships into their
synergistic, redundant, and unique components at the state level. On
the other hand, the most significant values for LTE are observed for
$q_2^+ > 0$, which is consistent with the causal relationship from
$q_1$ to $q_2^+$ occurring in this region. However, STE fails to
capture dependencies based on target states, as it projects the
results from LTE onto $q_2^+$. A similar conclusion can be drawn for
$q_3^+$ across all methods, although the most relevant values are now
observed for $q_3^+ < 0$. Once again, STE is unable to capture
target-dependent relationships.

The main conclusion from these results is that both LTE and STE differ
significantly from the state-dependent CTE and the method proposed in
this study. In particular, the maps generated by LTE and STE are
heavily influenced by rare events with very low
probabilities. However, as noted in the main text, this does not imply
that the results from LTE and STE are incorrect. Instead, they offer a
distinct local perspective on the total transfer entropy, emphasizing
contributions from individual state realizations rather than
probability-weighted averages.


\newpage

\section{Climate science application}
\subsection{Effect of the number of samples}

\begin{figure}[b!]
    \centering
    \subfloat[$N_{\rm{samples}}=780$]{
    \includegraphics[width=0.24\linewidth]{figures/climate_RCEtoW_Nsamples_780.pdf}
    \includegraphics[width=0.24\linewidth]{figures/climate_UCtoW_Nsamples_780.pdf}
    \includegraphics[width=0.24\linewidth]{figures/climate_RCEtoC_Nsamples_780.pdf}
    \includegraphics[width=0.24\linewidth]{figures/climate_UEtoC_Nsamples_780.pdf}}
    
    \subfloat[$N_{\rm{samples}}=520$]{
    \includegraphics[width=0.24\linewidth]{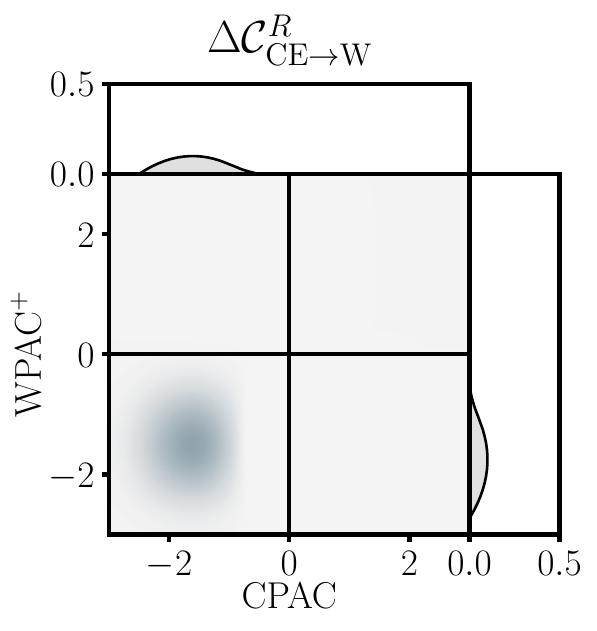}
    \includegraphics[width=0.24\linewidth]{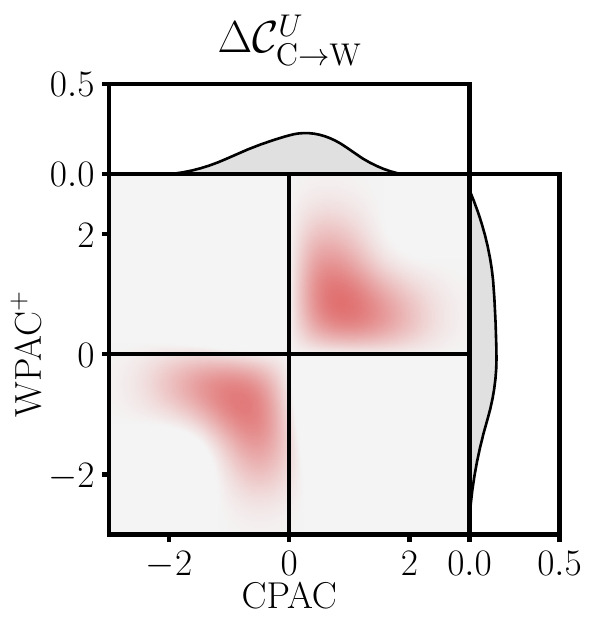}
    \includegraphics[width=0.24\linewidth]{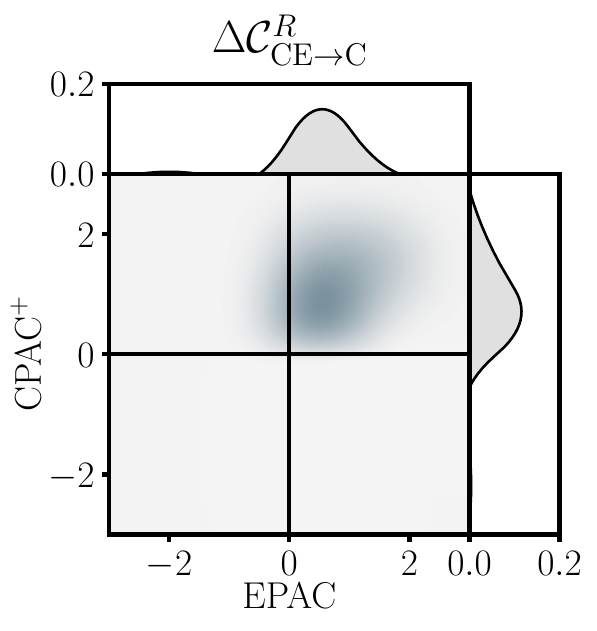}
    \includegraphics[width=0.24\linewidth]{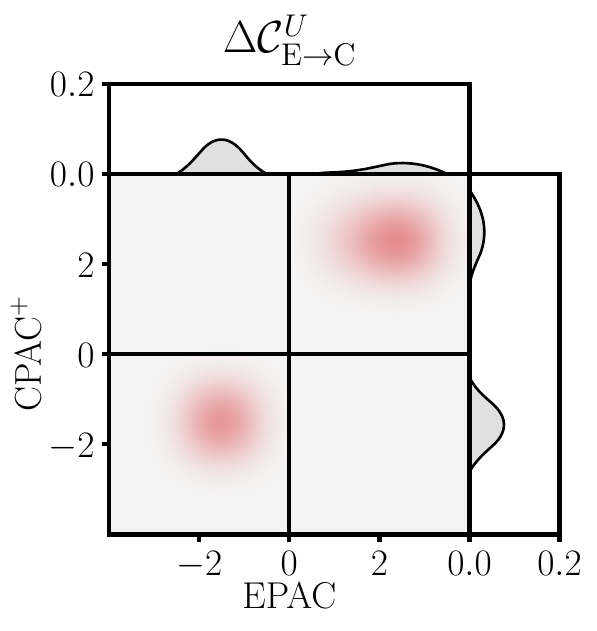}}
    
    \subfloat[$N_{\rm{samples}}=390$]{
    \includegraphics[width=0.24\linewidth]{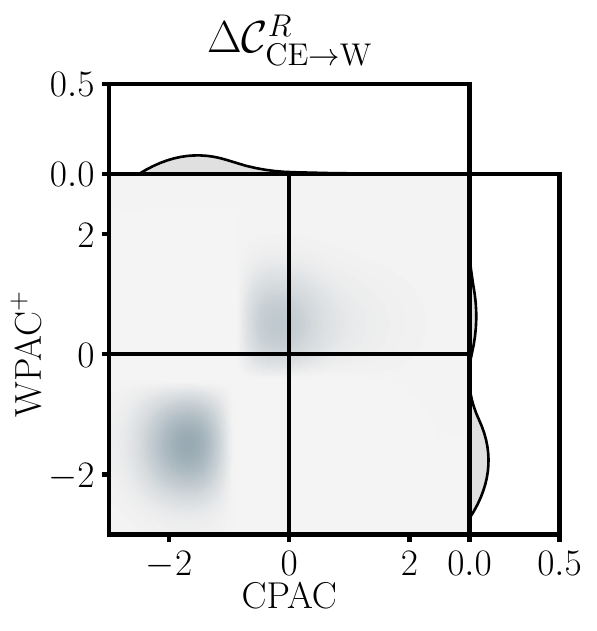}
    \includegraphics[width=0.24\linewidth]{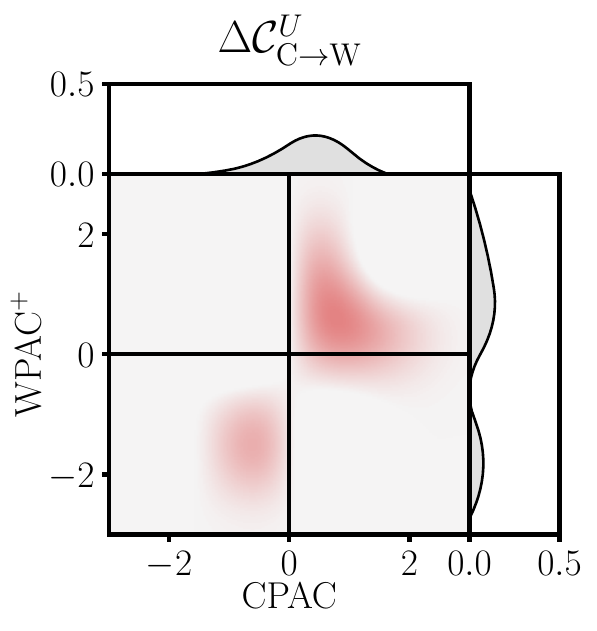}
    \includegraphics[width=0.24\linewidth]{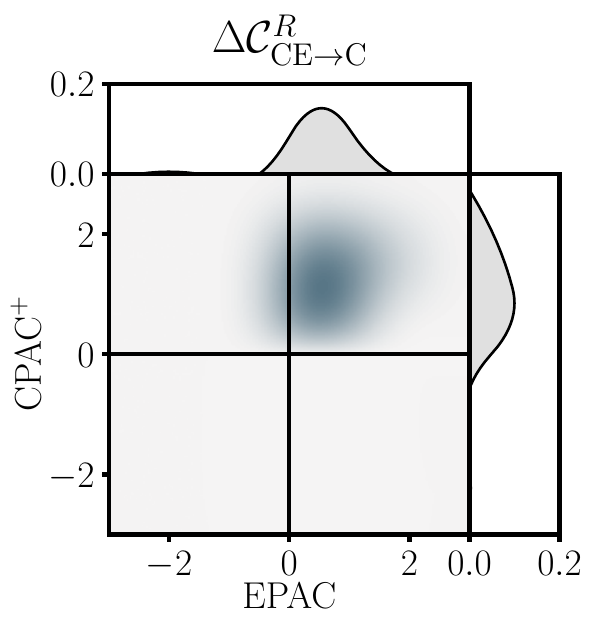}
    \includegraphics[width=0.24\linewidth]{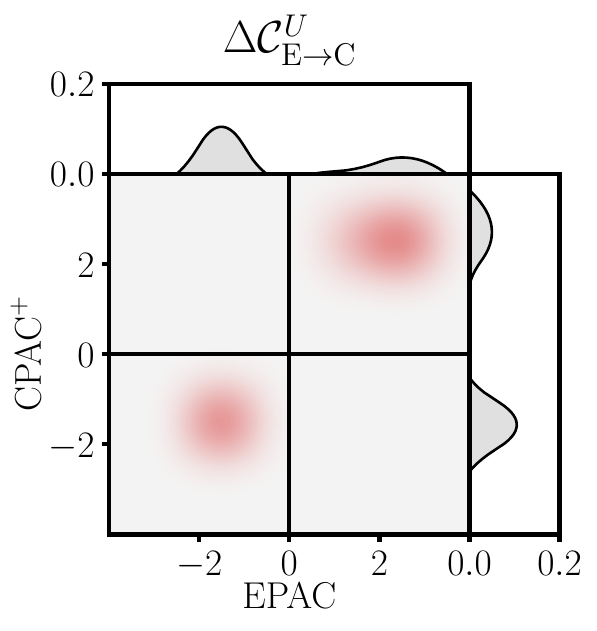}}
    
    \caption{Effect of the number of samples on the state-dependent
      causalities $\Iss_{CE\to W}^R$, $\Iss_{C\to W}^U$, $\Iss_{CE\to
        C}^R$, $\Iss_{E\to C}^U$ for three different sample sizes: (a)
      $N_{\rm{samples}}=780$, (b) $520$, and (b) $390$. The state-dependent 
      causalities are normalized with the mutual information.}
    \label{fig:climate-samples}
\end{figure}

In this section, we examine the impact of sample size on the climate
science application. The climate time series consist of regional
averages from reanalysis data spanning 1948–2012, totaling 780 monthly
observations. This is the same dataset used in
Ref.~\cite{runge2019pcmci}, and includes monthly surface pressure
anomalies in the West Pacific (WPAC), and surface air temperature
anomalies in the Central Pacific (CPAC) and East Pacific (EPAC).

To evaluate causality, we use a time lag of $\Delta T = 4$ months,
which was found to maximize cross-induced unique causality between the
variables. Figure~\ref{fig:climate-samples} shows the evolution of
state-dependent causal contributions across three different sample
sizes ($N_{\rm{samples}} = 390$, 520, and 780). In all cases, the
joint distribution is estimated using a $k$-NN density estimator with
$k=7$ neighbors.

Due to sample size constraints, this analysis focuses exclusively on
the redundant and unique causal contributions, as the estimation of
synergistic causality is considerably more sensitive to data
sparsity. To assess statistical significance, we conducted 1000 random
permutations, yielding a $p$-value of zero for the combined redundant
and unique causalities. In contrast, including the synergistic
component in the test substantially increases the $p$-value, which
highlights its reduced reliability under limited data and justifies
its exclusion from the analysis.

The first key observation from these results is that the overall
structure of the state-dependent causality maps remains consistent as
sample size decreases. This robustness suggests that the main
conclusions drawn from the full dataset are preserved, even under
moderate data limitations. The only noticeable deviations appear at
the lowest sample size, particularly for causal influences to
WPAC$^+$. Specifically, the redundant causality $\Iss^R_{CE \to W}$
exhibits a new, localized region of positive causal values near the
climatological mean of both CPAC and WPAC$^+$. For the unique
causality $\Inn^U_{C \to W}$, we observe a small reduction of causal
states in the region corresponding to WPAC$^+ < 0$ and CPAC $<
0$. This might be an indication that under reduced sample sizes, El
Niño-like states (positive CPAC anomalies) remain more prominently
causal, while La Niña-like states lose some significance. Notably, the
state-dependent causalities to CPAC$^+$ remain largely unchanged
across all sample sizes.